# On the Estimation and Use of Statistical Modelling in Information Retrieval

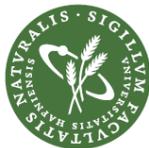

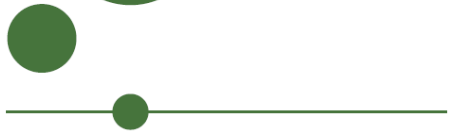

# Acknowledgements

I am deeply indebted to my two supervisors Christina Lioma and Jakob Grue Simonsen for the patience and guidance they have offered me during the past three years. I would also like to extend my thanks to Kalervo Järvelin, University of Tampere, with whom I shared many interesting and entertaining supervision meetings with while in Tampere. I would also like to express my sincere appreciation to my committee members, Professor Stéphane Marchand-Maillet, Professor Fabio Crestani and Associate Professor Aasa Feragen, for their comments which greatly helped improving this thesis.

*Copenhagen, 10 September 2016*                                    Casper Petersen.

# Abstract

Several tasks in information retrieval (IR) rely on assumptions regarding the distribution of some property (such as term frequency) in the data being processed. This thesis argues that such distributional assumptions can lead to incorrect conclusions and proposes a statistically principled method for determining the "true" distribution. This thesis further applies this method to derive a new family of ranking models that adapt their computations to the statistics of the data being processed. Experimental evaluation shows results on par or better than multiple strong baselines on several TREC collections.

Overall, this thesis concludes that distributional assumptions can be replaced with an effective, efficient and principled method for determining the "true" distribution and that using the "true" distribution can lead to improved retrieval performance.

# Resumé

Mange opgaver i information retrieval (IR) beror på formodninger vedrørende fordelingen af en attribut (såsom frekvensen af ord) i et datasæt. Denne afhandling argumenterer for at formodninger vedrørende fordelinger af attributter i IR datasæts kan føre til ukorrekte eller utilstrækkelige konklusioner og præsenterer en statistisk metode til at identificere den "rigtige" statistiske model der bedst passer en given distribution. Denne afhandling anvender efterfølgende denne metode til at udlede en ny familie af "ranking" modeller der tilpasser deres beregninger på baggrund af dataen der behandles. Eksperimentel evaluering på flere TREC datasæts viser at disse ranking modeller opnår tilsvarende eller bedre resultater end flere stærke baselines.

Overordnet konkluderer denne afhandling at formodninger vedrørende fordelinger af attributter i IR datasæt kan erstattes af en effektiv, hurtig og formel metode der kan identificere den "rigtige" statistiske model, og viser at brugen af den "rigtige" statistiske model kan føre til øget effektivitet.

# Contents

## Contents

# List of Figures

# List of Tables

# 1 Introduction

The 1998 blow up of the Long-Term Capital Management (LTCM) private hedge fund is said to have nearly destroyed the world's financial system [227]. The initial "success" of the hedge fund, operated by distinguished professors and Noble laureates in economics, was based on a supposedly flawless computational financial model [277] which allowed the hedge fund to invest billions of dollars with seemingly little risk. However, when LTCM crashed almost USD 100 billion were lost as a result of an error in their financial model [228]; an error that should only take place once every 800 trillion years. The problem with their financial model was that the assessment of the potential for overall losses was based on a Gaussian statistical model, which severely underestimates the chance of a cataclysmic event [81]. Had a more accurate statistical model been used, the loss incurred would have been an 8-year event [154], leading Jorion [228] to conclude that "surely the [statistical] model was wrong".

This "surely the [statistical] model was wrong" is a key point that this thesis will argue also needs to be addressed in information retrieval (IR). While catastrophic events are likely smaller in IR, the use of the "correct" statistical model has largely been overlooked by the IR community. This omission seems to rest on two main observations:

- Making distributional assumptions - that is, assuming that something is approximately distributed according to some statistical model - facilitates development of e.g. predictive models and is often made on the basis of mathematical convenience.[1] However, quoting Rissanen [365]:

  > "*We never want to make the false assumption that the observed data actually were generated by a distribution of some kind, say Gaussian, and then go on to analyze the consequences and make further decisions. Our deductions may be entertaining but quite irrelevant to the task at hand, namely, to learn useful properties from the data*"

  According to this view, a clear interpretation of data is only available under assumptions that are typically violated in practice [185], and caution should be exercised when undertaking inferential procedures on this basis.

- Making distributional assumptions does not seem to significantly affect the performance

---

[1]Strictly speaking, saying that $x$ is normally distributed is a reification of saying that the uncertainty in the values of $x$ is characterised by a normal distribution.

of IR tasks, such as assigning probabilities to unseen terms in language modelling or ranking documents (exceptions are reviewed in Section 2.4).

This thesis argues that the use of the "correct" statistical model in IR is important from both a theoretical and practical perspective:

- From a theoretical perspective, using the "correct" statistical model can improve IR effectiveness and/or efficiency, as we show in Chapter 5

- From a practical perspective, it allows accurate modelling of the phenomena of interest. This is important as formal frameworks are required or used across many applications of IR.

One key reason the use of the "correct" statistical model might be neglected by the IR community is the lack of formal framework that facilitates testing distributional assumptions of data in a principled manner. Developing such a formal framework and applying it to IR is the topic of this thesis.

## 1.1  Thesis Objective

This thesis examines the use and application of statistical model selection in IR. Specifically, this thesis is concerned with the following general research objective:

*How can a principled approach to statistical model selection be used in IR?*

and will focus on (i) the development of a principled approach to statistical model selection, and (ii) its application to IR. The principled approach combines a known and widely used method for distribution fitting, with principled statistical testing to determine, from a set of candidate distributions, the one which *most likely* is the "true" distribution relative to some data sample. The "true" distribution is then used derive new ranking models for ad hoc retrieval.

## 1.2  Research Questions

The following two research questions are examined:

1. **Research Question 1 (RQ1):** To what extent does commonly assumed statistical models fit common properties of IR datasets?

2. **Research Question 2 (RQ2):** Can the use of the correct statistical model improve ranking?

RQ1 is addressed in Chapter 4, and RQ2 in Chapter 5. A principled approach to model selection is the topic of Chapter 3.

## 1.3 Contributions

This thesis contributes knowledge on the topic of statistical model selection for IR in the following two ways:

1. New insights into the validity and limitations of commonly held assumptions about how properties of IR datasets tend to be distributed. Specifically, we contribute knowledge that commonly held assumptions regarding the distribution of properties in IR data are frequently *incorrect* when using ad hoc techniques, and that a principled approach to statistical model selection is a practical alternative.

2. New ranking models that adapt to the distribution of the data they process. Specifically, we contribute knowledge that because these ranking models "closely" approximate the distribution of the data, they (i) are optimised for individual collections, (ii) better conform to the theory on which they are based and (iii) can be used to improve ranking.

## 1.4 Published Work

Parts of this thesis have been peer-reviewed and published in:

**Power Law Distributions in Information Retrieval**
Casper Petersen and Jakob Grue Simonsen and Christina Lioma
*ACM Transactions on Information Systems,*34(2):1–37, 2016
Appendix A.1

**Adaptive Distributional Extensions to DFR Ranking**
Casper Petersen and Jakob Grue Simonsen and Kalervo Järvelin and Christina Lioma
*The 25th ACM International Conference on Information and Knowledge Management, 4 pages, (in press), 2016*
Appendix A.2

**Brief Announcement: Labeling Schemes for Power-Law Graphs**
Casper Petersen and Noy Rotbart and Jakob Grue Simonsen and Christian Wulff-Nilsen
*Proceedings of the ACM Principles on Distributed Computing (PODC),*(1):39-41, 2016
Appendix A.3

**Near Optimal Adjacency Labeling Schemes for Power-Law Graphs**
Casper Petersen and Noy Rotbart and Jakob Grue Simonsen and Christian Wulff-Nilsen
*Proceedings of the 43rd International Colloquium on Automata, Languages and Programming (ICALP),*(2):564-574, 2016
Appendix A.4

While working on this thesis, a number of interesting ideas materialised, which, albeit not directly relevant to the topic of this thesis, address interesting problems in IR and elsewhere. These ideas have been published in the following papers (in reverse chronological order):

**Deep Learning Relevance: Creating Relevant Information (as Opposed to Retrieving it)**
Christina Lioma, Birger Larsen, Casper Petersen and Jakob Grue Simonsen
*Proceedings of the 1st international Workshop on Neural Information Retrieval (Neu-IR) hosted by the 39th ACM SIGIR conference on research and development in information retrieval. 6 pages, 2016*
Appendix A.5

**Exploiting the Bipartite Structure of Entity Grids for Document Coherence and Retrieval**
Christina Lioma and Fabien Tarissan and Jakob Grue Simonsen and Casper Petersen and Birger Larsen
*Proceedings of the 2nd ACM SIGIR International Conference on the Theory of Information Retrieval (ICTIR)*,3(4):87-96,2016
Appendix A.6

**Entropy and Graph Based Modelling of Document Coherence using Discourse Entities: An Application to IR**
Casper Petersen and Christina Lioma and Jakob Grue Simonsen and Birger Larsen
*Proceedings of the ACM SIGIR International Conference on the Theory of Information Retrieval (ICTIR)*,1(1):191-200, 2015
Appendix A.7

**The Impact of using Combinatorial Optimisation for Static Caching of Posting Lists**
Casper Petersen and Jakob Grue Simonsen and Christina Lioma
*Proceedings of the Asian Information Retrieval Societies (AIRS)*,9460(3):420-425, 2015
Appendix A.8

**Comparative Study of Search Engine Result Visualisation: Ranked Lists Versus Graphs**
Casper Petersen and Christina Lioma and Jakob Grue Simonsen
*Proceedings of the 3rd European Workshop on Human-Computer Interaction and Information Retrieval (EuroHCIR)*, 1033(2):27-30, 2013
Appendix A.9

## 1.5   Prerequisites

This thesis is intended for a reader with a background in information retrieval corresponding to e.g., Croft et al. [132], Manning et al. [289] or Rijsbergen [364]. Knowledge of statistical modelling is not required as this will be introduced. Readers interested in a more thorough introduction to statistical modelling are referred to e.g. [24, 111, 258].

## 1.6   Structure of Thesis

The remainder of this thesis is organised into 6 chapters. Chapter 2 introduces the basics of statistical modelling required to understand the statistically principled method introduced in Chapter 3 which, itself, introduces and discusses several popular approaches to parameter estimation - or model fitting – and model selection. On the basis of this principled approach, this thesis presents two applications of it. Chapter 4 uses the principled approach to test common assumptions about the distribution of several properties of IR datasets. Chapter 5 uses the principled approach to derive new ranking models which adapt their computations to the statistics of the documents used for retrieval. Chapter 6 discusses our principled approach and findings, and Chapter 7 concludes this thesis and highlights future avenues of research.

# 2 Introduction to Statistical Modelling

This chapter introduces the basics of statistical modelling required to understand the principled method to statistical model selection introduced in Chapter 3. Because statistical modelling is used in many diverse areas, this chapter is *not* specific to IR, though applications of statistical modelling in IR are highlighted where possible. The interested reader is referred to e.g. [111] for further information on statistical modelling.

Section 2.2 introduces statistical modelling with focus on the two main schools of statistical modelling: Bayesian and frequentist. Section 2.3 introduces statistical models and gives examples of these. Section 2.4 gives examples of the use of statistical models in information retrieval (IR) with focus on two specific IR tasks: ranking and score distribution modelling.

## 2.1   Notation Used

The following notation is used throughout this thesis

- A probability space is a triple $(\Omega, F, \mathrm{Pr})$ where $\Omega$ is the non-empty sample space, $F$ are the subsets (or events) of $\Omega$ called the event space, and $\mathrm{Pr} : F \to \mathbb{R}$ is a probability measure on $F$ with $\mathrm{Pr}(\Omega) = 1$.[1]

- $X : \Omega \to \mathbb{R}$ is a random variable on $\Omega$ that maps events, or outcomes, in $\Omega$ to real values. Following convention, we denote all random variables with capital letters e.g. $X$, $Y$, $Z$. An event $\omega \in \Omega$ is a repeatable process with an observable random outcome.

- Denote by $x = \{x_1 = X_1, x_2 = X_2, ..., x_n = X_n\}$ a (potentially ordered) set of real-valued observations of a random variable $X$ (i.e. $x$ is a realisation of the sequence $X_i : i = 1, 2, ..., n$). In this thesis, $x$ can denote e.g. term frequencies, document lengths, citations received or some other property of an IR dataset. We will refer to $x$ as a sample (of data), and denote its size by $n = |x|$, where individual elements in $x$ are referred to as $x_i : i = 1, ..., n$.

- A discrete random variable $X$ on probability space $(\Omega, F, \mathrm{Pr})$ is a function $h : X \to \mathbb{R}$ such that the range of $X$ is finite or countably infinite and for $x \in \mathbb{R}, \{\omega \in \Omega : X(\omega) = x\} \in F$. A continuous random variable $X$ on probability space $(\Omega, F, \mathrm{Pr})$ if a function $X \to \mathbb{R}$ such that the range of $X$ is uncountable infinite.

---

[1] Formally, let $A$ be an event of $\Omega$. The event space, $F$, is then a $\sigma$-algebra closed under the countable set operations: (1) $\emptyset \in F$, (2) $A \in F \Rightarrow A^c \in F$ where $A^c$ denotes the complement of $A$, (3) if $A_1, A_2, ... \in F$ then $A_1 \cup A_2 \cup ... \in F$.

- The *probability mass function* (PMF), $f : X \to \mathbb{R}$ of a discrete random variable is defined as [168, p. 10]:

$$f(x) = \Pr(X = x) \tag{2.1}$$

  and fulfils $f(x) \geq 0 : \forall x \in \mathbb{R}$ and $\sum_x f(x) = 1$.

- The *probability density function* (PDF), $f : X \to \mathbb{R}$, of a continuous random variable $X$, for any two numbers $a$, $b$ such that $a < b$, is defined as [168, p. 10]:

$$f(x) = \Pr(a < X \leq b) = \int_a^b f(x) dx \tag{2.2}$$

  and fulfils $f(x) \geq 0 : \forall x$ and $\int_{-\infty}^{+\infty} f(x) dx = 1$. The overload of $f(x)$ is intentional as a PMF is a special case of a PDF for a discrete probability measure. It will be clear from the context when $f(x)$ refers to a PMF or PDF.

- Unless stated otherwise, a *probability distribution* in this thesis is taken to have an associated density (PDF) or mass (PMF) function.

- If $f(x)$ is conditioned on a real-valued vector $\theta \in \mathbb{R}^m : m \geq 1$, it is written as $f(x|\theta)$ [67, p. 14]. The notation $f(x|\theta)$ specifies the conditional probability of observing $x$ given the parameter(s) in vector $\theta$.

- $\Theta$ is a subset of $\mathbb{R}^k$ for some $k > 0$ referred to as the *parameter space*.

- $\theta = \{p_1, ..., p_m\} : \theta \in \mathbb{R}^m, \theta \subset \Theta, m \leq k$ is a *parameter* vector. All models considered in this thesis will be conditioned on $\theta$.

- The *cumulative distribution function* (CDF) $F(x) = \Pr(X \leq x)$ for a discrete random variable is defined as:

$$F(x) = \sum_{x_i \leq x} f(x_i) \tag{2.3}$$

  where $f(\cdot)$ is the PMF potentially conditioned on $\theta$. The CDF of a discrete random variable is non-decreasing and $\forall x : 0 \leq F(x) \leq 1$.

- The CDF $F(x) = \Pr(X \leq x)$ for a continuous random variable is defined as:

$$F(x) = \int_{-\infty}^x f(t) dt \tag{2.4}$$

  where $f(\cdot)$ is the PDF potentially conditioned on $\theta$. The CDF of a continuous random variable is non-decreasing and $\forall x : 0 \leq F(x) \leq 1$ and $\lim_{x \to -\infty} F(x) = 0$ and $\lim_{x \to \infty} F(x) = 1$.

- The empirical CDF $F_n(x) = \Pr(X \leq x)$ for a set of observations of size $n$ is a step funtion defined as:

$$F_n(x) = \frac{1}{n} \sum_{1 \leq i \leq n} \mathbb{1}(x_i \leq x) \tag{2.5}$$

  where $\mathbb{1}$ is an indicator function emitting 1 if $x_i \leq x$ and 0 otherwise.

- Denote by $F_0(x)$ the CDF of a statistical model, and by $F_n(x)$ the empirical CDF i.e. the CDF of the data. The CDF $F_0(x)$ will refer to Eqn. 2.3 if $x$ is discrete, and to Eqn. 2.4 if $x$ is continuous.

- The *complementary cumulative distribution* (CCDF) $\Pr(X \geq x) = 1 - F(x)$ is a function which is (i) decreasing, (ii) $\forall x : 0 \leq (1 - F(x)) \leq 1$, (iii) $\lim_{x \to -\infty}(1 - F(x)) = 1$ and (iv) $\lim_{x \to \infty}(1 - F(x)) = 0$.

- We denote by $\overline{F}_0(x)$ the CCDF of a statistical model, and by $\overline{F}_n(x)$ the empirical CCDF (ECCDF) of the sample.

- A set of random variables $\{X_i\} : i = 1...$ are *independent and identically distributed* (i.i.d.) if they are drawn independently from the same distribution [67, p. 26]. All random variables and their realisations are assumed to be i.i.d. in this thesis.

## 2.2  Statistical Modelling

Statistical modelling is a tool for those empirical sciences–such as IR [377]–where observations come embedded in noise, and "serves as a lens through which data can be viewed" [245]. The presence of noise means that subsequent data analysis should employ tools that (i) naturally accommodate the notion of noise but (ii) remain useful for inference or prediction. Statistical modelling is one such tool that is widely used for this purpose.

Two schools of statistics are considered foundational in statistical modelling: Bayesian and frequentist [193]. The Bayesian school argues that variation arises because of the observer and, consequently, attempts to quantify the observer's degree of belief through probabilities. The frequentist school argues that variation arises from random events in the real world and that probabilities relate to the frequencies of these events in the limiting case [444]. In other words, frequentist statistics examine the probability of the data given a statistical model, whereas Bayesian statistics examines the probability of a statistical model given the observed data. In Section 2.2.1 and 2.2.2, a bird's-eye view of these two schools of statistical modelling is given. The key differences between the two schools' approach are summarised in Table 2.1. The interested reader may consult the considerable body of literature available that advocates the use of a Bayesian [71, 333, 436] or a frequentist [300, 301, 302] approach.

| Property | Frequentist | Bayesian |
|---|---|---|
| Data | Repeatable random sample | Data are fixed |
| Model parameters | Fixed | Unknown and probabilistic |
| Prior information | Objective (data-driven) | Subjective |
| Estimation | Estimate model parameters from data | Update knowledge about model parameters by conditioning on observed data |
| Outcome | Point estimate | Posterior distribution |
| Scope of probability | Repeatable and testable | Not repeatable or testable |

Table 2.1: Key differences between Bayesian and frequentist approaches to statistical modelling. Information is synthesised from [24, Chap. 1][444].

In Chapter 3, we will introduce a principled approach to statistical modelling based on frequentist statistics.

### 2.2.1   Bayesian Statistical Modelling

Bayesian statistical modelling employs Bayes' rule, which relates the probability of a hypothesis $H$ before and after accounting for data/evidence $D$:

$$\Pr(H|D) = \frac{\Pr(D|H)\Pr(H)}{\Pr(D)} \tag{2.6}$$

where:

- $\Pr(H)$ is the *prior* probability that $H$ is true before considering the data [261, p. 169].

- $\Pr(D|H)$ is the *likelihood* of $D$ given $H$ expressed as a probability.

- $\Pr(D)$ is the probability of data when considering all hypotheses.

- $\Pr(H|D)$ is the *posterior distribution*, or updated belief, about $H$ after data are considered.

and we assume that $(\Omega, F, \Pr)$ denotes a fixed probability space where both the the evidence (e.g. data) and hypotheses (e.g. models) exist. In order to arrive at the posterior distribution, the prior distribution and likelihood must be specified [24, Chap. 1]:

- The prior probability $\Pr(H)$ represents a PDF or PMF (a conjugate prior) from the same family as the posterior $\Pr(H|D)$, the choice of which is mainly motivated by computational convenience or mathematical tractability [271]. $\Pr(H)$ may be specified using e.g. past experiences, intuition or experts.

- The likelihood $\Pr(D|H)$ is represented by a *statistical model*, which typically contains a number of unknown parameters. For Bayesian estimation of such parameters, $H$ would represent the (vector of) unknown parameters and the posterior reflects how the parameters should be updated given the likelihood of the data.

Using Bayes' rule for modelling has a number of advantages:

- Prior knowledge regarding the modelled phenomena can be encoded using $\Pr(H)$.

- Results can be tested for sensitivity to the chosen prior.

- Data can be used as it comes in. When a new data point arrives, the current beliefs are updated to reflect this change by re-calculating the posterior under this new evidence.

Using Bayes' rule, Bayesian statistical modelling proceeds as [24, p. 6]:

- **Step 1:** Model formulation: Express unknown quantities using the prior probability density function $\Pr(H)$ and specify the likelihood $\Pr(D|H)$.

- **Step 2:** Model estimation: Infer the unknown quantities based on the posterior $\Pr(H|D)$, the conditional probability distribution of the unknown quantities given the data. $\Pr(H|D)$ need not have a PMF or PDF.

- **Step 3:** Model selection: Evaluate how well the model fits and, if not satisfactory, return to Step 1.

The resulting posterior distribution can then be used for such activities as decision making or forecasting.

Criticism of the Bayesian paradigm includes, most noticeably, the use of a prior that allows analysts to inject their subjective opinions into a statistical investigations [183, 333][376, p. 377].[2] However, the influence of the subjective prior diminishes as the sample size grows. Furthermore, the prior is often selected on the basis of mathematical convenience and need not reflect any knowledge regarding the problem. This means that as the quality of the posterior is model-dependent, Bayesian modelling should be done carefully and typically requires specialised knowledge [24, Chap. 1].

### 2.2.2 Frequentist Statistical Modelling

In the frequentist paradigm (widely attributed to Ronald A. Fisher [165][3]), observed data comes from a hypothetical infinite population by repeated sampling. The purpose of statistical modelling, in Fisher's view, is to reduce observed data to relatively few quantities that adequately represent the hypothetical infinite population [165, p. 311].

Like Bayesian statistics, the frequentist paradigm relies on Bayes' rule in Eqn. 2.6, but does not make use of a prior $\Pr(H)$ and hence can be considered a method of statistical modelling that avoids the bias of the analyst. Instead, frequentist statistics relies only on the likelihood $\Pr(D|H)$ for estimation or inference. Identically to Bayesian statistics, the likelihood is represented as a PDF (or PMF) and the hypothesis as a parameter vector $\theta$ whose values are fixed, but unknown. However, an important concept for frequentist approaches is that of regularisation, which plays a role similar to priors in Bayesian modelling [123]. When estimating the probability of a hypothesis, one may find that other hypotheses are equally likely suggesting a high variance in the estimated hypothesis. Regularisation may be viewed as a method that reduces the variance of the estimated quantities at the cost of introducing some tolerable degree of bias [123]. For example, if $w(H|\theta)$ is the belief of $H$ given $\theta$, the regularised belief is $w(H|\theta) - \lambda\tau(\theta)$ where $\tau(\cdot)$ is a function that penalises less "good" values of $\theta$, and $\lambda$ controls the impact of the variance in $\theta$. Choices for $\tau$ include quadratic or absolute deviation [123]. As $\tau(\cdot)$ effectively integrates knowledge about "good" values of $\theta$, its purpose is very similar to the prior in Bayesian modelling, where the prior distribution of $\theta$ is a probability distribution that describes the knowledge about $\theta$ without regard to the data. Even without regularisation, the frequentist approach may be viewed as a Bayesian approach where a uniform prior is used.

The frequentist paradigm has a number of advantages:

- There is no real concept of priors, which, unlike the Bayesian approach, helps control for the experimenter's bias.

---

[2] Objective priors that attempt to mitigate this subjectivity have recently been suggested–see e.g. [183, 352].

[3] A closely related frequentist approach was proposed by Neyman and Pearson [183] (see e.g. [331]).

- It allows for the calculation of confidence intervals, which are different than Bayesian confidence intervals. Quoting [428] for the frequentist approach "there is a 95% probability that when I compute a confidence interval from data of this sort, the true value of the parameters will fall within it", whereas in a Bayesian approach "given our observed data, there is a 95% probability that the true value of $\theta$ falls within the credible region". The latter is a statement of probability about the parameter value given fixed bounds. The former is a statement regarding the probability about the bounds given a fixed parameter value.

- It is automatic in nature which is a major source of their popularity in contemporary statistics. In contrast, a problem sought modelled using the Bayesian approach must be carefully considered before application [153].

- It tends to be less computationally expensive than Bayesian approaches.

- It is arguably the most widely used approach for statistical modelling today [236].

The frequentist paradigm in statistical modelling is almost identical to the Bayesian:

- **Step 1:** Model formulation: Specify the statistical model representing the likelihood $\Pr(D|H)$. In Fisher's view, a (completely specified) parametric statistical model[4] effectively reduces the entire population to only the parameters of the model.

- **Step 2:** Model estimation: Estimate the likelihood of the data under $H$.

- **Step 3:** Model selection: Select the model that minimises some, typically information-theoretic, measure.

Criticism of the frequentist paradigm focusses on (i) most algorithms' inability to update their parameters [235, 236] and (ii) it is not clear how much data is needed before a good estimate of the parameter(s) is obtained. For (i), this means that if new data arrives, the entire process must be repeated. In contrast, a Bayesian approach needs only update its posterior. For (ii), this means that using small samples can lead to bad estimates of parameter values.

## 2.3   Statistical Models

Roughly, a *statistical model* is a simplification of a real-world (and often complex) phenomenon, and is a useful tool for quantifying evidence in stochastic data [24]. Unlike scientific models[5], statistical models only attempt to describe the relationship between the independent and dependent variable, with the assumption that the relationship extends past the measured independent values [206, Chap. 7]. Once a statistical model has been constructed, it can be used for prediction, control, information extraction, knowledge discovery and validation [245].

A statistical model is constructed from one or more samples of stochastic data with the purpose of approximating said data as accurately as possible, through the use of available data, prior information available to the modeller and what is to be accomplished by the analysis [245]. Thus, it is unlikely that the statistical model will behave in strict accordance with the observed phenomenon, but rather will (i) describe the most important property/properties of the phenomenon or (ii) only apply in certain situations or for a particular range of the data. Consequently, in any definition of the term "statistical model", one must entertain the (very real) possibility that models exist that are not compatible with any collected data.

---

[4]A completely specified parametric statistical model is a parametric statistical model with specific values for the parameters that completely describes the distribution of data [313, p. 369].

[5]A scientific model attempts to explain *why* variables are related [70].

A *parametric* statistical model, as used in this thesis, is formalised as [405]:

$$\mathcal{M}_\theta(x) = \left\{ f(x|\theta) : \theta \in \Theta \subset \mathbb{R}^m \right\} \tag{2.7}$$

If $x$ takes on discrete values, then $\mathcal{M}_\theta$ is a discrete statistical model. If $x$ takes on continuous values, then $\mathcal{M}_\theta$ is a continuous statistical model. For example, the quintessential continuous statistical model is the Gaussian (or normal) model with parameters $\theta = \{\mu, \sigma^2\}$ defined as:

$$\begin{aligned} \text{Gauss}(x|\theta) &= \left\{ f\left(x|\mu, \sigma^2\right) \right\} \\ &= \left\{ \frac{1}{\sqrt{2\pi\sigma^2}} e^{-\frac{(x-\mu)^2}{2\sigma^2}} : \mu \in \mathbb{R}, \sigma^2 \in \mathbb{R}^+ \right\} \end{aligned} \tag{2.8}$$

If $\mu$ (mean) and $\sigma^2$ (variance) are fixed, Gauss$(x|\theta)$ indexes a specific PDF [313, p. 369] which defines the probabilistic and parametric context in which prediction or discovery proceeds. As the parameter space of the Gaussian is unconstrained, a Gaussian model is a *set* or family of PDFs [185, 245] where each specific setting of $\mu$ and $\sigma^2$ makes some part of the data more probable than others. Thus, evidence and hypothesis, in the terminology of the Bayesian (Section 2.2.1) and frequentist (Section 2.2.2) view, corresponds to $x$ and a specific setting of the parameter values for the statistical model.

Each statistical model considered in this thesis is characterised by one or more parameters, termed either *location*, *scale* or *shape* parameters [168, p. 19]:

**Location parameter:** Controls the position of the distribution or density function along the horizontal axis. For example, if $X$ is distributed as a Gaussian with mean $\mu$ and variance $\sigma^2$, then $Y = X + k$ for some $k \in \mathbb{R}$ is from the same family as $X$ but with its mean $\mu$ shifted by $k$ to the right (if $k < 0$) or left (if $k > 0$).

**Scale parameter:** The multiplier by which $X$ is scaled. If $Y = Xk$ then $Y$ is from the same family as $X$, simply scaled by a factor of $k > 0$. For example, if $X$ is distributed as an exponential with $\lambda = 2$ and $Y = 5X$, then $Y$ is distributed as an exponential with $\lambda = 5 \cdot 2 = 10$.

**Shape parameter:** Controls the "shape" of the distribution or density and might stretch and shift the density, but is neither a location nor a shape parameter [373, p. 17].

Figure 2.1 shows the effect of varying these parameters for three different statistical models. It can be seen e.g. for the Gaussian model that increasing $\sigma^2$ flattens the distribution, and for the Exponential model that smaller values for $\lambda$ cause the distribution to drop off more rapidly.

Figure 2.1 also illustrates the concept of *nested* or *restricted* statistical models: given two statistical models $\mathcal{M}_\theta^f$ and $\mathcal{M}_\theta^g$ (where the superscript serves to denote different models), then if:

$$\mathcal{M}_\theta^f \subset \mathcal{M}_\theta^g \tag{2.9}$$

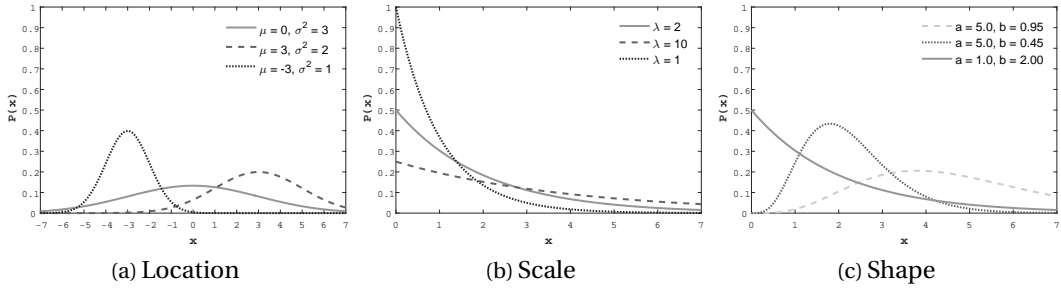

(a) Location             (b) Scale             (c) Shape

Figure 2.1: Effect of varying the location (2.1a–Gaussian model), scale (2.1b–exponential model) and shape (2.1c–Gamma model). Model parameters are given in the plots.

$\mathcal{M}_{\theta}^{f}$ is a nested, or restricted, model inside the full model $\mathcal{M}_{\theta}^{g}$. For example, the exponential model:

$$\begin{aligned} \text{Exp}(x|\theta) &= \left\{ f(x|\lambda) \right\} \\ &= \left\{ \lambda \exp^{-\lambda x} : \lambda \in \mathbb{R}^{+} \right\} \end{aligned} \tag{2.10}$$

is a restricted version of the gamma model:

$$\begin{aligned} \text{Gamma}(x|\theta) &= \left\{ f(x|a,b) \right\} \\ &= \left\{ \frac{1}{\Gamma(a)\,b^{a}} x^{a-1} \exp^{\frac{-x}{b}} : (a,b) \in \mathbb{R}^{+} \right\} \end{aligned} \tag{2.11}$$

by fixing $a = 1$ and where $\Gamma$ is the gamma function. Visually, this is shown by the solid lines in Figures 2.1b and 2.1c.

Underpinning each statistical model is a set of assumptions under which the data are analysed. Thus, assumptions are characteristics of the model and *not* of the data; when an assumption fails to be useful the error lies in the model and not the data, and any inferences drawn will be biassed and can even be rendered invalid [320]. Statistical models typically assume that $x$ is (a) i.i.d (see Section 2.1 in Chapter 1), (b) representative (i.e. the elements of $x$ were drawn with equal probability from some population/dataset) and (c) sufficiently large so the estimated parameter(s) or mode(s) (such as the mean or median) are close (in a statistical sense) to the real parameter(s) or mode(s). Each statistical model may rely on specific assumptions, however. Furthermore, because a parametric statistical model assumes that the underlying probability distribution is known up to a finite number of parameters [427, p. 8], the chosen parametric statistical model is likely wrong to begin with. The question of whether a "true" parametric statistical model exists is epistemological and is not one sought answered in this thesis. However, it may still be worthwhile to borrow weight from the parametric statistical model as long as the "true" data generating model is not too far away from the parametric class selected to model the distribution of our data. The argument could be made that a saturated model (a statistical model with as many parameters as data points) is "correct" as it accurately describes the data. Such saturated (or over-fitted) models, however, simply interpolate the data and do not describe the data more parsimoniously than the raw data does, and will typically generalise very poorly.

The Gaussian model is only one out of a large number of statistical models. A non-exhaustive set of statistical models commonly found in commercial and open-source software are shown

Tables 2.2 and 2.3. These are the models we consider in our approach. All statistical models in Tables 2.2 and 2.3 have support in either the non-negative or positive integers (for discrete models) or the real numbers (for continuous models). While many more models could be included, this set of models have been used to quantify a range of phenomena in IR e.g. [119, 120, 121, 274, 290, 346]. Specialised, or mixture, models such as the double Pareto [359] or the Poisson-Power law, used to quantify query lengths [31], can also be included. The statistical models shown in Table 2.2 and 2.3 are a mix of both continuous and discrete statistical models. Thus, for datasets with discrete data, such as IR datasets typically, discrete models (such as the geometric) are well-defined, but continuous statistical models are not. Despite this, continuous models can be used to describe discrete data for two reasons: (i) it provides a greater range of models with which to study data, and (ii) continuous statistical models can, in general, be discretised using continuity corrections, though care must be taken if the statistical properties of the original model, such as its moments, are to be preserved. A recent survey by Chakraborty [101] presents discrete approximations to most of the continuous distributions in Table 2.2. For more comprehensive listings of discrete and continuous statistical models available, see e.g. [168, 223, 224, 445].

Table 2.2: Statistical models used. Each model is identified by PDF/PMF and CDF. The domain of all parameters are given in Table 2.3

| Statistical model | Type | PDF/PMF | CDF | Notation |
|---|---|---|---|---|
| Exponential | Cont. | $f(x\|\mu) = \left\{\frac{1}{\mu}\exp\left(\frac{-x}{\mu}\right)\right\}$ | $F(x\|\mu) = 1 - \exp\left(\frac{-x}{\mu}\right)$ | exp: Exponential function. |
| Gamma | Cont. | $f(x\|a,b) = \left\{\frac{1}{b^a\Gamma(a)}x^{a-1}\exp\left(\frac{-x}{b}\right)\right\}$ | $F(x\|a,b) = \frac{1}{\Gamma(a)}\gamma\left(a,\frac{x}{b}\right)$ | $\Gamma$ : gamma function, $\gamma$ : Incomplete gamma function. |
| Gaussian | Cont. | $f(x\|\mu,\sigma^2) = \left\{\frac{1}{\sqrt{2\pi\sigma^2}}\exp\left(-\frac{(x-\mu)^2}{2\sigma^2}\right)\right\}$ | $F(x\|\mu,\sigma^2) = \frac{1}{2}\left(1+\mathrm{erf}\left(\frac{(x-\mu)}{\sqrt{2\sigma^2}}\right)\right)$ | |
| Generalized Extreme Value | Cont. | $f(x\|k,\mu,\sigma) = \left\{\left(\frac{1}{\sigma}\right)\exp\left(-\left(1+k\frac{(x-\mu)}{\sigma}\right)^{-\frac{1}{k}}\right)\left(1+k\frac{(x-\mu)}{\sigma}\right)^{-1-\frac{1}{k}}\right\}$ | $F(x\|k,\mu,\sigma) = \exp\left(-\left(1-\frac{k(x-\mu)}{\sigma}\right)^{\frac{1}{k}}\right)$ | $\phi$: Normal distribution. |
| Generalized Pareto | Cont. | $f(x\|k,\sigma,\theta) = \left\{\left(\frac{1}{\sigma}\right)\left(1+k\frac{(x-\theta)}{\sigma}\right)^{-1-\frac{1}{k}}\right\}$ | $1-\left(1+k\frac{(x-\theta)}{\sigma}\right)^{-\frac{1}{k}}$ | |
| Geometric | Disc. | $f(x\|p) = \left\{(1-p)^x p\right\}$ | $F(x\|p) = 1-(1-p)^{x+1}$ | |
| Inverse Gaussian | Cont. | $f(x\|\mu,\lambda) = \left\{\left(\frac{\lambda}{2\pi x^3}\right)^{\frac{1}{2}}\exp\left(\frac{-\lambda(x-\mu)^2}{2\mu^2 x}\right)\right\}$ | $F(x\|\mu,\lambda) = \phi\left(\sqrt{\frac{\lambda}{x}}\left(\frac{x}{\mu}-1\right)\right) + \exp\left(\frac{2\lambda}{\mu}\right)\phi\left(-\sqrt{\frac{\lambda}{x}}\left(\frac{x}{\mu}+1\right)\right)$ | erf: Error function. |
| Logistic | Cont. | $f(x\|\mu,\sigma) = \left\{\frac{\exp\left(\frac{x-\mu}{\sigma}\right)}{\sigma\left(1+\exp\left(\frac{(x-\mu)}{\sigma}\right)\right)^2}\right\}$ | $F(x\|\mu,\sigma) = \frac{1}{1+\exp\left(-\frac{x-\mu}{\sigma}\right)}$ | |
| Log-normal | Cont. | $f(x\|\mu,\sigma^2) = \left\{\frac{1}{x\sigma\sqrt{2\pi}}\exp\left(\frac{-(\ln x-\mu)^2}{2\sigma^2}\right)\right\}$ | $F(x\|\mu,\sigma) = \frac{1}{2} + \frac{1}{2}\mathrm{erf}\left(\frac{\ln x-\mu}{\sqrt{2\sigma^2}}\right)$ | |
| Nakagami | Cont. | $f(x\|\mu,\omega) = \left\{2\left(\frac{\mu}{\omega}\right)^\mu \frac{1}{\Gamma(\mu)}x^{2\mu-1}\exp\left(\frac{-\mu x^2}{\omega}\right)\right\}$ | $F(x\|\mu,\omega) = \frac{\gamma\left(\mu,\frac{\mu}{\omega}x^2\right)}{\Gamma(\mu)}$ | |
| Negative Binomial | Disc. | $f(x\|r,p) = \left\{\binom{r+x-1}{x}p^r(1-p)^x\right\}$ | $F(x\|r,p) = \sum_{i=0}^{x}\binom{r+i-1}{i}p^r(1-p)^i$ | $x!$: Factorial function. |
| Poisson | Disc. | $f(x\|\lambda) = \left\{\frac{\lambda^x}{x!}\exp(-\lambda)\right\}$ | $F(x\|\lambda) = \exp(-\lambda)\sum_{i=0}^{\lfloor x\rfloor}\frac{\lambda^i}{i!}$ | $\zeta$: Hurwitz zeta function. |
| Power law | Disc. | $f(x\|a,x_{min}) = \left\{\frac{x^{-a}}{\zeta(a,x_{min})}\right\}$ | $F(x\|a,x_{min}) = \frac{\zeta(a,x)}{\zeta(a,x_{min})}$ | |
| Rayleigh | Cont. | $f(x\|b) = \left\{\frac{x}{b^2}\exp\left(\frac{-x^2}{2b^2}\right)\right\}$ | $F(x\|b) = 1-\exp\left(\frac{-x^2}{2b^2}\right)$ | |
| Weibull | Cont. | $f(x\|a,b) = \left\{\frac{b}{a}\left(\frac{x}{a}\right)^{(b-1)}\exp\left(-\left(\frac{x}{a}\right)^b\right)\right\}$ | $F(x\|a,b) = 1-\exp\left(-\left(\frac{x}{a}\right)^b\right)$ | $\beta$: Beta function. |
| Yule–Simon | Disc. | $f(x\|p) = \left\{(p-1)\beta(x,p)\right\}$ | $F(x\|p) = 1-(x+1)\beta(x+1,p)$ | |

Table 2.3: Statistical models used. Each model is identified by its parameters, support and whether it is nested.

| Statistical model | Notation | Parameters | Support | Nested |
|---|---|---|---|---|
| Exponential | $\mathrm{Expl}(x;\lambda)$ | Scale | $x \in \mathbb{R}_0^+, \lambda \in \mathbb{R}^+$ | Yes (Wbl, Gamma, GP) |
| Gamma | $\mathrm{Gamma}(x;a,b)$ | Shape, Scale | $x \in \mathbb{R}_0^+, a,b \in \mathbb{R}^+$ | No |
| Gaussian | $\mathrm{Gauss}(x;\mu,\sigma^2)$ | Location, Scale | $x \in \mathbb{R}, \mu \in \mathbb{R}, \sigma^2 \in \mathbb{R}^+$ | No |
| Generalized Extreme Value | $\mathrm{Gev}(x;k,\mu,\sigma)$ | Shape, Location, Scale | $x \in [-\infty;+\infty] \; : \; k=0, \sigma \in \mathbb{R}^+, \{\mu,k\} \in \mathbb{R}$; $x \in [-\infty;\mu-\sigma/k] \; : \; k<0, \sigma \in \mathbb{R}^+, \{\mu,k\} \in \mathbb{R}$; $x \in [\mu-\sigma/k;+\infty] \; : \; k>0, \sigma \in \mathbb{R}^+, \{\mu,k\} \in \mathbb{R}$ | No |
| Generalized Pareto | $\mathrm{GP}(x;k,\sigma,\theta)$ | Shape, Scale, Location | $x \in [\theta,\infty) \; : \; k \geq 0, (k,\theta) \in \mathbb{R}, \sigma \in \mathbb{R}^+$; $x \in [0 \leq \frac{(x-\theta)}{\sigma} \leq -\frac{1}{k}] \; : \; k<0, (k,\theta) \in \mathbb{R}, \sigma \in \mathbb{R}^+$ | Yes (Neg, Bin) |
| Geometric | $\mathrm{Geo}(x;p)$ | Prob. of failure | $x \in \mathbb{N}_0^+, 0<p\leq 1$ | No |
| Inverse Gaussian | $\mathrm{InvGauss}(x;\mu,\lambda)$ | Location, Shape | $x \in \mathbb{R}^+, \mu,\theta \in \mathbb{R}^+$ | No |
| Logistic | $\mathrm{Log}(x;\mu,\sigma)$ | Location, Scale | $x \in \mathbb{R}, \mu \in \mathbb{R}, \sigma \in \mathbb{R}^+$ | No |
| Log-normal | $\mathrm{Logn}(x;\mu,\sigma^2)$ | Location, Scale | $x \in \mathbb{R}^+, \mu \in \mathbb{R}, \sigma^2 \in \mathbb{R}^+$ | No |
| Nakagami | $\mathrm{Naka}(x;\mu,\omega)$ | Shape, Scale | $x \in \mathbb{R}^+, \mu,\omega \in \mathbb{R}^+$ | No |
| Negative Binomial | $\mathrm{Nbin}(x;r,p)$ | # of trials, prob of success in a single trial | $x \in \mathbb{N}_0^+, \{r,p\} \in \mathbb{R}^+, r>0, p \in (0,1)$ | No |
| Poisson | $\mathrm{Poiss}(x;\lambda)$ | Scale | $x \in \mathbb{N}_0^+, \lambda \in \mathbb{R}^+$ | No |
| Power law | $\mathrm{Plaw}(x;\alpha,x_{\min})$ | Scale, cutoff | $x, x_{\min} \in \mathbb{N}^+, \alpha \in \mathbb{R}^+, \alpha > 1$ | No |
| Rayleigh | $\mathrm{Rayl}(x;b)$ | Scale | $x \in \mathbb{R}_0^+, b \in \mathbb{R}^+$ | Yes (Weibull) |
| Weibull | $\mathrm{Wbl}(x;a,b)$ | Scale, Shape | $x \in \mathbb{R}_0^+, a,b \in \mathbb{R}^+$ | No |
| Yule-Simon | $\mathrm{Yule}(x;p)$ | Scale | $x \in \mathbb{N}^+, p \in \mathbb{R}^+$ | No |

Three examples of statistical models commonly used in information retrieval (IR) are presented next.

### 2.3.1 Multinomial Model

The discrete multinomial model generalises the binomial (BIN) model with PMF:

$$
\begin{aligned}
\text{BIN}(x|\theta) &= \left\{ f(x|n,p) \right\} \\
&= \left\{ \binom{n}{x} p^x (1-p)^{n-x} : n \in \mathbb{N}^0, p \in \{0,1\} \right\}
\end{aligned}
\tag{2.12}
$$

when the outcomes of the experiment are not binary i.e. $p$ is not restricted to be 0 or 1. Notice here that $n$ is a parameter and not the cardinality of $x$. The multinomial counts the number of outcomes of a particular category when $n$ trials are performed that have $k$ possible outcomes, each with an associated probability $p_i : i = 1,...,k$. If $n_1,...,n_k$ are the total number of occurrences of outcomes 1 through $k$ in $n$ trials, then the probabilities $p_1,...,p_k$ are a multinomial model with parameters $(n_1,...,n_k,p_1,...,p_k)$. The multinomial model can be explained as follows: Given a complex device consisting of different components, each with a fixed probability of failing, the multinomial model can be used to evaluate the probability that any specific combination of components is damaged [168]. The PMF of the multinomial model is defined as:

$$
\begin{aligned}
\text{MN}(x|\theta) &= \left\{ f(x|n,p) \right\} \\
&= \left\{ \frac{n!}{x_1!,...,x_k!} p_1^{x_1} ... p_k^{x_k} : n \in \mathbb{N}^+, 0 < p_i \le 1, \sum_i p_i = 1 \right\}
\end{aligned}
\tag{2.13}
$$

where ! denotes the factorial function and $n!/x_1!,...,x_k!$ is the multinomial coefficient describing the number of ordered arrangements of $k_i : i = 1,...,k$ objects with probabilities $p_i : i = 1,...,k$. Parameter estimation of the multinomial model is possible by e.g. expanding Eqn. 2.13 into a telescopic product [216]. Applications of the multinomial model in IR include latent Dirichlet allocation [69], text categorisation [303] and language modelling [349]. The multinomial PMF of the lengths of 50,000 Twitter messages sampled in August 2015 from Twitter's public stream[6] is shown in Figure 2.2.

The assumptions underlying the multinomial model are:

- The $n$ trials are i.i.d.
- The probabilities $p_i$ remain constant across trials.

For information on the multinomial model see [168, pp. 135–136] and [437, pp. 95–98].

### 2.3.2 Poisson Model

The Poisson model is a widely used discrete statistical model for modelling counts or events that (i) occur within a certain period of time and (ii) at an average rate. The classic use of the

---

[6]https://dev.twitter.com/streaming/public.

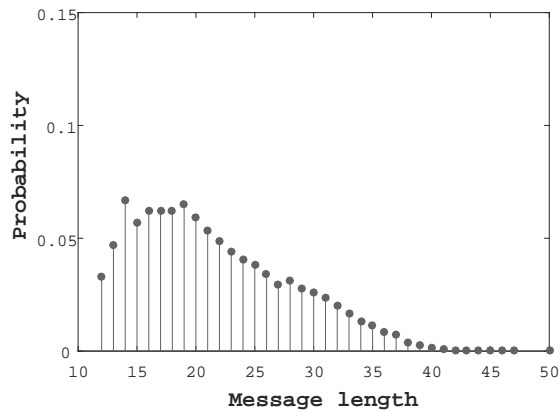

Figure 2.2: Multinomial PMF of the lengths of 50,000 Twitter messages mined in August 2015. Message lengths were calculated as the number of terms after splitting on white spaces. Message lengths are shown as impulses, as the multinomial model is a discrete statistical model.

Poisson model, by von Bortkiewicz [76, p. 25], approximated the number of Prussian army corps soldiers who were kicked to death by horses over a period of 20 years. The Poisson model has since been used in IR in various ways, for instance for generating queries [306], describing the number of visitors to a Web site per minute [390, p. 654] and estimating the number of file server virus infections during a 24-hour period [145, p. 232].

The Poisson model is defined as [168]:

$$\text{Poiss}(x|\theta) = \left\{ f(x|\lambda) \right\}$$
$$= \left\{ \lambda^x \frac{\exp^{-\lambda}}{x!} : \lambda \in \mathbb{R}^+ \right\} \tag{2.14}$$

where $\lambda$ is a scale parameter and ! is the factorial function. For large values of $\lambda$, the Poisson can be approximated by a Gaussian model with $\mu = \lambda$ and $\sigma^2 = \lambda$ [168]. Two Poisson densities with different $\lambda$ values are shown in Figure 2.3.

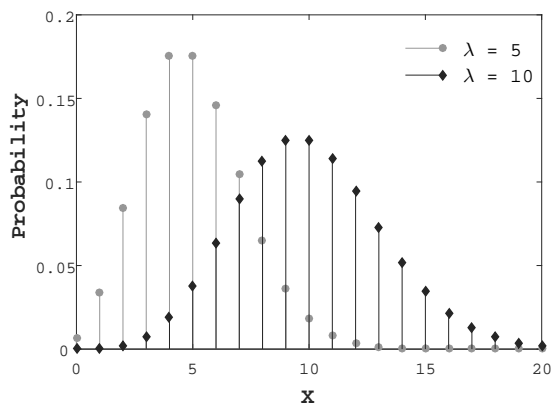

Figure 2.3: Poisson PMFs for $\lambda = 5$ (circles) and $\lambda = 10$ (diamonds). Data points are shown as impulses, as the Poisson is a discrete statistical model.

The assumptions underlying the Poisson model are:

- Events occur independently of each other.

- The average rate of event occurrences is fixed.

- The time period where events occur is fixed.

For further information on the properties of the Poisson model see [168, pp. 152–156] or [437, pp. 134–137].

### 2.3.3 Weibull Model

The continuous Weibull model [440] is commonly used in reliability engineering to describe the distribution of a device's lifetime: decreasing, constant and increasing failure rates, corresponding to the "burn-in", "random" and "wearout" (as a result of ageing) phases of a device's life [241].

The Weibull model generalises the exponential model (i.e. the exponential model is a restricted model of the Weibull), which assumes that the probability of failure of a device as a function of its lifetime remains constant. Instead, using a Weibull model, a device's probability of failure initially decreases (the device is new), then remains constant for a period of time before increasing (the device becomes old) with time. The Weibull model is defined as [168]:

$$
\begin{aligned}
\mathrm{Wbl}(x|\theta) &= \left\{ f(x|a,b) \right\} \\
&= \left\{ \frac{b}{a}\left(\frac{x}{a}\right)^{(b-1)} \exp^{-\left[\frac{x}{a}\right]^{b}} : (a,b) \in \mathbb{R}^{+} \right\}
\end{aligned}
\tag{2.15}
$$

where $a$ is the scale parameter and $b$ is the shape parameter. The Weibull model has been used in IR to quantify the distribution of dwell times on Web pages [274] and modelling comments in discussion threads [262]. A three- or five-parameter Weibull model is also common [168]. A Weibull distribution with different parameter settings is shown in Figure 2.4.

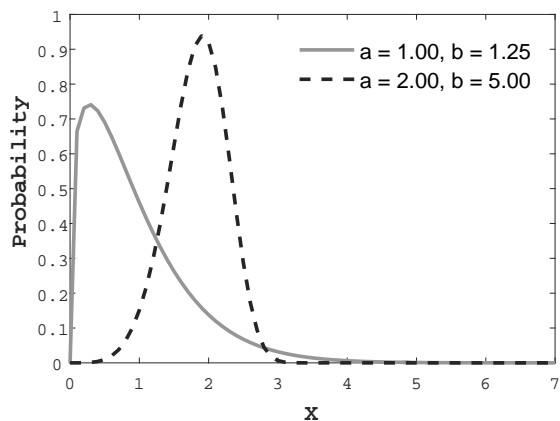

Figure 2.4: Weibull PDFs for $a = 1, b = 1.25$ (solid line) and $a = 2, b = 5.0$ (dashed line). Data are shown as a curve as the Weibull is a continuous statistical model.

The assumptions underlying the Weibull model are:

- Events cannot occur simultaneously.

- Events are independent.

For further information on the properties of the Weibull model see [168, pp. 193–201] or [437, pp. 152–153].

## 2.4 Statistical Model Use in Information Retrieval

Statistical models have long been used in IR. In text categorisation, McCallum and Nigam [303] compare the multi-variate Bernoulli and multinomial models. The former specifies that a document is a vector of binary attributes indicating what terms do and do not occur. The latter specifies that a document is represented by a set of term occurrences from the document. Both models are evaluated using five datasets (Yahoo!, Industry Sector, Newsgroups, WebKB and Reuters) in a classification task. Their results show that the multinomial achieves a classification accuracy between 11% and 28% over the multi-variate Bernoulli model on all datasets, when the vocabulary size is above 1,000.

Liu et al. [274] propose to model the dwell time on a Web page across all visits to understand user browsing behaviours using a Weibull distribution. Their results show that the dwell time is no more than 70 seconds on 80% of the Web pages used (Web pages taken from a custom dataset), that users "screen" Web pages in the early stage of browsing the Web page and that the Weibull is a more accurate distribution than the exponential. Based on their results, the authors propose to predict Web page dwell time based on document features: HTML tags (93 in total), Content (top-1000 most frequent terms) and Dynamic (9 features including the time required to layout and render a Web page and the byte-size of the rendered page). Their prediction results show that dynamic features are almost consistently the best feature to use.

Lee and Salamatian [262] investigate the characteristics of on-line commenting behaviours using data from a digital camera review site, Myspace and an on-line portal service from Korea. The authors find that the lifetime of a discussion thread can be accurately modelled using a Weibull model though no other model is considered.

In language modelling, a query $q$ is assumed to be generated by a model estimated using a document $d$. While most research assumes a multinomial model, Mei et al. [306] propose to use Poisson distributions. In their model, each term in the vocabulary is assumed to be sampled from $n$ independent homogeneous Poisson distributions. The likelihood of some text $\mathbf{w} = \{w_1, ..., w_n\}$ (be that a document or query) being generated is then:

$$\Pr(\mathbf{w}|\lambda) = \prod_{i=1}^{n} \frac{\exp^{-\lambda_i \cdot \mathbf{w}} (\lambda_i \cdot |\mathbf{w}|)^{c(w_i, \mathbf{w})}}{c(w_i, \mathbf{w})} \tag{2.16}$$

where $\lambda = \{\lambda_1, ..., \lambda_n\}$ are the Poisson parameters for $w$ and $c(w_i, \mathbf{w})$ is the frequency of $w_i$ in $\mathbf{w}$. The Poisson language model is then:

$$\Pr(q|d) = \prod_{w \in V} \Pr(c(w, q)|\lambda_d) \tag{2.17}$$

where $\lambda_d$ is the document language model and Eqn. 2.17 is defined for the entire vocabulary $V$. Experimental evaluation of Eqn. 2.17 compared to the multinomial, shows that the Poisson language model statistically significantly outperforms a Jelinek-Mercer smoothed multinomial language model when using term-dependent two-stage smoothing on four Text REtrieval Conference (TREC) collections.

Badue et al. [45] propose to model workload characterisation for Web search from a system perspective, by analysing the distribution of query interarrival times and per-query execution time that reach a cluster of index servers using an exponential and hyperexponential distribution. Their results show that both the interarrival times and query service times are best quantified using a hypergeometric distribution using a sample of Web pages and queries from TodoBR. They conclude that exponential-based models tend to underestimate system performance, leading to unrealistic results.

Because the characterisation of user churn - the number of user entering or leaving a network - is important for both the design and application of a peer-to-peer system, Stutzbach and Rejaie [415] investigate three popular peer-to-peer (P2P) services (Gnutella, Kad and BitTorrent). Experimenting with both an exponential, log-normal and Weibull distribution, they find that (i) inter-arrival times are best quantified using a Weibull distribution, and (ii) session lengths are best quantified by a Weibull or log-normal distribution. The authors also attempt to fit a power law distribution but find this provides a poor fit. They suggest that their results may affect the selection of key design parameters, the overlay structure and the resiliency of the overlay.[7]

Zhou et al. [461] combine relevance-weighting based on document lengths with classic BM25. The relevance-weighting functions are derived from kernel density estimation of document lengths, which are transformed to find the distributions of relevant and non-relevant documents before using maximum likelihood estimation to obtain the distribution parameters. Using the relevance-weighting functions substantially outperforms tuned BM25 when evaluated on four standard TREC test collections.

Hui et al. [212] propose ranking models, called NG models, that integrate different statistical models into DFR ranking models. In their approach, a set of statistical models is first fitted to the term collection frequencies of TREC disks 1 & 2, TREC disks 4 & 5, WT10g and .GOV2 and their fit evaluated using Pearson's correlation coefficient. Any parameters of the statistical models are determined using gradient search on half of the queries of each dataset. The NG models are evaluated against an untuned Dirichlet-smoothed query language model, BM25 and the standard Poisson DFR model with Laplace normalisation. The results show that all baselines perform statistically significantly better than any of the proposed models. The authors next propose simplified versions of their NG models which show statistically significant improvement over the standard NG models and, in some cases for WT10g and .GOV2, better performance than BM25.

Ranking models and score normalisation are two tasks in IR which rely on statistical models. Ranking models are a core part of any IR system whose purpose is to rank documents in decreasing order of, typically, relevance. A number of ranking models exist that, implicitly or explicitly, have distributional assumptions are reviewed in Section 2.4.1. A second important task in IR is score normalisation. Score normalisation is relevant in e.g. data fusion, as scores provided by component IR systems may not be comparable. The key to these methods is to be able to discriminate between relevant and non-relevant documents, and substantial effort has been devoted to modelling the distributions of relevant and non-relevant documents for this purpose. Section 2.4.5 reviews combinations of statistical models used to quantify the distribution of relevant and non-relevant documents.

---

[7]An overlay is the connectivity formed by peers participating in the P2P network.

### 2.4.1 Ranking Models

A core task of an IR system is to retrieve the most relevant documents in response to some input such as a user query.[8] To retrieve the most relevant documents, each document is scored using a ranking model $R(q,d)$ where $q$ is a query and $d$ is a document. The highest scoring documents are then returned to the user, using the implicit assumption that relevance is a monotonically increasing function of the score. Hence, a document with a higher score is also seen as more relevant to the specific information need/query. Ranking models have been extensively researched over the years and examples include exact match models (e.g. the Boolean [132, Chap. 7] and best-match models such as the vector space model [375], probabilistic models (e.g. language models [349], Divergence From Randomness (DFR) models [19] or the BM25 model [369]) and machine learning models (e.g. learning-to-rank [84, 85, 95, 275]).

### 2.4.2 2-Poisson Indexing Model

The 2-Poisson model is a probabilistic model of indexing rather than retrieval [16]. Its purpose is to select key words which are likely informative of an arbitrary document, and use those key words for creating an index. Such key words are referred to as speciality terms: an element of a technical or scientific vocabulary that does not occur statistically independently and constitutes a good index term [197]. In contrast, a non-speciality term is *not* a member of a technical or scientific vocabulary and occurs independently [197]. For non-speciality terms, the Poisson model (see Section 2.3.2) was proposed by Harter [197, 198] to quantify their frequency distribution. Harter found, however, that any single Poisson was unlikely to approximate the frequency distribution of speciality terms. Assuming (i) that the probability of document relevance is a function of the extent a topic is treated in the document and (ii) that the number of term tokens is indicative of this extent, Harter suggests that speciality terms distinguish two classes of documents, depending on the extent that the topic named by the speciality term is treated in documents. The 2-Poisson indexing model, originally formulated by Bookstein and Swanson [73], quantifies these two classes as a linear combination of two Poisson distributions:

$$\Pr(X = k | \lambda_1, \lambda_2) = \pi \frac{\lambda_1^k \exp^{-\lambda_1}}{k!} + (1 - \pi) \frac{\lambda_2^k \exp^{-\lambda_2}}{k!} \qquad (2.18)$$

where $\lambda_1$ (Class 1) and $\lambda_2$ (Class 2) are the means of the two Poisson models with $\lambda_1 \neq \lambda_2$, $\pi$ is the proportion of documents in the collection belonging to Class 1, $(1 - \pi)$ is the proportion belonging to Class 2, and $P(k)$ is the probability that a document contains $k$ occurrences of a given term. Any variation in the number of tokens (within-document term frequency) of speciality terms belonging to the same class is due to random fluctuations. Eqn. 2.18 implicitly quantifies the distribution of non-speciality terms if either class is empty i.e. if $\pi = 0$ or $\pi = 1$.

In other words, the 2-Poisson model assumes that two distinct populations of documents exist for each term [197, p. 204]: one population will consist of $\pi$ "elite" documents where the frequency of a term $t$ occurs more densely than in all other documents and are distributed by $\lambda_1$, and the other population will consist of $(1-\pi)$ non-elite documents where term frequencies are distributed by $\lambda_2 \neq \lambda_1$.

---

[8]Documents are used as an umbrella term to describe text, images, videos or other modalities of information.

The probability of a term $t$ appearing $k$ times in document $d$ part of the "elite" set of documents is given by [198]:

$$
\begin{aligned}
\Pr(d \in \text{Elite set}|X = k) &= \frac{\Pr(X = k, d \in \text{Elite set})}{\Pr(X = k)} \\
&= \frac{\pi \frac{\lambda_1^k \exp^{-\lambda_1}}{k!}}{\pi \frac{\lambda_1^k \exp^{-\lambda_1}}{k!} + (1-\pi) \frac{\lambda_2^k \exp^{-\lambda_2}}{k!}}
\end{aligned}
\tag{2.19}
$$

which estimates the relative level of treatment of $t$ in $d$. Combined with the $z$-measure [197]:

$$
z = \frac{\lambda_1 - \lambda_2}{\sqrt{\lambda_1 + \lambda_2}}
\tag{2.20}
$$

which estimates the overall effectiveness of $t$ as an index term, a measure of a term's 'indexability' is given by:

$$
\beta = \Pr(d \in \text{Elite set}|X = k) + z
\tag{2.21}
$$

which are ranked to determine what key words to index [198].[9]

Despite the theoretical justifications, the 2-Poisson model performed poorly and rejected (using the $\chi^2$-test) the 2-Poisson hypothesis for 62% of the terms tested by Harter. Srinivasan [408] later generalised the 2-Poisson to the $n$-Poisson model, effectively allowing $n$ classes to be involved. Margulis [292] later found support for $n$-Poisson model for 70% of the terms using improved parameter estimation and longer documents.

One reason the 2-Poisson may have performed poorly is that it was based on empirical evidence of the distribution of term frequencies in a curated dataset of 650 abstracts. Contemporary datasets, however, are noisy, orders of magnitude larger and domain-specific. Thus, to justify the use of the 2-Poisson model in contemporary datasets, it must be verified that the within-document term frequencies of terms in the elite and non-elite set of documents are (i) distributed by a Poisson distribution with (ii) different means. If this cannot be verified, it means that Eqn. 2.18 gives inaccurate estimates of $\Pr(X = k|\lambda_1, \lambda_2)$. For example, assuming a geometric distribution provides the best fit to the within-document term frequencies in both the elite and non-elite set of some term $t$, the resulting "2-Geometric" model is:

$$
\Pr(X = k|p_1, p_2) = \pi\Big((1-p_1)^{k-1} p_1\Big) + (1-\pi)\Big((1-p_2)^{k-1} p_2\Big)
\tag{2.22}
$$

where $(1 - p_i)^{k-1} p_i$ is the geometric distribution with parameter $p_i : i \in \{1, 2\}$. Evaluating the 2-Poisson and 2-Geometric model for the same input produces very different results. For example, setting $k = 10, \lambda_1 = p_1 = 0.8, \lambda_2 = p_2 = 0.4, \pi = 0.4$, the 2-Poisson model returns $\Pr(X = k|\lambda_1, \lambda_2) \approx 7.98e - 09$ and the 2-Geometric returns $\Pr(X = k|p_1, p_2) \approx 0.0016$ which is several orders of magnitude higher. Thus, calculating the $\beta$ value for this term using both the 2-Poisson and 2-Geometric model, where we assume that $z$ is fixed and only the top$-k$ ranked terms are considered for use as indexing terms, it is more likely that $t$ would be considered a good indexing term under the 2-Geometric model than under the 2-Poisson model. As this could substantially impact what keywords are indexed, and, subsequently, what can be retrieved, this suggests that it is important to use the best possible statistical model.

---

[9]Harter [197, 198] does not specify how the $\beta$ values are ranked or if only terms above/below some cutoff of $\beta$ values are considered good indexing terms.

### 2.4.3 Language Modelling

A language model (LM) is, in its most basic form, a probability distribution over terms in a language where each term, $t$ is assigned a non-zero probability denoting its probability of occurrence in the "language". A "language" here is defined as a non-empty finite sequence of symbols or terms. Given a query $q$ and document $d \in C$ for some collection $C$, $d$'s LM, $\theta_d$, is a probabilistic model that estimates the probability that $q$ was generated by $d$. In other words, each document is viewed as a sample from the language, and its relevance to $q$ is estimated as the probability that $q$ was generated from this sample.

Using language models for retrieval in IR was initiated by Ponte and Croft [349] who developed a retrieval model based on the multiple-Bernoulli statistical model. Because the multiple–Bernoulli model is computationally expensive[10] [42] and ignores term-frequencies - a useful feature for retrieval [374] - a number of authors, most notably Hiemstra [203] and Song and Croft [403], introduced the multinomial model as a representation of documents which has since become the de facto choice [132, p. 254]. Liu and Croft [276] and Zhai [457, Chap. 6] provide a comprehensive survey of areas–relevance feedback (e.g. [131, 260, 348]), distributed IR (e.g. [396, 453]), query ambiguity (e.g. [133, 134]) and ad hoc retrieval (e.g. [200, 204, 231, 249, 269, 307, 309])–where LMs have been used successfully.

Formally, ranking documents relative to a query $q$ is estimated through Bayes' rule:

$$\begin{aligned} \Pr(d|q) &= \frac{\Pr(q|d)\Pr(d)}{\Pr(q)} \\ &\propto \Pr(q|d)\Pr(d) \end{aligned} \tag{2.23}$$

where $\Pr(d)$ denotes any prior knowledge that $d$ is relevant to any query and $\Pr(q)$ is a normalisation constant. The generic LM in Eqn. 2.23 contains two probability distributions: the distribution of $\Pr(q|d)$ and the distribution of $\Pr(d)$. The probability $\Pr(q|d)$ is usually estimated by assuming query terms are i.i.d:

$$\Pr(q|d) = \prod_{q_i \in q \cap d} \Pr(q_i|d) \tag{2.24}$$

corresponding to a unigram query-likelihood LM. The term-independence assumption is paramount to most retrieval models [132, p. 345] and, despite not reflecting reality [89, Chap. 9], has proved effective for retrieval. The probability $\Pr(q_i|d)$ can then be estimated using e.g. maximum likelihood estimation:

$$\Pr(q_i|d) = \frac{f_{q_i,d}}{|d|} \tag{2.25}$$

---

[10]The multiple-Bernoulli model is given by [457, p. 29]: $\Pr(q|d) = \prod_{q_i \in q \cap d} \Pr(q_i|d) \prod_{q_i \in q, q_i \notin d} \left(1 - \Pr(q_i|d)\right)$; a direct implementation would be equal to $|T| \times |D|$ where $T$ is the number of terms in the vocabulary, and $|D|$ is the number of documents in the used collection. Historically, the multiple–Bernoulli model is the basis for the Binary Independence Model [132, Chap. 7].

where $f_{q_i,d}$ is the frequency of query term $q_i$ in $d$. The probability $\Pr(d)$ is commonly assumed to be uniform for all documents, but research has shown that incorporating prior knowledge regarding $\Pr(d)$ (that is, query-independent information) into the LM can improve retrieval performance. Examples where assuming a non-uniform distribution of $\Pr(d)$ have improved retrieval include age dependent document priors in link structure analysis [200], language models for informational and navigational Web search [230], to exploit link structure of Wikipedia Web pages to improve ad hoc retrieval [231], entry page search [249] and literature search [307].

Hauff and Azzopardi [200], for example, use a "scale-free" distribution of $\Pr(d)$ to estimate popularity scores for a document based on its age and link information. A scale-free distribution of $\Pr(d)$ is a distribution of the form $\Pr(x|\alpha) = \{x^{-\alpha} : \alpha > 1\}$. Their approach is based on the assumption that document popularity scores, defined as "a smoothed ratio of actual over expected number of in-links and normalised to a range between 1 and 3", derived from the WT2g test collection are distributed according to scale-free distribution, but no evidence is presented which supports this claim. The issue with this assumption is shown in Figure 2.5 for synthetic data, where the scale-free distribution is shown as a solid line and the "true" distribution of popularity scores is shown as a dashed line. In this example, the "true" distribution is taken to be an exponential. It can be seen that the scale-free distribution assigns smaller probabilities compared the true distribution for all popularity scores below approximately 2.2 and much higher probabilities to popularity scores above. This discrepancy in the modelling indicates that more accurate, and possibly better, results are obtainable by using the "true" distribution.

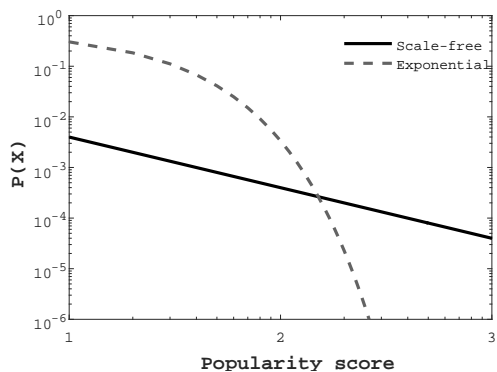

Figure 2.5: Distribution of "popularity" scores for an assumed scale-free distribution (solid line) and a hypothetical "true" exponential distribution (dashed line).

## 2.4.4   Divergence from Randomness

Divergence from Randomness (DFR) is a framework developed by Amati [17] for deriving probabilistic nonparametric ranking models for IR [17, 19]. Here nonparametric means that parameters are determined from the data, instead of the statistical model. DFR stipulates that:

> *The more the divergence of the within-document term-frequency from its frequency within the collection, the more the information carried by the term t in the document d.*

DFR is a generalisation of Harter's 2-Poisson model [17]. According to DFR, a term's weight in a document is the product of two information content functions, $inf_1$ and $inf_2$:

$$inf_1 \cdot inf_2 \tag{2.26}$$

The first informative content function, $inf_1$, is defined as:

$$inf_1 = -\log_2(P_1) \tag{2.27}$$

where $P_1$ is a PDF or PMF which gives the probability of having, by pure chance, $k$ occurrences of a term $t$ in some document. The larger $inf_1$ is, the less its within-document term frequency of $t$ is distributed in accordance with the chosen $P_1$ and hence the higher the informative content. The second informative content function, $inf_2$, is defined as:

$$inf_2 = 1 - P_2 \tag{2.28}$$

where $P_2$ is the probability of accepting a term $t$ as a good descriptor of the current document when the document is compared to its "elite" set and is based on the observation [17] that, while rare, when informative terms appear, they tend to occur highly frequently. Consequently, $1 - P_2$ is the risk of accepting a term $t$ as a good descriptor of the current document.

In DFR, $P_1$ is the models against which a term's divergence is measured to assess its informativeness. This means that selecting the "right" distribution for $P_1$ becomes important to accurately discriminate between informative and non-informative terms. Consequently, using the "right" distribution is crucial for separating informative from non-informative terms [19, 121] which may lead to increased retrieval effectiveness. Choosing $P_1$ has been investigated in previous research [19, 119, 120, 121], but no evidence is offered that suggests that the proposed distributions are a good fit. Thus, also here, evidence-based selection of $P_1$ may lead to more accurate results and, possibly, higher performance.

### 2.4.5 Modelling Score Distributions

Contemporary IR systems calculate a score for each document in the collection in response to a query. Documents are then ranked (sorted in decreasing order of their score) and the top-$k$ documents returned for some $k \geq 1$. For single IR systems, this is the de facto method for retrieving information, but for advanced applications such as distributed IR or filtering, it becomes necessary to merge scores from *multiple* IR systems to produce a final ranked list.

As an example, a distributed IR (e.g. federated or P2P search) system consists of a broker which accepts queries from users, and forwards them to $m$ IR systems, each of which has been allocated some part of a collection $C$. Each of the $m$ IR systems scores the queries against their sub-collection and returns the results to the broker which creates the final ranking after fusing the results from the $m$ systems. Because each IR system may not use the same method to score documents in their allocated sub-collection (and because the characteristics of each sub-collection may vary considerably), the scores can vary considerably. For example, one system may score each document in the interval $[0;1]$ while another may score each document in the interval $[-10;+10]$. The broker now faces the meta-search problem: how to combine the ranked results from each system to a final ranking–a problem faced by most conventional search engines [34].[11]

---

[11] Combining results is only one part of the meta-search problem. See [395] for details.

Score normalisation maps disparate scores into a common interval and is an important step for the broker [92, 288]. Early research [169, 263] used simple range-based normalisation techniques in an attempt to improve the ranked list produced by the broker. For example, Lee [169] proposed to normalise a score $s_i$ by:

$$\hat{s}_i = \frac{s_i - \min(s)}{\max(s) - \min(s)} \tag{2.29}$$

where $\hat{s}_i$ is the normalised score and $s$ is the set of all scores. The $\hat{s}_i$ normalisation and other normalisations such as MIN, MAX, COMBSUM and COMBMNZ [169] were found to work reasonably well for distributed IR, fusion and filtering [34]. Despite their success, these techniques have been criticised [34, 288] for not considering the distribution of scores. Using the score distribution allows the use of their statistical properties which may better generalise to future unseen documents, rather than using just the values of the scores [28]. Modelling the score distribution is important, as it has been observed that the distribution of relevant documents is different from that of non-relevant documents [287, 288].

As an example, Figure 2.6 shows synthetic score distributions for "relevant" and "non-relevant" documents, generated specifically to illustrate overlapping score distributions; the two curves

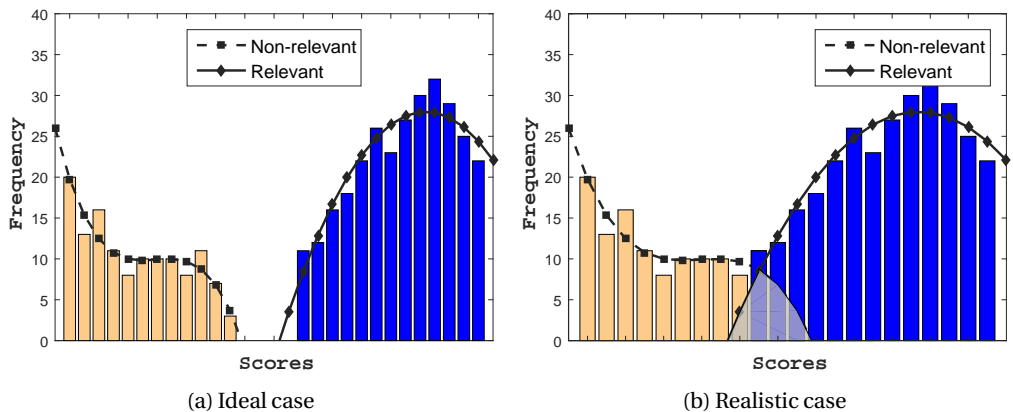

(a) Ideal case          (b) Realistic case

Figure 2.6: Empirical distributions of scores for non-relevant (light grey/yellow) and relevant (dark grey/blue) documents using synthetic data in the ideal (2.6a) and realistic (2.6b) case for a single hypothetical query. Example reproduced from [147].

(dashed for non-relevant and solid for relevant) represent some PDF or PMF.

Assuming binary relevance, the probability of relevance can be calculated [30] given the two densities (and their mix-weight in areas where they overlap; see Fig. 2.6b). Precision, recall and other performance metrics can also be calculated at any threshold [33]. This makes modelling score distributions as part of score normalisation a critical task for distributed retrieval or fusion/meta-search [32].

A key challenge of score distribution modelling therefore is to select the distributions that best fit the non-relevant and relevant data. In Figure 2.6a the two score distributions are separated (their curves do not intersect) and based on the scores alone, it is possible to distinguish non-relevant documents from relevant ones. In reality, the two score distributions likely intersect, as in Figure 2.6b, where the shaded area denotes the region of intersection where both non-relevant and relevant documents have similar scores. Doloc-Mihu and Raghavan [148] investigated how to minimise this overlap by using kernel methods to increase separation between distributions for adaptive image retrieval. They found that using a Gaussian distribution (Eqn. 2.8) to model relevant images, and an exponential distribution (Eqn. 2.10) to model non-relevant images (represented as colour histograms) gave the best results.

While several combinations of models have been investigated over the years–including Gaussian–Gaussian with equal [416] and unequal variance [417]; two exponentials [417]; 2-Poisson [72]; mixture of Gaussians and gamma [233] and two (shifted) gamma models [57]–the use of a Gaussian and exponential model to quantify the distribution of scores of relevant and non-relevant documents is currently the most common choice for score distribution modelling [28, 233, 368]. The use of a Gaussian/exponential model has been used for combining outputs of search engines [287] and information filtering [29, 125] that includes adaptive filtering where only documents above some score threshold are considered for dissemination [460]. For example, in combining the results of several search engines, Manmatha et al. [287] empirically demonstrate that the Gaussian/exponential models can approximate the score distributions for relevant and non-relevant data taken from the TREC-3, TREC-4 and TREC-6 ad-hoc tracks on a per-query basis.[12] In the absence of relevance information, the authors propose a Gaussian/exponential mixture model to calculate the posterior probabilities of relevance given the scores via Bayes' rule. They find that using their model to combine the output of multiple search engines, such as INQUERY and SMART, substantially improves performance over using a single search engine and that results are generally better when relevance assessments are available.

While score distribution modelling is practical for automatic query performance prediction [135]; outlier detection [172]; information filtering [30, 125]; combining search engine results [287]; distributed IR [32, 57]; and video [443] and image retrieval [148], research on the theoretical properties of scoring distributions also exists. Rijsbergen [364, p. 122] remarks that for the SMART search engine, there was no evidence to support the observations that the distributions of scores for relevant and non-relevant documents were the same or even Gaussian, as originally formulated by Swets [416]. Evidence in support of Swets' model was given by Madigan et al. [284], who also find that the Gaussian/exponential combination is not valid from the perspective of extreme value theory.

Robertson [368] formulates the convexity hypothesis, which is a condition on the probability of relevance of a document with a specific score: the higher the score, the higher the probability of relevance [147]. He tests an exponential/exponential, Gaussian/Gaussian, Poisson/Poisson, gamma/gamma and a Gaussian/exponential model to quantify the distribution of relevant/non-relevant documents. He finds that the Gaussian/exponential model (the model combination used by e.g. Manmatha et al. [287]) violates the convexity hypothesis, which implies that a suitable reordering of the documents in the range of scores where convexity is not observed will improve the IR system.

Arampatzis and Robertson [34] consider suitable hypotheses to apply separately to the distribution of scores of relevant and non-relevant documents for popular combinations of distri-butions. The strong scoring distribution hypothesis states that the score distributions - when modelled using some statistical model of relevant and non-relevant documents - should approximate Dirac's delta function, shifted to lie on the maximum/minimum score for relevant/non-relevant documents. The idea behind the hypothesis is to achieve separation between the two distributions (i.e. minimise their overlap; see Fig. 2.6b). On this basis, the authors find that the Gaussian and gamma model are good candidates for modelling the score distribution of relevant documents and that the Gaussian, exponential, Poisson and gamma can all be used to model the scores for non-relevant documents. This suggests a rejection not only of the two exponential model [417] and the 2-Poisson [72] model, but also of the two Gaussian model [30]. Recognising that an approximation to Dirac's delta function is not needed for perfect ranking, the weak scoring distribution hypothesis states that the score distributions should be able to achieve full separation for some parameter values. The weak scoring distribution hypothesis does not reject any of the combinations of models suggested.

---

[12]The authors also experiment with Poisson/Poisson, exponential/Poisson and exponential/Gamma models, but find that these in general do not fit the data well [287, p. 3].

### 2.4.6 Predictive Modelling

The goal of predictive modelling is to apply a statistical model to data with the purpose of predicting future observations [393]. For information retrieval, examples may include predicting (i) the size of an inverted index, (ii) the volume of a query stream or (iii) the number of citations a paper will receive. Predictive modelling serves a number of necessary scientific functions [393] and it is paramount that a predictive model is as "correct" as possible to assess the distance between theory and practice. For example, Mao and Lu [290] attempt to predict the popularity of articles by mining clicks of each article as its access changes over time from PubMed query logs. They fit six different regression models (including the power law and log-normal distribution) to the time series of article accesses to quantify the correlation between the historical data and the predictions outputted by each regression model. Their results show that click trends of PubMed articles can be substantially better predicted for new articles when modelled using a log-normal instead of a power law.

A second example is predicting the size of an inverted index using $\gamma$-coding [289, Chap. 5]. Using Zipf's law which states that the frequency of any term is inversely proportional to its rank $i$ to quantify the distribution of terms in a collection, there is a constant $c'$ such that:

$$\mathrm{cf}_i = \frac{c'}{i} \tag{2.30}$$

where $\mathrm{cf}_i$ is the collection frequency of the $i^{\text{th}}$ term. By choosing a different constant $c$, the fractions $c/i$ become relative frequencies and sum to 1 i.e. $c/i = \mathrm{cf}_i/|T|$ where $|T|$ is the cardinality of the set of terms in the collection, the constant $c$ becomes:

$$c = \frac{1}{H_M} \tag{2.31}$$

where $H_M$ is the $M^{\text{th}}$ harmonic number, and $M$ is the number of distinct terms in the collection. Setting $M = 400,000$ and approximating $H_M$ with $\log(M)$, $c$ becomes:

$$c = \frac{1}{H_M} \approx \frac{1}{\log M} \approx \frac{1}{13} \tag{2.32}$$

meaning that the $i^{\text{th}}$ term has a relative frequency of $1/(13i)$.

The expected average number of occurrences of $t$ in a document of length $L$ is:

$$L\frac{c}{i} \approx \frac{200 \cdot \frac{1}{13}}{i} \approx \frac{15}{i} \tag{2.33}$$

By stratifying the vocabulary of the inverted index into blocks, such that, on average, term $i$ occurs $15/i$ times per document, the size of a collection with 400,000 documents is estimated to be approximately 224MB - about one-fourth of the original size. However, this estimate can be more than twice as big [289] if the term collection frequency distribution is not (approximately) Zipfian.

### 2.4.7 Real Life Examples

The misuse of statistical models has had serious financial repercussions. For example, the 100 billion–USD hedge fund Long Term Capital Management blowup was the result of an event ten standard deviations away from the mean in a Gaussian distribution, which should take place once every 800 trillion years [228]. However, the assessment of the potential for overall losses was based on a Gaussian model, which severely underestimates the chance of a cataclysmic event [81]. Using a heavy-tailed distribution showed that the loss incurred was an eight-year event and not an 800 trillion–year event [154]. Similarly, the "Black Monday" stock market crash on October 19, 1987, should have been a $10^{-68}$ event using a Gaussian model [154]. As the author suggests, "surely the model was wrong" [228].

Babbel et al. [43] give examples on how the consideration of non-Gaussian distributions could have changed U.S. court case outcomes from employment discrimination to criminal prosecution. For example, in the defence of a man charged with the murder of his wife, the prosecutor showed that the probable driving time between his departure from work and the commission of the crime at his home in no way suggested a Gaussian model, as otherwise claimed by his defence. Statistical models are also important in a number of areas where it is critical to use the best-fitting distribution such as hydrology [199], where extreme, but very rare, events must be accurately modelled to e.g. determine how high a dam must be to limit flooding or how high off the water a bridge must be to remain above high water mark, and in fatigue analysis of wind turbines [378] in order to schedule (expensive) maintenance.

## 2.5 Summary

This chapter introduced statistical modelling with emphasis on statistical models: families of parameterised probability density/mass functions which naturally accommodate data variance and are used to quantify the distribution of data. Statistical models are widely used in IR, and several examples were presented. The key observation from these examples is that assumptions or unsubstantiated choices of the distribution of some property (i.e. term frequencies or Web site "popularity scores") frequently guide modelling and other efforts in IR. Because it is not clear, however, if such assumptions or choices are justified and what the potential advantages of using the "true" distribution are, the next chapter introduces a statistically principled approach to fit and compare multiple statistical models before selecting the best-fitting, statistical model.

# 3 A Principled Approach to Statistical Model Selection

This chapter introduces a principled approach to statistical model selection: given some sample of i.i.d. data $x$ and a non-empty set of statistical models, determine the *single* statistical model that best quantifies, or fits, the distribution of $x$.

This chapter is organised as follows. Section 3.1 motivates the need for model selection through several examples in information retrieval (IR) and real life. Section 3.2 lists the statistical models used in our principled approach. Sections 3.3 and 3.4 review several approaches for estimating the parameters of statistical models, and for selecting the best-fitting statistical model. Section 3.5 presents our principled approach to statistical model selection and Section 3.6 summarises the chapter. This chapter makes use of the notation introduced in Section 2.1.

## 3.1 Introduction

Choosing a suitable statistical model is critical to all statistical work with data [111, p. 1]. A suitable statistical model may be e.g. (i) the most parsimonious model, (ii) the statistical model with the fewest parameters, (iii) the statistical model that provides the best approximation to the distribution of $x$ or (iv) the statistical model that strikes the best balance between simplicity (fewer parameters to estimate, leading to lower variability at the cost of model bias) and complexity (more parameters to estimate, leading to higher variability but lower model bias) [111, pp. 9–14]. The approach proposed here focusses on selecting the statistical model that provides the best approximation, or fit, to the distribution of $x$.

Choosing a suitable statistical model can be stated as follows. Suppose there are $K$ statistical models indexed by $k$, $\mathcal{M}_k$, each parameterised by $\theta_k \in \Theta_k$ with corresponding log-likelihood functions $\mathcal{L}_k(\theta_k|x)$ (introduced in Section 3.3.2) and an i.i.d. sample of data $x$. According to the $M$-closed framework [63], the assumption is made that one of the $K$ models is the "true" model. While the assumption that (i) a "true" model exists (see Section 2.3) and (ii) is among the $K$ models under consideration may be controversial, it is used here to frame the problem. The goal of model selection is then, given the set of $K$ models, to select which one is the "true" model. Because it is unlikely that one of the $K$ models is the "true" model [163, p. 27], in practice, model selection seeks to determine which of the $K$ models is the best approximation to the "true" model. In this chapter, we denote such a model as the best-fitting model.

The remainder of this chapter introduces a principled approach to statistical model selec-tion. Our approach is based on the *frequentist* paradigm as it is (i) objective [71, Chap. 1] and (ii) the most widely used statistical paradigm in IR [18].  Referring back to Section 2.2.2 (and paraphrased below), our approach consists of three steps:

- **Step 1:** Model formulation: Specify the statistical model representing the likelihood of the data.

- **Step 2:** Model estimation: Estimate the likelihood of the data under the specified statistical model.

- **Step 3:**  Model selection:  Select the statistical model that minimises some, typically information-theoretic, measure.

Model formulation is the topic of Section 3.2.  Model estimation is introduced in Section 3.3, where several popular approaches to estimating the optimal parameter values of a parameterised statistical model are discussed.  Section 3.4 introduces and discusses approaches to model selection.

## 3.2   Model Formulation

The purpose of model formulation is to specify the statistical model representing the likelihood of the data.  In this thesis, we apply model selection to the models listed in Tables 2.2 and 2.3 in Chapter 2.  The likelihood of the data under each of these models is obtained using one of the model estimation techniques discussed in the next section.

## 3.3   Model Estimation

Given an i.i.d. sample of data $x$ and a parameterised statistical model $\{f(x|\theta) : \theta \in \Theta\}$, *model estimation* is the problem of estimating the values $\hat{\theta} \in \mathbb{R}^m$ that best "explain" $x$. While several parameter estimation techniques exist, such as (generalised) method of moments [195], maximum spacing estimation [357], hill estimator [150, 207] and minimum $\chi^2$ [62], this section focusses on two estimation techniques: maximum likelihood estimation (MLE) and maximum a posteriori (MAP). Both techniques are widely used [124, 312, 323, 383] and represent different views on parameter estimation. MLE is a frequentist approach to parameter estimation, which returns an estimate for each parameter depending on how "likely" the parameters "explain" the sample.  In contrast, MAP is a Bayesian approach that encodes prior belief regarding the distribution of parameters into the estimation process. Despite our approach being frequentist, we include MAP to highlight that other approaches can be used. Regardless of the method used, we refer to a model whose parameters have been estimated as a *fitted* model.

The following nomenclature will be used:

- An *estimator*, when applied to data, constructs a (point) estimate [266, Chap. 4]. All estimators/estimates in this thesis are written with a circumflex over the symbol i.e. $\hat{\theta}$.

- A *point estimator*, or statistic, of a population parameter is a single value that represents the true population parameter.

- An estimator is *consistent* if the constructed estimator $\hat{\theta}$ converges to the true value $\theta_0$ as the sample size goes to infinity.

- An estimator is *unbiased* if the deviation of the expectation ($\mathbb{E}$) of $\hat{\theta}$ from the true value is 0 i.e. $\mathbb{E}[\hat{\theta}] - \theta_0 = 0$ [266, Chap. 4].

- An estimator is *efficient* if it is the minimum variance unbiased estimator i.e. $Var(\hat{\theta}) \leq Var(\hat{\theta}^*)$ where $Var(\hat{\theta})$ denotes the variance of the optimal parameters, and $Var(\hat{\theta}^*)$ is the variance of all other estimates of $\theta$.

Note that consistency and unbiassedness of an estimator are not equivalent. Consistency relates to the sampling distribution of the estimator as the sample size increases. Unbiassedness relates to the expectation of the sampling distribution of the estimator. To motivate the use of MLE and MAP, we first introduce ordinary least square which is the most common method used to estimate power law exponents [56]. We focus on power laws as these statistical models are the topic of Chapter 4.

### 3.3.1 Ordinary Least Square

Ordinary least squares (OLS) is a widely used estimator to determine the function $h : \mathbb{R} \to \mathbb{R}$ that best characterises the linear relationship between an independent (or explanatory) variable $x$ and a dependent (or response) variable $y$. More specifically, given a list of $n$ pairs of observations $(x_i, y_i)$, find the line that best relates $y_i$ to $x_i$ for all $i$. In its typical formulation, this translates into finding the coefficients $a$, $b$ in:

$$\hat{y}_i = a + x_i b + \epsilon_i : i = 1, 2, \dots, n \tag{3.1}$$

where $\hat{y}_i$ is the approximation to $y_i$ using the values for $a$ (the $y$-intercept or regression constant), $b$ is the slope for the line represented by Eqn. 3.1 and $\epsilon_i$ is the error term for observation $x_i$.[1] As $\epsilon = \{\epsilon_i\}$ is assumed to be normally distributed [56] with mean $\mu = 0$ and $\sigma^2$, it will be left out of the formula from here on. While OLS itself does not require the assumption that error terms are normally distributed, calculation of confidence intervals, $p$-values and various significance tests are based on the assumption of normally distributed errors.

The smaller the residual (the difference between $y_i$ and $\hat{y}_i$ for some $i$), the better the approximation for data point $x_i$. Thus, the goal is to find the line that minimises the sum of the residuals. One expression that minimises this sum is:

$$\text{minimise} \quad \sum_{i=1}^{n} (y_i - \hat{y}_i) \tag{3.2}$$

but the problem is that residuals of different signs can cancel each other out. An example is shown in Figure 3.1 where the dashed line corresponds to the fitted line, and the solid line to the observations. Because the magnitude of the residuals for $x_1$ and $x_3$ is identical but has opposite signs they cancel out, and because the residual for $x_2$ is 0, the fit would be considered optimal by Eqn. 3.2. Thus, Eqn. 3.2 is not appropriate for OLS. Taking the absolute value of the residuals in Eqn. 3.2 solves the problem, but requires linear programming techniques to solve [1]. A third expression that minimises the residuals is to square the residuals. This avoids the residuals cancelling out and is simple to obtain compared to using linear programming [1].

---

[1] $\epsilon_i$ jointly comprises all other factors that affect $y_i$.

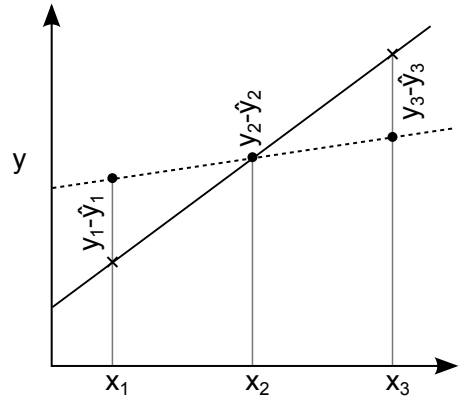

Figure 3.1: Minimising the residuals in OLS. The dashed line corresponds to the fitted line.

OLS's estimate of $a$ and $b$ is defined as the value $\delta$ that minimises the squared residuals:

$$\delta = \sum_{i=1}^{n} \left( y_i - \hat{y}_i \right)^2 = \sum_{i=1}^{n} \left( y_i - (a + x_i b) \right)^2 \tag{3.3}$$

As the quadratic expression is minimum when its first-order derivatives vanish, the optimal solution for $a$ is:

$$\frac{\delta}{\partial a} = 2na + 2b \sum_{i=1}^{n} x_i - 2 \sum_{i=1}^{n} y_i = 0 \tag{3.4}$$

and for $b$:

$$\frac{\delta}{\partial b} = 2b \sum_{i=1}^{n} x_i^2 + 2a \sum_{i=1}^{n} x_i - 2 \sum_{i=1}^{n} y_i x_i = 0 \tag{3.5}$$

Solving both equations gives [2, 51]:

$$a = \hat{y} - b\hat{x} \tag{3.6}$$

and

$$b = \frac{\sum_{i=1}^{n} (y_i - \hat{y})(x_i - \hat{x})}{\sum_{i=1}^{n} (x_i - \hat{x})^2} \tag{3.7}$$

where $\hat{x}$ and $\hat{y}$ are the estimated mean values of $x$ and $y$, respectively. If the Gauss–Markov Theorem is fulfilled and the Gauss–Markov conditions are met [356], OLS gives the best linear, unbiassed (and maximum likelihood) estimates of $\alpha$ and $\beta$ [2, 51].

However, the simplicity of OLS masks its sensitivity to outliers as a consequence of squaring the differences in the denominator in Eqn. 3.7. If $x_i$ is an outlier, the resulting contribution from the term $(x_i - \hat{x})$ will inflate the final result of the denominator, as the difference between e.g. $x_i = 50$ and $\hat{x} = 10$ is only 40 whereas the difference between $x_i = 50^2$ and $\hat{x} = 10^2$ is 2,400.

This presence of outliers in a dataset will result in a substantially lower estimate for $b$ when using OLS. Figure 3.2a shows an example where OLS was used to fit a line to a dataset consisting of discrete power law numbers with $\alpha = 2.5$ generated using the R package "`poweRlaw`". The OLS estimate of $\alpha = -1.67$ is much smaller than the $\alpha = 2.5$ value used to generate the dataset. The inset in Figure 3.2a shows how OLS, for 500 synthetic power law datasets (of size $n = 20,000$), consistently underestimates the exponent.

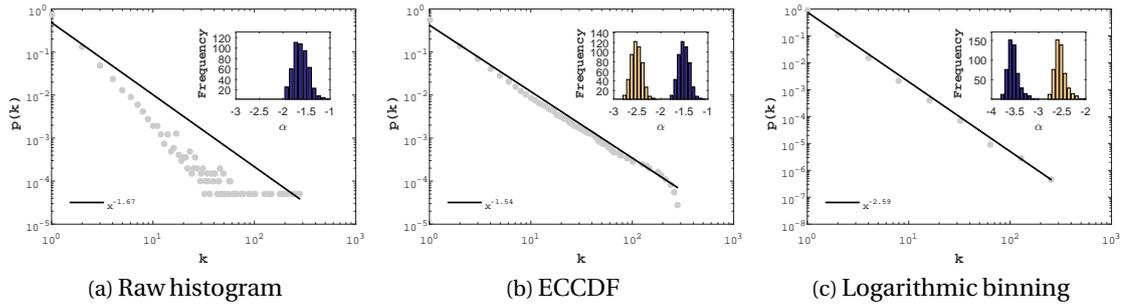

(a) Raw histogram  (b) ECCDF  (c) Logarithmic binning

Figure 3.2: Ordinary least squares regression to the same data using the three graphical methods introduced. The dataset consists of power law generated discrete random numbers with $\alpha = 2.5$. Insets show the estimated $\alpha$ values for $n = 500$ datasets generated identically as the one shown. Fig. 3.2b and 3.2c show two $\pm 1$ shifted distributions: the dark/grey (blue/red) distribution shows the parameter value estimated by linear least squares without/with correction, respectively.

Dealing with the outliers in the tail of the distribution by using the ECCDF (see Figure 3.2b) provides a qualitative better fit. However, when using the ECCDF, Newman [328] shows that the technique underestimates $\alpha$ by 1. This uncorrected exponent is shown in Figure 3.2b, and the distribution of 500 exponents (same data as used for the raw histogram) is shown in dark grey in the inset. The light grey distribution, also in the inset, shows the corrected distribution where the exponents have 1 added. Figure 3.2c shows that using logarithmic binning also provides a qualitatively better fit than the raw histogram. However, using logarithmic binning *overestimates* the value of $\alpha$ by $-1$ [328]. Identically to Figure 3.2b, the inset of Figure 3.2c shows the distribution of the uncorrected (dark grey) and corrected (light grey) exponent values.

While the OLS is a special case of MLE provided the Gauss–Markov conditions and theorem are met, this is not true in many cases [322, Chap. 21] and thus other methods relaxing this assumption are required. The next sections review two such methods: MLE and MAP.

### 3.3.2 Maximum Likelihood Estimation

Maximum likelihood estimation (MLE) is a standard approach to parameter estimation and inference in statistics [323] that provides a consistent and unbiassed point estimate of each parameter of the considered statistical model under certain mild conditions [184]. Intuitively, MLE tunes the model parameters $\theta$ so that the product of density/mass values at the sample observations is maximised [21] i.e. so that the probability of observing the data is maximised.

This tuning can be explained as follows. Because any choice of value(s) for $\theta$ will make some data more probable than other data, the purpose of the MLE tuning is to determine the values for $\theta$ that most likely generated the observed data [323]. For $\theta$ containing $k$ parameters, this requires searching the $k$-dimensional hyperplane spanned by each parameter in $\theta$. This is accomplished by the *likelihood function* which conditions $\theta$ on the data and then searches for the optimal values of $\theta$. More formally, given a statistical model $\mathcal{M}_\theta(x) = \{f(x|\theta)\}$, the likelihood function $\mathcal{L}$ is defined as:

$$\mathcal{L}(\theta|x) = f(x|\theta) \tag{3.8}$$

where the set notation has been dropped for clarity. By conditioning $\theta$ on $x$, the likelihood function measures the support provided by the data for each possible value of the parameter. Assuming that the data $x$ are i.i.d., Eqn. 3.8 becomes:

$$\mathcal{L}(\theta|x) = \prod_{i=1}^{n} f(x_i|\theta) \tag{3.9}$$

Because the product of probabilities in Eqn. 3.9 may cause numeric underflow, the log-likelihood function:

$$\mathcal{L}(\theta|x) = \sum_{i=1}^{n} \log(f(x_i|\theta)) \tag{3.10}$$

is often used instead, as the logarithm is a monotonically increasing in $\theta$, and hence the same values of $\theta$ maximise both Eqn. 3.9 and 3.10. Henceforth, whenever we write $\mathcal{L}$ we refer to the log-likelihood function.

The *maximum likelihood estimate* is then the parameter vector $\hat{\theta}$ maximising Eqn. 3.10:

$$\hat{\theta} = \underset{\theta \in \Theta}{\operatorname{argmax}}\, \mathcal{L}(\theta|x) \tag{3.11}$$

In many cases, the definition of the likelihood function $\mathcal{L}$ is divided with the sample size $n$ (see Section 2.1). In doing so, the likelihood function tends to converge in probability to a constant function, meaning that $\hat{\theta}$ becomes increasingly concentrated near the true parameter value $\theta_0$, which accounts for the consistency of the maximum likelihood estimator [464, p. 40][184, p. 524].

MLE is closely linked to the *Kullback–Leibler* (KL) divergence–a measure of the divergence of the density of a parametric statistical model $f(x|\theta)$ to the "true", but unknown, density of $x$, $g(x)$ [176]. The KL divergence for discrete probability distributions $f$ and $g$ is defined as:

$$\mathrm{KL}(g(x), f(x|\theta)) = \sum g(x) \left[ \log \frac{g(x)}{f(x|\theta)} \right] = \mathbb{E}\left[ \log \frac{g(x)}{f(x|\theta)} \right] \tag{3.12}$$

where $\mathbb{E}$ denotes the expectation. $\mathrm{KL}(g(x), f(x|\theta)) \geq 0$ with equality only if $g(x) = f(x|\theta)$. Notice, that Eqn. 3.12 is non-symmetric i.e. $\mathrm{KL}\big(g(x), f(x|\theta)\big) \neq \mathrm{KL}\big(f(x|\theta), g(x)\big)$. For example, if $f(y|\theta_f)$ and $g(y|\theta_g)$ are normal distributions with $\theta_f = \{\mu = 0, \sigma^2 = 0.5\}$, $\theta_g = \{\mu = 3, \sigma^2 = 2\}$ and e.g. $y$ ranges from $-2$ to $2$ in unit steps, then $\mathrm{KL}(g(x|\theta_g), f(x|\theta_f)) = 116.21$ and $\mathrm{KL}(f(x|\theta_f), g(x|\theta_g)) = 10.35$. A symmetric version of the KL divergence can be found in e.g. [37, p. 267].

The best approximation of $f(x|\theta)$ to $g(x)$ can be found by minimising:

$$\underset{\theta}{\operatorname{argmin}} \quad \mathrm{KL}\big(g(x), f(x|\theta)\big) \tag{3.13}$$

Then, since:

$$\text{KL}\big(g(x), f(x|\theta)\big) = \sum g(x)\left[\log\frac{g(x)}{f(x|\theta)}\right]$$
$$= \sum g(x)\log g(x)\mathrm{dx} - \sum g(x)\log f(x|\theta) \quad (3.14)$$

where the first term does not depend on $\theta$, Eqn. 3.13 has the same solutions as:

$$\underset{\theta}{\text{argmin}} \quad -\sum g(x)\log f(x|\theta) \quad (3.15)$$

which is equivalent to minimising the (negative) log-likelihood of Eqn. 3.10 [382, p. 276]. For each value of the parameter vector $\theta$, applying the strong law of large numbers [111, p. 25]:

$$n^{-1}\mathscr{L}(\theta|x) \xrightarrow{a.s.} \sum g(x)\log f(x|\theta)\mathrm{dx} = \mathbb{E}\big[\log f(x|\theta)\big] \quad (3.16)$$

where a.s. means converges almost surely (i.e. with probability 1) if the sum is finite. Thus:

$$\mathbb{E}\big[\log f(x|\theta)\big] = \mathbb{E}\big[\log f(x|\theta)\big] - \mathbb{E}\big[\log g(x)\big] + \mathbb{E}\big[\log g(x)\big]$$
$$= \mathbb{E}\left[\log\frac{g(x)}{f(x|\theta)}\right] + \mathbb{E}\big[\log g(x)\big] \quad (3.17)$$
$$= \text{KL}\big(g(x), f(x|\theta)\big) + \mathbb{E}\big[\log g(x)\big]$$

where the second term is independent of $\theta$. Thus, minimising the negative log-likelihood converges to the sum of the entropy under the true distribution $g(x)$, plus the KL divergence between the true distribution and the candidate probability distribution. Therefore, minimising the KL divergence:

$$\hat{\theta} = \underset{\theta}{\text{argmin}}\ \text{KL}\big(g(x), f(x|\theta)\big) \quad (3.18)$$

is equivalent to maximising the log-likelihood (Eqn. 3.11) which, again, is equivalent to finding the density $f(x|\theta)$ that is closest to the true density $g(x)$.

A necessary condition for $\hat{\theta}$ to exist is [323]:

$$\frac{\partial\mathscr{L}(\theta|x)}{\partial\theta_i} = 0 : i = 1,...,m \quad (3.19)$$

where $\theta_i$ indexes the $i^{\text{th}}$ parameter in $\theta$ and $\partial$ denotes the partial derivative. Furthermore, to ensure $\mathscr{L}(\theta|x)$ is a maximum estimator, the shape of the log-likelihood function in the neighbourhood of $\hat{\theta}$ must be convex [323]:

$$\frac{\partial^2\mathscr{L}(\theta|x)}{\partial^2\theta_i} < 0 : i = 1,...,m \quad (3.20)$$

meaning that $\mathscr{L}$ is of differentiability class $C^2$. In cases where $\mathscr{L}(\theta|x)$ cannot be solved directly, numerical approximation algorithms such as the Newton–Raphson algorithm can be used to optimise $\theta$ by improving upon an initial set of user-supplied values for $\theta$. Such optimisation algorithms, however, do not guarantee to find a set of parameter values that uniquely maximise the log-likelihood function [323]. In these scenarios, the algorithms converge to a local maximum. The estimated parameters may converge to a local maximum if (i) $\mathscr{L}(\theta|x)$ is not convex [68] or (ii) there is no available method which guarantees to provide an initial guess of $\theta$ that is within the region of the global maximum.[2] Notice that Eqn. 3.10 assumes a global maximum exists. In situations where $\mathscr{L}(\theta|x)$ converges to a local maximum (or is suspected of converging to a local maximum), a common approach is to restart/reseed the optimisation algorithm with a new guess for $\theta$ [323]. Illustrative examples of using MLE can be found in e.g. [184, 201, 323].

### 3.3.3 Maximum A Posteriori Estimation

Maximum a posteriori (MAP) estimation seeks to maximise the posterior distribution given the data:

$$\hat{\theta} = \underset{\theta \in \Theta}{\operatorname{argmax}} \Pr(\theta|x) \tag{3.21}$$

and is an extension of MLE, where prior belief of the parameters is encoded [201] by using Bayes' rule (restated here from Section 2.2.1):

$$\Pr(\theta|x) = \frac{\Pr(x|\theta)\Pr(\theta)}{\Pr(x)} \tag{3.22}$$

where $\Pr(\theta|x)$ is the posterior distribution, $\Pr(x|\theta)$ is the likelihood, $\Pr(\theta)$ is the prior distribution and $\Pr(x)$ is the probability of the evidence.

Applying Bayes' rule to Eqn. 3.21 [201]:

$$\begin{aligned} \hat{\theta} &= \underset{\theta \in \Theta}{\operatorname{argmax}} \left[ \frac{\Pr(x|\theta)\Pr(\theta)}{\Pr(x)} \right] \\ &\propto \underset{\theta \in \Theta}{\operatorname{argmax}} [\Pr(x|\theta)\Pr(\theta)] \\ &= \underset{\theta \in \Theta}{\operatorname{argmax}} [\mathscr{L}(\theta|x)\Pr(\theta)] \end{aligned} \tag{3.23}$$

where, in the last line, we have used $\Pr(x|\theta) = f(x|\theta)$. In Eqn. 3.23, the prior information is used to make a direct estimate by maximising $\theta$. This makes MAP a special case of Bayesian estimation, which places a probability distribution over the possible parameters [201]. This also means that MAP is a generalisation of MLE, and when $\Pr(\theta)$ is uniform $\hat{\theta}_{\mathrm{MLE}} = \hat{\theta}_{\mathrm{MAP}}$. Furthermore, MLE and MAP will converge to the same parameter values for $\theta$ as the sample size tends to infinite. MAP has the added advantages of encoding [201]:

- The maximum a posteriori value (as MAP) for the data-generating parameters.

- The expectation of parameter estimates.

- The variance of the parameter estimates indicative of estimation confidence.

but comes at the cost of calculating the probability of the evidence, $\Pr(x)$, in Eqn. 2.6 as a fixed point estimate is no longer sought. By marginalising over $\theta$, the probability of evidence is given by:

$$\Pr(x) = \int_{\theta} \Pr(x|\theta)\Pr(\theta)d\theta \tag{3.24}$$

---

[2]See e.g. [67] for other numerical approximation algorithms.

which is rarely analytically tractable, as the integral is of a highly variable function over, typically, a high-dimensional parameter space and there might be unknown variables [201]. Monte Carlo simulations [325, 326] or Gibbs sampling [362] can be used to approximate the integral [266, Chap. 4], but conjugate priors - families of distributions that have the same functional form as $\Pr(x|\theta)$ and have convenient mathematical properties that substantially simplify calculations [164] - are typically preferred [164].

The use of conjugate priors can be seen as regularising the estimated parameters by treating them as if they were drawn from a random process; this, in turn, avoids the high variance often associated with MLE estimates [201]. Furthermore, an additional advantage of MAP (and Bayesian estimation) is its ability to obtain good parameter estimates from less data compared to MLE, which must get all its information from the data. Despite this, there are many cases where using MAP (and Bayesian estimation) remains infeasible despite the considerable advancements in computational power over the last few decades [164].

## 3.4 Model Selection

Model selection is the problem of using data (evidence) to select one model from a list of models $\mathcal{M}_1, ..., \mathcal{M}_K$ [438] and is critical to all statistical work with data [111, 465] as several competing models may vie for supremacy [315, 114],[397, pp. 439] and [162, pp. 52-53].[3]

A quantitative measure of a model's fit to data is called a *goodness-of-fit* (GOF) test, the purpose of which is to determine to what extent a fitted model adequately describes the observations in the data [208, 299]. If the discrepancy between the fitted model and the data is sufficiently small, one can say the model is "correct" [208].

GOF tests can be classified as *absolute*, *relative* and *parsimonious* [296, 419]. Below is a non-exhaustive list for each category:

**Absolute GOF tests** do not use an alternative model as basis for comparison but consider each statistical model's fit in isolation. Members of this class include $\chi^2$ [342], Akaike's Information Criterion (AIC) [11], Bayesian Information Criterion (BIC) [385], the Kolmogorov–Smirnov test (KS), the Lilliefors test [270], Clarke's test [115], the Cramér–von–Mises (CM) statistic and the Anderson–Darling statistic (AD).

**Relative GOF tests** compare the fit of a statistical model to a baseline or *null* model. The null model is assumed to be the "true", but unknown model that generated the observed data. Members of this class include the likelihood ratio (LR) test and its variations such as Vuong's LR test [434], the Tucker–Lewis index (TLI), the Nonnormed Fit Index (NFI) and the comparative fit index (CFI)[299].

**Parsimonious GOF tests** favour statistical models with fewer parameters over statistical models with many parameters. Members of this class include minimum description length (MDL) [186], parsimonious CFI (PCFI) and parsimonious NFI (PNFI).

---

[3]Feller, in criticising the use of the logistic model, writes: "...not only the logistic distribution but also the normal, the Cauchy, and other distributions can be fitted to the same material with the same or better goodness of fit.".

It is beyond the scope of this chapter to address all of the GOF tests listed above. Consequently, the GOF tests focussed on here are those that (i) are readily available in most statistical software packages, (ii) are among the least controversial [113] and (iii) are among the most widely used GOF tests [25, 36, 180, 297, 386]: AICc, MDL, Vuong's LR, KS and AD.

The following concepts will be used in the discussion of the GOF tests.

- A *test statistic* is a standardised value calculated for some sample of data $x$.

- A *test*, in this chapter, is a function that maps values of the test statistic to either 0 or 1, where "0" implies acceptance of the null hypothesis $H_0$ and "1" implies a rejection of the null hypothesis $H_0$ and an acceptance of the alternative hypothesis $H_1$.

### 3.4.1 Akaike's Information Criterion with Finite Sample Correction

Given a sample of i.i.d. data $x$, Akaike's Information Criterion (AIC) [11] is an information-theoretic interpretation[4] of the likelihood function given by:

$$\text{AIC}(x) = -2\mathscr{L}(\hat{\theta}|x) + 2|\hat{\theta}| \tag{3.25}$$

which is an asymptotically unbiased estimator of the Kullback–Leibler divergence between the true, but unknown, data-generating density and the hypothesized density [253, 350]. The AIC can be applied to nested and non-nested models equally well, has no null hypothesis [87, Chap. 7] and is a descriptive measure rather than a significance test [381].

Because Eqn. 3.25 was derived under the condition that the sample size is infinite, in the presence of finite-sized samples, AIC has a non-vanishing bias that depends on the number of fitted parameters. This limits its use particularly in instances where the sample size is not much larger than the number of fitted parameters of the candidate model [214]. Consequently, a number of alternative AIC formulations and bias adjustments to the AIC, such as [11, 39, 65, 194, 214, 385, 435], have been proposed in the literature. These attempt to penalise less parsimonious models by using the number of parameters in the likelihood function and, in some cases, the sample size. In Eqn. 3.25, for example, the term $2|\hat{\theta}|$ is the penalty term. We refer collectively to such alternatives as AIC with finite sample correction (AICc). Examples of AICc include:

1. Akaike [11] and Schwartz [385] suggest:

$$\text{AICc}(x) = -2\mathscr{L}(\hat{\theta}|x) + |\hat{\theta}|\log(n) + \varphi \tag{3.26}$$

   where $\varphi$ is a constant. Schwartz [385] also suggests a version without $\varphi$.

2. Hurvich and Tsai [214] suggest setting $\varphi = (2|\hat{\theta}|(|\hat{\theta}| + 1))/(n - |\hat{\theta}| - 1)$ in Eqn. 3.26. which vanishes in the limit of large $n$ as $\log n \in o(\sqrt{n})$. Specifically, if $n$ is large with respect to $|\hat{\theta}|$, the bias adjustment term is negligible and there is little difference between AIC and AICc. This implies that they are asymptotically equivalent in the presence of large finite-sized samples, and AICc is an efficient estimator [258, p. 13].

---

[4]For the full derivation see [40].

3. Bhansali and Downham [65] suggest:

$$\text{AICc}(x) = -2\mathscr{L}(\hat{\theta}|x) + \varphi|\hat{\theta}| \qquad (3.27)$$

where $\varphi$ is a constant in the interval $\varphi \in [1;4]$. Atkinson [39] suggests a variation with no such restriction on $\varphi$.

4. Hannah and Quinn [194] suggest:

$$\text{AICc}(x) = 2\mathscr{L}(\hat{\theta}|x) - 2\varphi|\hat{\theta}|\text{loglog}(n) \qquad (3.28)$$

where $\varphi > 1$ is a constant. The specific value of $\varphi$, for which the authors give no advice on how to select, is the most important as the term $\text{loglog}(n)$ will remain small even for very large sample sizes.

5. Wagenmakers and Farrell [435] suggest Akaike weights which can be interpreted as conditional probabilities for each statistical smodel as:

$$w_i(\text{AIC}) = \frac{\exp\left\{-\frac{1}{2}\Delta_i(\text{AIC})\right\}}{\sum_{k=1}^{K}\exp\left\{-\frac{1}{2}\Delta_k(\text{AIC})\right\}} \qquad (3.29)$$

where $\Delta_i(\text{AIC})$ is the difference in AIC with respect to the best candidate model.

A lower AICc value indicates a good fit of the statistical model to the data. From an entropy perspective, this means that the model with the lowest AICc value removes the smallest amount of information. In other words, AICc is a "badness-of-fit" measure [234, p. 116]. As none of the AICc versions are normed, it is not possible to interpret an AICc value in isolation. Hence, using AICc for model selection requires first obtaining the AICc value of all models of interest and then selecting the model with the lowest AICc value.[5] Further information on AIC and related entropy measures are given in [87].

### 3.4.2 Minimum Description Length

The *minimum description length* (MDL) [365], is a recent approach to model selection which, unlike e.g. the AICc, does not assume the presence of a "true" data generating model [185].

In MDL, the key principle is that regularity in data may be used to compress or encode the data (e.g. to describe the data in full using fewer symbols compared to using a literal description), so that it may be transmitted in the most economical way. In terms of model selection, MDL selects the model that best encodes the data.

In Figure 3.3, for example, the data points are equidistantly distributed on a straight line. Based on this regularity, rather than transmitting the data, one can transmit (i) the location of the first data point, (ii) the direction of the line, (iii) number of points and (iv) the distance between points.

---

[5]Some authors [83, 87] recommend computing the AIC differences, rather than the raw AIC values, for all candidate models, as the AIC is on a relative scale.

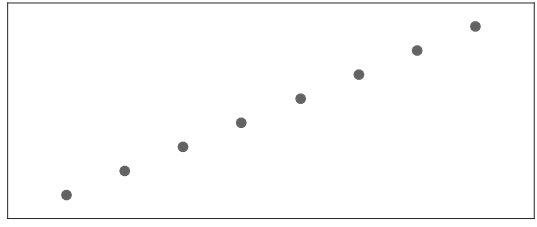

Figure 3.3: Points on a straight line. Example reproduced from [433, p. 57].

Encoding proceeds by first specifying the model to be used, and then specifying the data using this model [139]. Assume that $L : f \to \mathbb{N}^+$ is a function that, given a probability distribution $f(\cdot|\theta)$ of some statistical model $\mathcal{M}_\theta$, outputs the length of the binary string required to encode $f(\cdot|\theta)$. While MDL does not require a parameterised probability distribution, it is used here to be coherent with the preceding text and chapters of this thesis. Furthermore, let $L_f : \mathbb{R}^n \to \mathbb{N}^+$ be the length of the binary string required to encode the data $x$ using $f(\cdot|\theta)$. The MDL principle selects the model that minimises the sum of $L$ and $L_f$ i.e. the total length of their binary strings:

$$L_{\mathrm{MDL}} = \min_{f(\cdot|\theta) \in \mathcal{M}_\theta} \left[ L\big(f(\cdot|\theta)\big) + L_{f(\cdot|\theta)}(x) \right] \tag{3.30}$$

Equation 3.30 is more commonly known as the two-part version of MDL [185]. In Eqn. 3.30, $L\big(f(\cdot|\theta)\big)$ represents the complexity or richness of the selected probability distribution, and $L_{f(\cdot|\theta)}(x)$ represents how well the probability distribution is able to compress/encode the data i.e. its goodness-of-fit [139]. Similar to AICc, $L_{\mathrm{MDL}}$ is parsimonious as it trades-off goodness-of-fit with model complexity using $L\big(f(\cdot|\theta)\big)$ as penalty term (where AICc uses e.g. $2|\hat{\theta}|$). Encoding the data using $f$ is best done using Shannon-Fano coding [159, 388] whereby $L_{f(\cdot|\theta)}(x) = -\log_2 f(x|\theta)$ where $f(x|\theta)$ is the probability mass or density of $x$ under $f(\cdot|\theta)$. Thus, finding an efficient code or finding a good probability distribution are equivalent statements in MDL. Finding a good encoding for $L\big(f(\cdot|\theta)\big)$, however, is difficult [186], but Rissanen [366] showed that whenever a member $f(\cdot|\theta) \in \mathcal{M}_\theta$ fits the data well i.e. that $L_{f(\cdot|\theta)}(x)$ is small, then $L_{\mathcal{M}_\theta}(x)$ is also small. The resulting one-part MDL is thus given by:

$$L_{\mathrm{MDL}} = \min_{\mathcal{M}_\theta \in \mathcal{Z}} \left[ L_{\mathcal{M}_\theta}(D) \right] \tag{3.31}$$

where $\mathcal{Z}$ is a set of statistical models instead of probability distributions. As Eqn. 3.31 does not account for the richness of a statistical model $\mathcal{M}_\theta \in \mathcal{Z}$, a parametric complexity term is added. The complexity term, called COMP, estimates a statistical model's complexity by how many data sequences an element (probability distribution) of that statistical model [185] can fit. Then, model selection using MDL can be stated as:

$$L_{\mathrm{MDL}} = \min_{\mathcal{M}_\theta \in \mathcal{Z}} \left[ L_{\mathcal{M}_\theta}(D) + \mathrm{COMP}(\mathcal{M}_\theta) \right] \tag{3.32}$$

which again involves a trade-off between model fit and model richness. The "best" model is then the model $\mathcal{M}_\theta \in \mathcal{Z}$ that minimises Eqn. 3.32. Further information regarding MDL can be found in e.g. [185, 293, 264].

### 3.4.3 Likelihood Ratio Test

The *likelihood ratio* (LR) test is widely used to compare nested statistical models. Given two statistical models $\mathcal{M}_{\theta_1}^1$ and $\mathcal{M}_{\theta_2}^2$ fitted to $x$, where $\mathcal{M}_{\theta_1}^1$ is the restricted model parameterised by $\theta_1$ and $\mathcal{M}_{\theta_2}^2$ is the full model parameterised by $\theta_2$, the LR test can be stated as:

$$
\begin{aligned}
H_0 &: g = \mathcal{M}_{\theta_1}^1, \quad \mathcal{M}_{\theta_1}^1 \text{ is the true model} \\
H_1 &: g = \mathcal{M}_{\theta_2}^2, \quad \mathcal{M}_{\theta_2}^2 \text{ is the true model}
\end{aligned}
\tag{3.33}
$$

where $g$ denotes the "true" distribution of $x$. Thus $H_0$ stipulates that the restricted model is the "true" model, and $H_1$ that the full model is the "true" model. Using their maximised log-likelihoods $\mathcal{L}^1(\hat{\theta}_1|x)$ and $\mathcal{L}^2(\hat{\theta}_2|x)$, the test statistic is [166]:

$$
D = \frac{\mathcal{L}^1(\hat{\theta}_1|x)}{\mathcal{L}^2(\hat{\theta}_2|x)}
\tag{3.34}
$$

with the added restriction that the values in $\hat{\theta}_1$ are within the parameter space of $\hat{\theta}_2$ and not on its boundary. As $D$ gets smaller, the observed results get further away from $H_0$. Taking the logarithm of $D$ and multiplying by $-2$ gives:

$$
D = -2\log\left(\frac{\mathcal{L}^1(\hat{\theta}_1|x)}{\mathcal{L}^2(\hat{\theta}_2|x)}\right) = -2\big(\log(\mathcal{L}^1(\hat{\theta}_1|x)) - \log(\mathcal{L}^2(\hat{\theta}_2|x))\big)
\tag{3.35}
$$

whose distribution is asymptotically $\chi^2$ distributed, with degrees of freedom equal to the number of free parameters between the full and restricted model [304, pp. 53–58]. Despite their widespread use, LR tests implicitly assume that at least one of the models tested is correct [350] which can lead to incorrect results when models are *misspecified*, however. A model is misspecified if, assuming the true distribution $g$ has density $p$, then $p \notin \mathcal{M}_\theta$ [392].[6] This generally means that the model does not account for some important non-linearities [446, p. 275]. For example, if the true model is a second-degree polynomial, and the estimated model is a first-degree polynomial, the estimated model misspecifies the true model as it does not account for any non-linearity.

### 3.4.4 Vuong's Likelihood Ratio Test

When $\mathcal{M}_{\theta_1}^1$ is not nested by $\mathcal{M}_{\theta_2}^2$, the distribution of $D$ is no longer $\chi^2$ and the LR test cannot be used [141, p. 43]. Instead, information criteria like AIC, BIC and Vuong's LR test [434] can be used. Vuong's LR test is based on the KL divergence and can be used to compare both nested, overlapping and non-nested models.[7] We focus here on non-nested models. Vuong's LR test [434] also attempts to resolve how to determine the better model under the assumption that neither model is the "true" model (similar to MDL).[8] Given the element-wise log-likelihoods $\mathcal{L}^1(\hat{\theta}_1|x_i)$ and $\mathcal{L}^2(\hat{\theta}_2|x_i)$ for $i = 1,...,n$ evaluated at their ML estimates for two statistical models $\mathcal{M}^1_{\theta_1}$ and $\mathcal{M}^2_{\theta_2}$, Vuong's LR test characterises the samples variance in the individual log-likelihood ratios by:

$$\omega^2 = \text{var}\left[\frac{\mathcal{L}^1(\hat{\theta}_1|x_i)}{\mathcal{L}^2(\hat{\theta}_2|x_i)}\right] : i = 1,...,n \qquad (3.36)$$

where var denotes the variance. Conceptually, if one model is nested in the other, then the variability in these ratios should be close to zero. Conversely, if the models are non-nested the variability in the element-wise likelihood ratios will not be close to zero [308]. Notice that because the variance is taken w.r.t. the ML estimated parameters, Eqn. 3.36 makes no assumptions that either model is correct. Testing for nested models using Eqn. 3.36 is then:

$$H_0 : \omega^2 = 0, \quad \mathcal{M}^1_{\theta_1} \subset \mathcal{M}^2_{\theta_2} \qquad (3.37)$$

$$H_1 : \omega^2 > 0, \quad \mathcal{M}^1_{\theta_1} \not\subset \mathcal{M}^2_{\theta_2} \qquad (3.38)$$

where the sample estimate of $\omega^2$ is given by [308, 434]:

$$\hat{\omega}^2 = \sqrt{\frac{1}{n}\sum_{i=1}^{n}\left[\frac{\mathcal{L}^1(\hat{\theta}_1|x_i)}{\mathcal{L}^2(\hat{\theta}_2|x_i)}\right]^2 - \left[\frac{1}{n}\sum_{i=1}^{n}\frac{\mathcal{L}^1(\hat{\theta}_1|x_i)}{\mathcal{L}^2(\hat{\theta}_2|x_i)}\right]^2} \qquad (3.39)$$

---

[6] George Box stated that "all models are wrong, but some are more useful than others". Thus, in a broad sense all models are misspecified.

[7] Two models $\mathcal{M}_{\theta_1}^1$ $\mathcal{M}_{\theta_2}^2$ are overlapping if $\mathcal{M}_{\theta_1}^1 \cap \mathcal{M}_{\theta_2}^2 \neq \emptyset$ and $\mathcal{M}_{\theta_1}^1 \subsetneq \mathcal{M}_{\theta_2}^2$ and $\mathcal{M}_{\theta_2}^2 \subsetneq \mathcal{M}_{\theta_1}^1$.

[8] Prescott [351] argues that because a model is only an approximation they cannot be regarded as a null hypothesis to be statistically tested.

Eqn. 3.37 can be used to directly test if two fitted statistical models are nested. If $H_0$ is not rejected, we cannot rule out that the compared models are nested given the sample used. In this scenario, we may prefer the most parsimonious model. If $H_0$ is rejected, Vuong proposes to compare the two models using a non-nested LR test with the associated hypotheses:

$$H_0 : \mathbb{E}\left[\frac{\mathcal{L}^1(\hat{\theta}_1|x)}{\mathcal{L}^2(\hat{\theta}_2|x)}\right] = 0, \quad \mathcal{M}^1_{\theta_1} \text{ is equal to } \mathcal{M}^2_{\theta_2} \qquad (3.40)$$

$$H_1 : \mathbb{E}\left[\frac{\mathcal{L}^1(\hat{\theta}_1|x)}{\mathcal{L}^2(\hat{\theta}_2|x)}\right] > 0, \quad \mathcal{M}^1_{\theta_1} \text{ is better than } \mathcal{M}^2_{\theta_2} \qquad (3.41)$$

$$H_2 : \mathbb{E}\left[\frac{\mathcal{L}^1(\hat{\theta}_1|x)}{\mathcal{L}^2(\hat{\theta}_2|x)}\right] < 0, \quad \mathcal{M}^2_{\theta_2} \text{ is better than } \mathcal{M}^1_{\theta_1} \qquad (3.42)$$

where $\mathbb{E}$ is the expectation w.r.t. the "true" data-generating model and the subscript on $x$ has been dropped. $H_0$ states that the KL divergence between $\mathcal{M}^1_{\theta_1}$ and the true model is equal to the KL divergence between $\mathcal{M}^2_{\theta_2}$ and the true model, respectively. $H_1$ and $H_2$ states, respectively, that one model is closer to the true model than the other. The test statistic under $H_0$ (Eqn. 3.40) is then [434]:

$$Z_V = \sqrt{n}\frac{LR(\hat{\theta}_1,\hat{\theta}_2)}{\hat{\omega}^2} \qquad (3.43)$$

which is a consistent and unbiassed estimator of the asymptotic KL divergence [253]. In Eqn. 3.43, $LR(\hat{\theta}_1,\hat{\theta}_2)$ is the difference in the models' maximised average log-likelihood:

$$LR(\hat{\theta}_1,\hat{\theta}_2) = \sum_{i=1}^{n}\mathcal{L}^1(\hat{\theta}_1|x_i) - \sum_{i=1}^{n}\mathcal{L}^2(\hat{\theta}_2|x_i) \qquad (3.44)$$

Vuong shows, under some regular conditions, that the asymptotic distribution of Eqn. 3.43 is a Gaussian with $\mu = 0$ and $\sigma^2 = \hat{\omega}^2$.[9] Thus, a $p$-value can be calculated by comparing the test statistic to a Normal distribution. Genius and Strazzera [176] have claimed that probabilistic model selection using Vuong's LR test has a distinct advantage over more traditional model selection criteria such as AIC or BIC, which are deterministic [176]. For statistical model selection, the model that is closest to the true distribution is the "best" one [434].

---

[9]See assumptions A1 through A6 in [434] for the regular conditions.

### 3.4.5 Kolmogorov–Smirnov Test

Assume that a population has a specific CDF $F_0$. Given a sample from that population, one expects the empirical CDF, $F_n$, should be "close" to the specified CDF. If it is not close, we may suspect that $F_0$ is not the correct CDF. The Kolmogorov–Smirnov (KS) test [244, 398] is a nonparametric GOF test [25] which quantifies the notion of closeness as its test statistic is based on the maximum difference between the empirical CDF and the CDF of the hypothesised statistical model. The KS test is defined as:

$$H_0 : F_0(x) = F_n(x), \quad x \text{ is distributed according to } F_0$$
$$H_1 : F_0(x) \neq F_n(x), \quad x \text{ is not distributed according to } F_0$$

(3.45)

The KS test statistic is given by:

$$D_{\{x_1,...,x_n\}} = \sup_x |F_n(x) - F_0(x)|$$

(3.46)

where sup is the supremum, $|\cdot|$ is the absolute value and $x$ ranges over the domain of definition of both $F_0(\cdot)$ and $F_n(\cdot)$.[10]

As an example, Figure 3.4 shows $F_n(x)$ of a random set of 120 twitter messages (their lengths) and two MLE fitted models (Gaussian and exponential). The Gaussian distribution (Figure 3.4a) qualitatively approximates the data well evidenced by a maximum distance of 0.043. In contrast, the exponential (Fig. 3.4b) does not provide as good an approximation with maximum distance of 0.261.

The KS test has a number of desirable properties [161]:

- It is distribution-free (that is, the KS test does not rely on assumptions that the data are drawn from a given probability distribution) provided (i) data are i.i.d. samples from a continuous distribution and (ii) are *not* fully specified i.e. the parameters of $F_n(\cdot)$ are not estimated from the data. If data are i.i.d. samples from a *discrete* model with support exactly on the positive integers, Arnold and Emerson [35] give an alternative formulation.

- The KS test is exact provided there are no ties in the data. An exact test calculates $p$-values based on the exact distribution of the test statistic [88, p. 403]. In contrast, asymptotic tests calculate $p$-values based on the assumption that the data, given a sufficiently large sample size, asymptotically conform to a particular distribution [425].

- It is insensitive to the sample size, easy to compute and readily understood graphically.

and disadvantages:

---

[10] Formally, let $A \subset \mathbb{R}$. A number $u$ is an upper bound for $A$ if $\forall a \in A, u \geq a$. A number $v \in \mathbb{R}$ is a supremum (or least upper bound) for $A$ if (i) $v$ is an upper bound on $A$ and (ii) for all upper bounds $v'$ on $A$, $v \leq v'$.

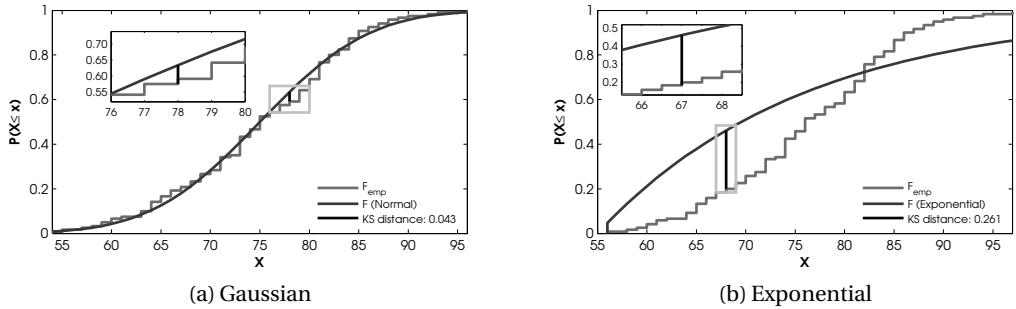

<div align="center">(a) Gaussian          (b) Exponential</div>

Figure 3.4: Examples of maximum distances used to calculate the KS statistic (Eqn. 3.46). Data consists of the lengths of 120 randomly selected twitter messages (shown as a step function). MLE fitted models are shown as smooth curves. Insets zoom in on the location of the maximum distances. (3.4a) An MLE fitted Gaussian distribution with maximum distance $D_{[x_1,...,x_n]} = 0.043$. (3.4b) An MLE fitted exponential distribution with maximum distance $D_{[x_1,...,x_n]} = 0.261$.

- The KS test assumes that $F_0$ is completely specified in advance (for example, Gaussian distribution with $\mu = 0$ and $\sigma^2 = 1$). If any parameters of $F_0$ are estimated from the sample data, the confidence levels used to reject or accept $H_0$ will be more conservative (tend to give larger $p$-values) [270]. As an example, Figure 3.5 shows the frequency distribution of 100,000 KS test statistics calculated from samples of size $n = 1000$ drawn from a standard Gaussian distribution. Super-imposed is the frequency distribution for the same 100,000 KS test statistics, each of size $n = 1000$, when the parameters of the standard Gaussian are ML estimated for each individual sample. The shape of the distributions clearly differs.

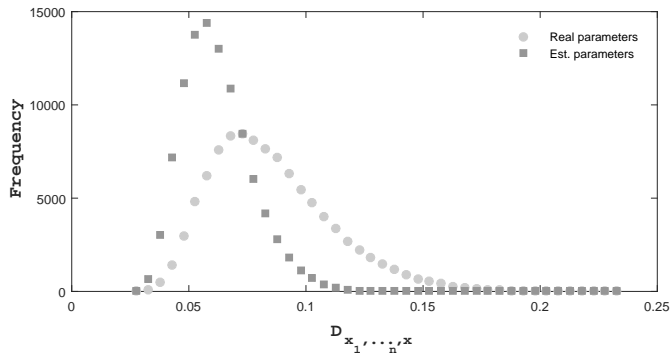

Figure 3.5: Distribution of 100,000 KS test statistics calculated from $n = 1,000$ samples drawn from a standard Gaussian distribution (circles) and the same distribution when the parameters of the Gaussian are estimated for each sample.

Suppose now that a maximum $D_{[x_1,...,x_n]} = 0.05$ is fixed and a confidence level is calculated. Using the distribution where the Gaussian parameters are not estimated, gives a 95.25% confidence level that the samples can be described by a Gaussian. When $\mu$ and $\sigma^2$ are estimated from each sample, the confidence level drops to 78.49%. The difference in confidence levels is close to 17% which can lead to a wrong conclusion i.e. if $H_0$ is false, we may fail to reject it when using the KS test and estimated parameters [238].

- It is more sensitive near the centre of the distribution than at the tail [161] where fluctuations tend to occur due to data sparsity.

While parameters should not be estimated from sample data for the KS test, the use of e.g. bootstrap resampling can be used to estimate the critical values [44]. For model selection using the KS test, and assuming that the true statistical model is one of the candidate models considered, the single model which cannot reject $H_0$ is the correct model. In practice, the model with the lowest $D_{\{x_1,...,x_n\}}$ will be selected. This corresponds to the statistical model whose maximum deviation from $F_n(x)$ is the smallest.

### 3.4.6 Anderson–Darling Test

The Anderson-Darling (AD) is used to test if $x$ came from a population with a specific distribution. The null and alternative hypothesis tested by AD are [23]:

$$H_0 : F_n(x) = F_0(x), \quad x \text{ is distributed by } F_0$$
$$H_1 : F_n(x) \neq F_0(x), \quad x \text{ is not distributed by } F_0 \tag{3.47}$$

Like the KS test, the AD test is based on measuring the discrepancy between the theoretical distribution function $F_0$ and $F_n$, the empirical CDF of the data. The AD test statistic is a special case of the Cramér–Von Mises (CVM) GOF test which is based on the squared integral of the difference between $F_0$ and $F_n$:

$$\text{CVM} = n \int_{-\infty}^{\infty} [F_n(x) - F_0(x)]^2 \, w[F_0(x)] \, dF_0(x) \tag{3.48}$$

where $w(\cdot)$ is a suitable weight function giving weight to the squared area between the $F_n(\cdot)$ and the line $y = x$ [294]. Setting $w(\cdot) = 1$ gives the original CVM test statistic. The AD test statistic is given in Eqn. 3.48 where $w(x)$ is given by [22]:

$$w[F_0(x)] = \frac{1}{F_0(x)(1 - F_0(x))} \tag{3.49}$$

which assigns weights inversely proportionally to the discrepancy between $F_0$ and $F_n$. This weighting function places higher weight on discrepancies in the tails of the distributions which are necessarily becoming smaller as both approach 1 [160]. The AD test statistic is given by:

$$A_n = n \int_{-\infty}^{\infty} \frac{[F_n(x) - F_0(x)]^2}{F_0(x)(1 - F_0(x))} \, dF_0(x) \tag{3.50}$$

Anderson and Darling [23] showed that Eqn. 3.50 be written as:

$$A_n = -n - \frac{1}{n} \sum_{j=1}^{n} (2j-1) \big[ \log\big(F_0(x_j)\big) + \log\big(1 - F_0(x_{n-j+1})\big) \big] \tag{3.51}$$

where $x_1 < x_2 < ,..., < x_n$ is the ordered sample. In contrast to the KS test which is most sensitive to values in the middle of the range of the distribution [285, p. 62], $A_n$ receives more contribution from the tail of the distribution [25]. If the estimated $A_n$ statistic exceeds the critical value at a particular certain level the null hypothesis can be rejected. The critical levels for $A_n$ are dependent on the specific distribution being tested. While this allows for a more sensitive test, critical values were known initially only for a handful of distributions, thus limiting the use of the test. However, recent research has provided critical values for other distributions, including the power law [128, 410, 411]. Furthermore, the AD has been found to be a more powerful test than the KS test [155].

## 3.5   A Principled Approach to Statistical Model Selection

The principled approach to statistical model selection presented in this section is an extension of the work by Clauset et al. [117]. Our approach consists of the three steps given below and combines several of the methods introduced in the previous sections.

---

**Principled Approach to Statistical Model Selection**

1. **Estimation step**: Estimate the parameters of each statistical model in Table 2.2 using MLE (Section 3.3).

2. **Comparison step**: Compare each estimated statistical model to all other models using a likelihood ratio (LR) test. The LR test used depends on whether the models compared are nested:

   **Nested**  If the pair of models being compared are nested, use the LR test (Section 3.4.3) to compute their ratio and $p$-value.

   **Non-nested**  If the pair of models being compared are not nested, the LR (and $p$-value) are calculated using Vuong's LR test (Section 3.4.4).

   If the $p$-value resulting from the comparison of any two models is below 0.05, the sign of the LR is a reliable indicator of which of the models is the best fit to the data. Compute also the AICc value (Section 3.4.1) for each model.

3. **Selection step**: Select the statistical model which "wins" most pairwise comparisons. A statistical model "wins" a comparison if it is the better of the two and $p < 0.05$.

---

Our principled approach is based on Fisher's seminal work [165], whereby the standard method to select the statistical model that best approximates the distribution of $x$ is to (a) estimate the values of the parameters of each candidate statistical model which would most likely have generated $x$; (b) use a statistical significance test to identify which candidate model minimises some, typically information-theoretic distance measure, to an unknown "true" model.

Although our approach makes specific choices for what methods to use for parameter estimation and choice of statistical tests by which the best-fitting model is selected, other combinations of methods could also be used. The particular combination we use combines standard approaches for parameter estimation (MLE) and statistical testing (Vuong's LR test and AICc), and is similar to the one used by Clauset et al. [117]. The differences between our method and that of Clauset et al. is delayed to Section 3.5.1. The output of the approach is what we refer to as a Vuong table. As an illustration, the Vuong table resulting from applying our principled approach to the citations of the iSearch dataset [280] is shown in Table 3.1. The citation dataset in iSearch contains the number of citations received by an article in the physics domain and is approximately power-law-distributed [358]. A description of the iSearch dataset can be found in Section 4.4.1.

Table 3.1: Vuong table for the iSearch citations dataset. The best-fitting model (dark grey) is the Generalized Extreme Value with MLE parameters: shape = 4.889, scale = 4.494, loc = 1.919. The best-fitting discrete model (light grey) is the Yule with MLE parameter: scale = 1.518. The lower triangular part of the table is blacked out as the results are symmetric.

Table 3.1 may be read as follows. Each column contains a *p*-value and the value of Vuong's LR test for a specific model. For example, all values in the last column are the results from comparing the Yule–Simon model to all other models. A *p*-value below .05 means the sign of the LR test is a reliable indicator of which model (i.e. the models identified at a specific row and column) is the best. Reading the table column-wise, a *negative* LR means the model identified by the column is the preferred. Reading the table row-wise, a *positive* LR means the model identified by the row is preferred. In both cases, the model identified by the sign is only preferred if *p* < .05. For example, comparing the exponential with the Yule distribution, shows that *p* < .05 and LR= −39.4. The *p* value can be changed, but is fixed to .05 as this is a commonly used significance level. The closer the LR value is to 0, the less difference there is between the compared models. The last row in the Vuong table contains the AICc value of each model. Columns and rows highlighted in light grey correspond to the best-fitting *discrete* model. Columns and rows highlighted in medium grey correspond to the best-fitting statistical model *overall*. For example, in Table 3.1 the Yule distribution is the best-fitting discrete model, and the GEV is the best-fitting distribution overall. Finally, dark grey cells correspond to nested models (see Table 2.3) whose LR and *p*-value are calculated using the standard LR test.

### 3.5.1 Differences Between Our Approach and Clauset's

The original approach [117] is concerned with testing empirical data for discrete "power laws" i.e. statistical models of the form:

$$PL(x|\theta) = \{f(x|\alpha, x_{\min})\}$$
$$= \left\{ \frac{x^{-\alpha}}{\zeta(\alpha, x_{\min})} : \alpha > 1 \right\} \qquad (3.52)$$

where $\alpha$ is the power law exponent, $\zeta$ is the Hurwitz zeta function [215] and $x_{\min}$ is a lower threshold of the data above which the power law "holds".[11] The power law is then assumed to be the "true" model against which alternative statistical models are evaluated for a better fit. Using their method, the authors evaluate a large number of datasets and find that the power law is only valid in a few cases. Our approach is similar to that of Clauset et al. but contains a number of differences.

---

[11] The introduction to power laws is postponed to Chapter 4.

The first difference between our approach and the original is that our approach *generalises* that of Clauset et al. by not assuming that the power law is the "true" model. That is, in our approach, each statistical model is, in turn, assumed to be the "true" model and all candidate models are compared to this model. Because each of the $K$ statistical models is compared to all remaining $K - 1$ models, the $k^{\text{th}}$ model that "wins" the most comparisons can be said to be the "true" model. This is a valid generalisation as it is essential $K$ applications of Clauset's method where each application assumes a different "true" model.

A second difference is that Clauset et al. [117] perform a Monte Carlo goodness of fit test of the null hypothesis that data follows a power law. This Monte Carlo step is typically used on the - small - datasets obtained by truncating the datasets to the region where a power law fits best (i.e., above $x_{\min}$). We omit the Monte Carlo step in our approach as it suffers from similar problems as many ordinary goodness-of-fit tests [367, 384] as it only rejects models, and will do so as sample sizes become large. This is especially damaging for IR data, which tend to be large.

A third difference is that we do not consider determining $x_{\min}$ i.e. the lower cutoff above which the data best fit a power law. In the original approach, this lower cutoff is determined by iteratively removing data points below some candidate value $x_i$ and calculating the KS test. The $x_i$ with the smallest KS test is then set to $x_{\min}$. Our reason for not estimating $x_{\min}$ is two-fold. Firstly, the running time of their method is quadratic in the difference between the largest and smallest value in each dataset [344] making it infeasible for datasets with outliers such as many IR datasets. In our approach, the running time (for the statistical models whose MLE parameters do not have a closed-form solution) for a statistical model is dominated by the Nelder-Mead (NM) simplex algorithm [327] which is an derivative-free, unconstrained optimisation technique, used to find the MLE parameters. The NM simplex is the native method used by MATLAB where our principled approach was developed. Convergence results of the NM simplex algorithm exist for one and two dimensions [257] on strictly convex functions with bounded level sets, but, in the general case, no convergence results exist and the NM simplex may fail to converge or converge slowly [126, p. 7][345]. Secondly, the method aggressively discards data. A toy example of the impact $x_{\min}$ has on estimating where the power law "begins" is shown in Figure 3.6 showing the distribution of a toy set of observations $x$ and a hypothetical $x_{\min}$. Because only data to the right of $x_{\min}$ are considered "power law", the method discards most of the data. Consequently, if removing so much data is required for a particular model to fit "closely", an entirely different model should likely be considered.

As a concrete example, we use their method to estimate $x_{\min}$ for three different IR datasets: the Excite query log [407], the iSearch [280] and the Congressional Record (CR) dataset from TREC-5. In the Excite query log, each data point is the length (number of terms) of a query; in the iSearch dataset, each data point is the length (number of terms) of each document and in the CR dataset, each data point is the frequency of a given term in the collection. The results show that 99.99% of the data points in the Excite queries, (ii) 96.97% of the data points in the iSearch dataset and (iii) 99.78% of the data points in the CR dataset are discarded by their method.

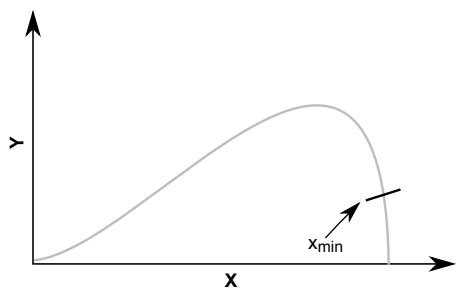

Figure 3.6: The impact of $x_{\min}$ on distributions of data that are not power law-distributed.

Repeating the experiment on the same datasets used by Clauset et al. [117] shows that between 81.2% and 94.4% of all data is discarded. Similar observations are made by Garcia et al. [173]. For these reasons, our approach does not estimate a lower cutoff.

A fourth difference is that our approach considers nearly twice as many models as Clauset et al. and can easily accommodate many more. Finally, our approach also uses AICc [214] when selecting the best model. Like Vuong's LR test, the AICc is based on KL divergence and will select the same model in the limit of large sample sizes, namely the model that minimises the information loss w.r.t. the "true" model. Hence, if the two tests disagree it indicates that the sample may not be large enough to rule out one model and that this too could be considered when selecting the right model.

A fifth difference is that the implementation of Clauset et al. contains errors during the estimation process. These are caused by the built-in non-linear minimisation routine in R, which replaces NaNs produced during optimisation with the largest possible number. A NaN is not-a-number. These are produced by the optimisation procedure when the values sought estimated stray into regions of the parameter space that causes e.g. numerical underflow. The values are replaced to force the optimiser back in a feasible region though this may lead to (i) extended estimation time and (ii) locally optimal parameter values. Practically, this inflates the value of the LR tests used which for statistical models that approximate some data $x$ equally well, may lead to the wrong model being selected.[12] Our re-implementation avoids these errors.

## 3.6   Summary

Model selection is important in many areas including IR. However, model selection is rarely performed in IR where the prevalent approach is to assume a specific underlying distribution either on the basis of mathematical convenience or an ad hoc approach. This chapter presented a statistically principled approach to perform model selection which is an extension of the method by Clauset et al. [117]. Our approach uses MLE to fit statistical models to datasets before using Vuong's LR test [434] and AICc to select the best-fitting model.

The next two chapters evaluate our approach on two different IR tasks. In the first task, we test if properties of IR datasets such as term frequencies are (approximately) power law distributed. In the second task, we use model selection to derive improved ranking models for ad hoc retrieval.

---

[12]We have made Clauset et al. aware of these errors.

# 4 Revisiting Commonly Assumed Distributions for IR Datasets

This chapter is concerned with testing commonly held assumptions regarding the distribution of the properties in IR datasets. A very common assumed distribution of such properties in IR is the *power law*, and therefore this chapter focusses on testing power law assumptions of properties in IR datasets.

Section 4.1 introduces the power law statistical model and motivates the study of power laws using examples from IR. Section 4.2 reviews common graphical methods used to observe power laws and demonstrates why these are inaccurate. Section 4.3 reviews the literature in IR where such graphical methods (among others) are used to observe approximate power laws, and areas in IR where power laws have been used to improve IR tasks. Section 4.4 applies the principled method of statistical model selection proposed in Chapter 3 to multiple properties of IR datasets reported to be approximately power law distributed. Section 4.5 investigates the efficiency of our principled approach and Section 4.6 concludes the chapter.

## 4.1 Introduction

Many properties of standard IR datasets are assumed to be distributed according to some statistical model. Of these, the *power law* model has attracted a lot of attention [344] and properties such as term-frequencies [49, 103, 104, 318, 422], query frequencies [47, 50, 144] and citations [20, 105, 307] are widely assumed (or claimed) to be approximately power law-distributed. Despite the large number of observations and assumptions regarding power laws made in IR, most of these are frequently made using informal methods such as qualitative inspection of the distribution of the data. The problem is that informal methods are not a rigorous test for the presence of a power law, and research in different areas has shown that many assumed/observed power laws using such informal methods are unlikely to be power laws [56, 117, 180, 328, 414].

For testing power law assumptions, we use the principled approach presented in Chapter 3. Section 4.1.1 reviews the basics of power laws and motivates, through multiple examples, the need for using a principled approach to test approximate power law distributions.

### 4.1.1   Basics of Power Laws

A discrete *power law* is a statistical model of the form:

$$\text{PL}(x|\theta) = \left\{ f(x|\alpha, x_{\min}) \right\}$$
$$= \left\{ \frac{x^{-\alpha}}{\zeta(\alpha, x_{\min})} : \alpha > 1 \right\} \tag{4.1}$$

where $\alpha$ is the power law exponent, $\zeta$ is the generalised or Hurwitz zeta function [117]:

$$\zeta(\alpha, x_{\min}) = \sum_{n=0}^{\infty} (n + x_{\min})^{-\alpha} \tag{4.2}$$

and $x_{\min}$ is a threshold above which the power law "optimally" fits the data [117].

Power laws appear, perhaps confusingly, under many aliases in the literature, such as "scale-free"; "fat/heavy-tailed"; "80-20"; "Lévy distribution"; "allometric scaling law"; "Zipf's law"; "Lotka's law"; "Pareto distribution" and "Heaps' law" [205, 310]. These often refer to specific instances of a power law. For example, Zipf's law–a power law where $\alpha \approx 1$ [7, 268, 295]–has been reported [252, 463] to describe the distribution how word frequencies in e.g. present-day American-English, adopted in models for describing how language evolves [127, 332, 347] and used for examining website requests [14, 78, 137, 178]; Heaps' law–a power law where $0.4 \lesssim \alpha \lesssim 0.6$ [462] (for English texts)–has been used in approximate text retrieval [48] and for estimating the vocabulary size of large text collections [90]. However, as all aliases refer to the same basic form of Eqn. 4.2 [205], such distinctions will be ignored and referred to collectively as "power laws".

A power law implies heterogeneity: a small number of something occurs extremely often whereas many more instances are much less frequent [4]. Figure 4.1 shows two examples of data that have been reported to be approximately distributed as a power law. Figure 4.1a shows the probability of incoming links to Web sites in the complete ND.EDU domain [13]; Figure 4.1b shows the probability of receiving $k$ citations by a scientific paper published in 1981 and between 1981 and June 1997 [328, 358]. Both examples show that there is a high probability of observing either websites with few incoming links or papers receiving few citations. In contrast, Web sites with many incoming links and papers receiving thousands of citations have probabilities that are orders of magnitude lower.

Many areas of science are replete with examples such as those in Figure 4.1 from economics [27, 170, 402] to sociology [242, 316], computational epidemiology [156] (Fig. 4.2a), knowledge diffusion [105] (Fig. 4.2b), microbiology [64, 221, 371, 423] (Fig. 4.2c) and ecology [291] (Fig. 4.2d).[1] Figure 4.2 shows four examples using data from different scientific areas, all of which are reported to be approximately power law-distributed.

One reason behind the large number of power law observations is how easily they can be identified (see [152, Chap. 18] and [191, 205]). Denote by $C = 1/\zeta(\alpha, x_{\min})$ and let $y$ be a power law distributed variable (of the form of Eqn. 4.1):

$$y = C x^{-\alpha} \tag{4.3}$$

---

[1] In fact, power laws have become so popular that the front page of the New Scientist magazine in April 2002 read: "How can a single law govern our sex lives, the proteins in our bodies, movie stars and supercool atoms? Nature is telling us something...".

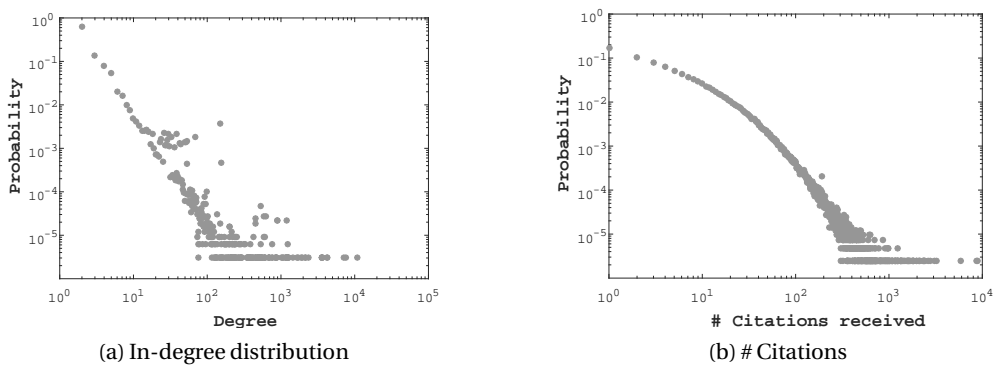

(a) In-degree distribution

(b) # Citations

Figure 4.1: Examples of data reported to be approximately power law-distributed. (a) Probability of incoming links to websites in the complete `nd.edu` domain [13]. (b) Probability of receiving $k$ citations by a scientific paper between publication and June 1997 [328, 358].

Taking the logarithm on both sides of Eqn. 4.3, we obtain the linear form [117]:

$$\begin{aligned}
\log(y) &= \log(Cx^{-\alpha}) \\
&= \log(C) - \alpha \log(x)
\end{aligned} \tag{4.4}$$

implying that a power law should appear as a straight line on double-logarithmic axes. All plots in Figures 4.1 and 4.2 were constructed by their authors using this method. Thus, probing for a power law is straightforward [117]:

1. Measure the quantity of interest $x$

2. Construct the histogram of $x$

3. Plot the histogram on double-logarithmic axes (a log-log plot)

If the topmost points of each "bar" in the histogram fall approximately on a (single) straight line, then $x$ must be drawn from a power law distribution.

On this basis, consider again the plots in Figure 4.1 and Figure 4.2, all of which were reported to be power law-distributed and hence should follow a straight line on double-logarithmic axes. Qualitative inspection of the plots, however, suggests they all deviate from a straight line to various degrees. Figure 4.1a seems best approximated by a straight line for degrees $k < 50$ but much less so for higher degrees. The data in Figure 4.1b exhibit curvature between citations of degree $1 \leq k \lesssim 100$ and seems poorly fitted by a straight line. Similarly, the initial plateau of degrees in both Figure 4.2a ($1 \leq k \lesssim 30$) and 4.2b ($1 \leq k \lesssim 10$) makes either distribution poorly fitted by a (single) straight line. The data in Figure 4.2c appears scattered and not well approximated by a straight line. Finally, the data in Figure 4.2d appears to substantially deviate from a straight line for any range of the data.

Given the qualitative differences between the plots and their deviation from a straight line, it seems unlikely that a power law can hold for all, or any, of the datasets from such disparate

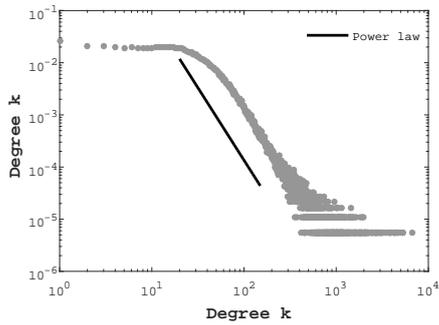

(a) Degree distribution of a network propagating an infectious disease throughout a population

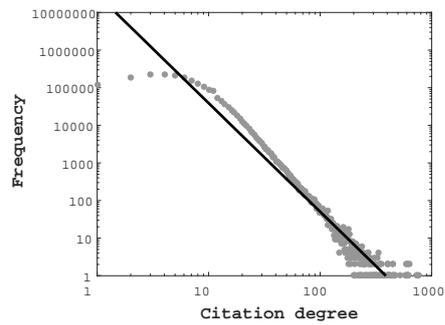

(b) Patents cited

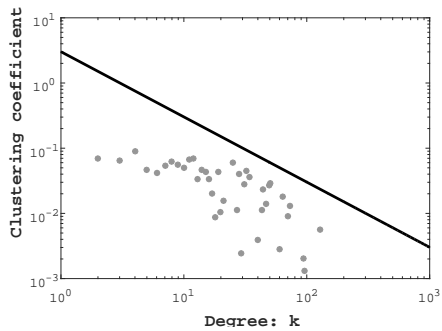

(c) Clustering coefficients of a protein-to-protein connection network

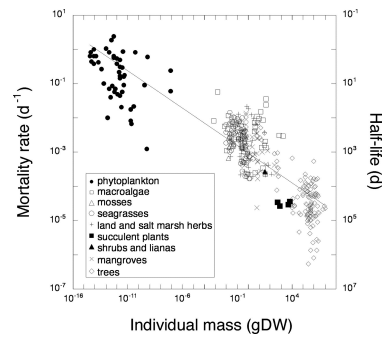

(d) Plant mortality rate and mass

Figure 4.2: Examples of reported approximate power law-distributions in published research. All plots are reproduced as in their respective papers except Figure 4.2d. A straight line corresponds to the reported power law (though not always necessarily indicative of the range of the data where the power is observed). (4.2a) Degree distribution of a network used to simulate the progress of a disease within a host and of transmission between hosts [156]. (4.2b) Number of patents cited in patents granted between January 1963 and December 1999 [105]. (4.2c) Clustering coefficient distribution of the human protein-protein interaction (PPI) network (a PPI network consists of connections between proteins as the results of biochemical events.) [371]. (4.2d) Relationship between plant mortality rate and the individual mass of plants [291].

realms, which, in addition, raises a legitimate concern regarding the validity of power law observations made in the literature [268].

Specifically, several researchers [56, 117, 237, 180, 328, 414] have cautioned against the perceived "universality" of power laws and raised severe criticism against the following three mainstream methods for detecting power laws:

- Using log-log plots to observe power laws [117, 237, 328].

- Using linear regression or linear least-squares to fit a straight line to the data and estimate the power law exponent $\alpha$ [56, 117, 180].

- Absence of testing if *alternative* statistical models provide a better fit to the data [117, 162].

On the basis of the shortcomings of the ad hoc methods most commonly used to probe for power laws and the many areas of IR where (approximate) power laws are observed, this chapter investigates the following research question:

1. **Research Question 1 (RQ1):** To what extent does commonly assumed statistical models fit common properties of IR datasets?

To answer RQ1, we apply the principled approach from Chapter 3 to term frequencies, document lengths, query frequencies, query lengths, citation frequencies and syntactic unigram frequencies; all of which have been reported to be (approximately) power law distributed:

**Term frequency:** The probability that a term occurs $n = 1, 2 \ldots$ times in a dataset, is approximately power law distributed [104, 103, 318].

**Query frequency:** The probability that a query occurs $n = 1, , 2 \ldots$ times in a query log/stream, is approximately power law distributed [50, 47, 144].

**Query length:** The probability that a query has $n = 1,2\ldots$ terms, is approximately power law distributed [118], though only the tail is power law distributed according to Arampatzis and Kamps [31].

**Document length:** The probability that a document has $n = 1,2\ldots$ terms, is approximately power law distributed [380, 409] .

**Citation frequency:** The probability that a paper is cited $n = 1,2\ldots$ times, is approximately power law distributed [358].

**Syntactic unigram frequency:** The probability that a syntactic unigram (a term and its parts of speech) occurs $n = 1,2\ldots$ times in a dataset, is approximately power law distributed [272].

The next section begins by reviewing graphical methods. These are commonly used qualitative approaches to identify power laws which are widely used in IR.

## 4.2  Graphical Methods for Identifying Power Laws

A graphical method is a visualisation of quantitative data that is widely used in exploratory data analysis to e.g. detect outliers, uncover underlying structure or otherwise obtain insight into datasets [217]. Graphical methods include scatter plots, histograms (such as log-log plots), linear regression and time series analysis [429].

While it is a conceptually simple approach to analysing a dataset, visual inspection of such constructed log-log plots is not a rigorous test for the presence of a power law [56, 117, 268]. Indeed, the use (and abuse) of graphical methods to detect power laws is unfortunate, as previous research has shown that many assumed/observed power laws based on such observations are unlikely to be power laws [56, 117, 180, 328, 414].

Three graphical methods appear to dominate the literature when power laws are observed and are based on the observation from Eqn. 4.4. Examples of these graphical methods are shown in Figure 4.3, and comprise:

1. **Graphical Method 1 (GM1):** Log-log plot of the "raw" histogram of the data (see e.g. [9, 46, 79, 108, 137, 142, 178, 209, 211, 239, 335, 337, 430]).

2. **Graphical Method 2 (GM2):** Log-log plot of the ECCDF $\overline{F}_n(x)$ (see e.g. [75, 96, 219, 404]).

3. **Graphical Method 3 (GM3):** Log-log plot of the logarithmically binned histogram of the data (see e.g. [5, 6, 12, 13, 20, 221, 412]).

These methods are introduced in the following sections.

### 4.2.1   GM1: Raw Histogram

A histogram summarises the distribution of a univariate data set. This summarisation is typically done by splitting the data set into a finite set of equal-sized intervals called *bins* and averaging the data that falls in specific bins [324, 310]. A "raw" histogram defines each bin to be a unique value in the data set. Thus, if there are $n$ unique values in a data set there will be $n$ bins. The count in each bin is therefore the number of values in the dataset taking that particular unique value. The bins and counts are then plotted on double-logarithm axes and visually asserted to be an approximate straight line. Raw histograms (when plotted on double logarithmic axes) often appear cone-shaped: a narrow stem, (corresponding to the head of the distribution) that, as the probability of observing some quantity begins to decline, they start to fan out. This "fanning" is sometimes, but incorrectly referred to as "noise" but is a consequence of sampling error: as the number of samples becomes small, the statistical fluctuations become large [328], thus causing the samples to waver. This phenomenon is shown in Figure 4.3a for the distribution of in-degrees in a sample of Web pages: the more incoming links a Web page has (i.e., the higher its degree), the less likely such a node is observed. To generate the probability distribution of the histogram, each bin is divided by the cardinality of the data set.

### 4.2.2   GM2: Empirical Complementary Cumulative Distribution Function

Using the ECCDF (see Chapter 2) avoids fluctuations at the "tail" of the distribution and facilitates the evaluation of the power law exponent [341, 442]. While many authors advocate using the CCDF when using graphical methods to probe for power laws [117, 237, 328], this method has been criticised for making it difficult to distinguish "noise" from true features of the distribution [310]. Figure 4.3b shows an example of the ECCDF. The "gap" in the ECCDF (highlighted using a box) is caused by the outliers in Figure 4.3a (also using a box).

### 4.2.3   GM3: Logarithmic Binning

To avoid the paucity of data associated with using the raw histogram, logarithmic binning is a practical alternative [310]. Unlike the binning in the raw histogram, each bin in logarithmic binning encompasses a range of values. Let $\delta w$ denote the width of each bin. Then, a bin of (constant) logarithmic width means that the logarithm of the upper edge of bin ($x_{i+1}$) is equal to the logarithm of the lower edge of that bin ($x_i$) plus $\delta w$ [442]. Formally,

$$\log(x_{i+1}) = \log(x_i) + \delta w \tag{4.5}$$

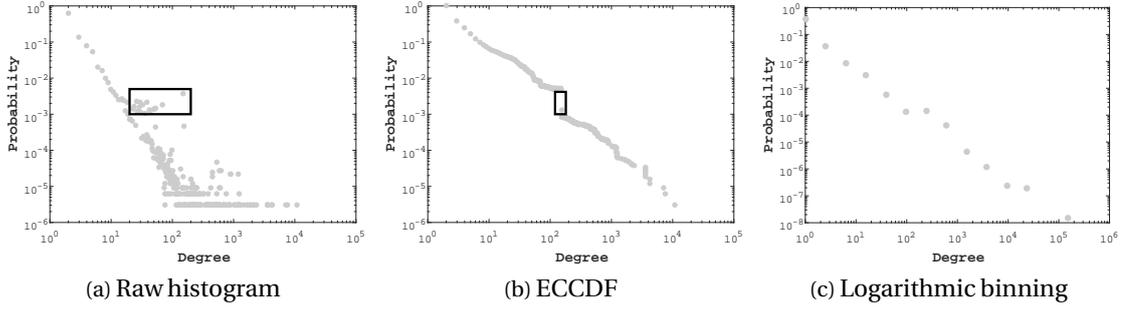

(a) Raw histogram         (b) ECCDF         (c) Logarithmic binning

Figure 4.3: Commonly used graphical methods for power law observations for the data used in [13]. (4.3a) Raw histogram (normalised). (4.3b) Empirical complementary cumulative distribution function. (4.3c) Logarithmic binning. The plot is not identical to the one in the paper by Albert et al. [13]. Personal communication with the authors revealed that the base number for the logarithmic bins was between 2 and 3. The plot here used a base number of 2.5.

Thus,

$$\begin{aligned} x_{i+1} &= \exp^{(\log(x_i) + \delta w)} \\ &= x_i \cdot \exp^{\delta w} \end{aligned} \tag{4.6}$$

with the number of observations in bin ($x_i$) equal to the density of observations in the bin times the bin-width [442]. Another version of logarithmic binning varies the width $\delta w$ of each bin such that bin ($x_{i+1}$) is a multiple of bin $x_i$ [328] and normalises the count in $x_i$ by its width.[2] Logarithmically spaced bins (for discrete data) where the width varies as a multiple may be created as:

$$[c^i, c^{i+1} - 1] : i = 0, \dots, k \tag{4.7}$$

where $c$ is the base (typically a positive integer $\geq 2$), $i$ denotes the number of the given bin and $[\cdot]$ denotes the interval with $c^i$ and $c^{i+1} - 1$ as endpoints. Common choices for $c$ are 2 and 10. For example, using $c = k = 2$ would create bins $\left[ [2^0, 2^1 - 1], [2^1, 2^2 - 1], [2^2, 2^3 - 1] \right]$. As $k$ grows, bins become wider, causing bins in the tail of the distribution to get more samples, which effectively smooths the data in areas where samples are sparse. This effect of logarithmic binning can be seen by comparing Figure 4.3c to Figure 4.3a. As data are smoothed, however, information is lost–including in the areas of the distribution where data are not sparse. To avoid this, Milojević [310] suggests using *partial* logarithmic binning where only data above some threshold for "statistical noise" are smoothed. This threshold, however, is determined empirically which can substantially influence the resulting binning procedure. Furthermore, because the value of each bin is the average of the counts of all values falling in the bin, the procedure is sensitive to outliers.

---

[2]Because bins will appear evenly spaced when plotted on logarithmic axes; the approach was called "logarithmic binning" [328].

Another problem related to logarithmic binning, is that any subsequent regression performed in the logarithmic domain imposes an exponential penalty on the errors. Consider again the equation:

$$y = Cx^{-\alpha} \tag{4.8}$$

along with the log-transformed regression model used to describe the power relationship in Eqn. 4.8 [330, 399, 413]:

$$log(y) = \log(C) - a\log(x) + \epsilon \tag{4.9}$$

where $\epsilon$ now has zero mean in logarithmic units instead of in arithmetic units prior to log-transforming. Because addition of logarithms in Eqn. 4.9 results in multiplication in the arithmetic equivalent, $\epsilon$ must be included if the regression model is back-transformed to the arithmetic domain [330, 399]:

$$y = Cx^{-\alpha}\exp^{\epsilon} \tag{4.10}$$

where, unless $\epsilon = 0$ (perfect fit of the model to the data), will bias $y$ by $\exp^{\epsilon}$. Helsel [202] shows that the mean of items in logarithmic units is more than 2,000% in error when back-transformed to estimate the mean in arithmetic units. A number of correction factors have been proposed to correct for the bias. The most widely used correction factor is the multiplication of Eqn. 4.10 by $\exp^{s^2/2}$ where $s^2$ is the error variance for Eqn. 4.9 [399]. The reason is that exponentiation of a standard Gaussian results in a log-normal distribution with $\mu = \exp^{(\mu+\sigma^2/2)}$. Thus, the mean of a log-normal distribution cannot be obtained by exponentiating the mean of a Gaussian [413]. Other correction factors include the ratio estimator [400] and minimum variance unbiased estimators [399].

A key problem with probing for power laws through visual inspection using either graphical method is that many statistical models show up as approximately straight lines on log-log plots. Below are given two examples of why caution should be exercised when using visual inspection of log-log plots to probe for power laws. The two examples emphasise that (i) *alternative* models can also give rise to a straight line on log-log plots after being *fitted* and (ii) alternative models may approximate the distribution of the data better. The first example uses *ranked* data and the second highlight why visual inspection of a log-log plot is not a rigorous test.

## Example 1

Zipf's law inversely relates a term's rank $r$ in an English text of $N$ terms to its frequency [295]. That is, given $n$ data points $x = \{x_i\} : 1 \le i \le n$, order them in decreasing order $x_1 \ge x_2 \ge ... \ge x_r \ge ... \ge x_n$ of their frequency in $x^3$. Given such *ranked* data, Zipf's law is given by [463]:

$$x_r = \frac{C}{r^{\alpha}} \tag{4.11}$$

or equivalently

$$x_r = Cr^{-\alpha} : \alpha > 1, r \in \mathbb{Z}^+ \tag{4.12}$$

which follows the same form as Eqn. 4.4:

$$\log(x_r) = \log(C) - \alpha\log(x) \tag{4.13}$$

---

[3]While this example uses so-called rank-frequency plots, Newman [328] states that because "the cumulative distribution $P(x)$ is proportional to the rank $r$ of a term" then rank-frequency plots are cumulative plots.

and, consequently, may be said to be a power law. However, Eqn. 4.12 is special case of the Yule–Simon distribution [295, 397, 456]:

$$x_r = Cr^k b^r : k \in \mathbb{R}_{<0}, r \in \mathbb{Z}^+ \tag{4.14}$$

with corresponding linear form:

$$\begin{aligned} \log(x_r) &= \log(Cr^k b^r) \\ &= \log(C) - k\log(r) + r\log(b) \end{aligned} \tag{4.15}$$

which also produces a straight line on double logarithmic plots. However, variants of the log-normal [268]:

$$\log(x_r) = \log(C) - \alpha\log(r) - b(\log(r))^2 \tag{4.16}$$

and Weibull distribution [268]:

$$\log(x_r) = \log(C) - \alpha\log(r) - be^{\beta\log(r)} : 0 < \beta < 1 \tag{4.17}$$

also follow an approximate straight line on double logarithmic axes. Figure 4.4 shows Eqns. 4.13, 4.15, 4.16 and 4.17 on a log-log plot for synthetic data. For all equations, $a$ and $b$ were chosen so the different plots could be visually distinguished. $C$ was fixed to 10.

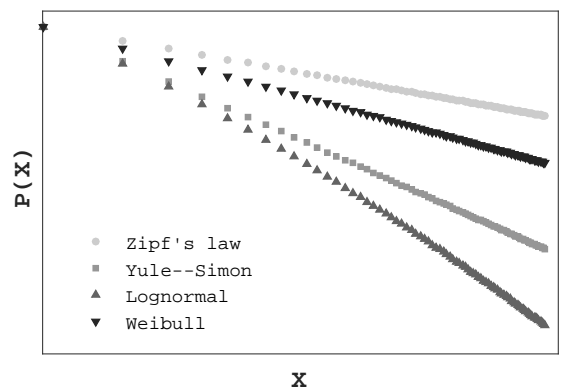

Figure 4.4: Alternatives to Zipf's law which, for ranked data, produce a straight line on log-log plots. The lines are generated using Eqns. 4.13, 4.15, 4.16 and 4.17.

Figure 4.4 shows that for ranked data at least, claiming that some quantity $x$ is approximately power law-distributed is not straight-forward and, in fact, several competing statistical models may also, or better, quantify the distribution of the data.

## Example 2

Consider the probability distributions in Figure 4.1, which were both reported to be approximately power law-distributed [13, 358]. Figure 4.5 shows the same plot but with three fitted statistical models superimposed: a power law (Eqn. 4.1), a Yule–Simon (Eqn. 4.15) and a lognormal (Eqn. 4.16).[4]

---

[4]These models were fitted using the principled approach in Chapter 3.

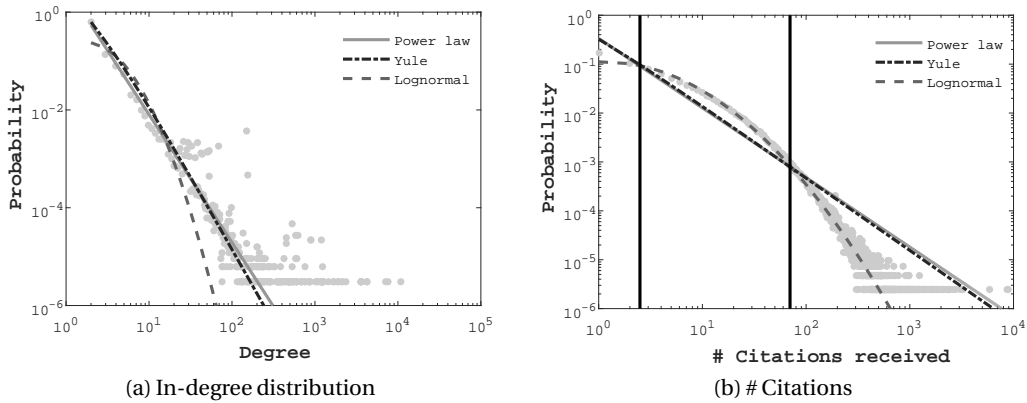

(a) In-degree distribution      (b) # Citations

Figure 4.5: Power laws and alternative statistical models fitted to the data from [13] and [328, 358]. The solid line is the power law. The dot-dashed line is a Yule–Simon model and the dashed line is a log-normal.

Figure 4.5a shows that both the power law and the Yule–Simon model quantify the degree distribution equally well: both models appear to approximate the most frequent degrees well, but neither model appears able to approximate nodes with higher degrees. Specifically, the probability of observing a website of degree $k \approx 75$ is the same as observing a website with degree $k \approx 10{,}000$ (and websites of many different degrees in that range). Similarly, the log-normal model appears to quantify the distribution of the most frequent degrees well (although it tends to overestimate the probability of them) but severely underestimates the probability of websites of degree $k \gtrsim 20$.

In contrast, Figure 4.5b shows that, visually, the log-normal provide a better fit to the distribution of the number of citations received, than both the power law and Yule–Simon. The latter two models underestimate the probability of papers that receives between $3 \lesssim k \lesssim 70$ citations (the part of the empirical distribution between the vertical lines) and subsequently begins to substantially overestimate the probability of papers that receive $\approx 100$ or more citations.

Taken together, the two examples suggest that visual inspection is not a sound approach for determining that something is approximately power law-distributed. However, many observations of approximate power laws in IR are carried out using these approaches, which brings into question many of these findings. This is the exact gap we propose to address in this chapter using the principled approach presented in Chapter 3. To further frame the potential use of our approach, the next section reviews several areas of IR where power laws, largely observed using graphical or similar ad hoc methods, have been reported and applied. Many of the works cited also use e.g. OLS to estimate the power law exponent which we showed in Chapter 3 to be error-prone.

## 4.3 Power Law Observations and Applications in IR

Power laws are frequently reported in IR [56, 315]. This section is divided into three different, but complementary, subsections focussing on the observation and application of power laws in IR:

- **Observations of power laws:** Papers that report power law-distributed properties of IR

data in one or more of the following ways: (i) visual inspection of log-log plot, (ii) reporting power law exponent, (iii) fitting regression line to the data.

- **Application of power laws:** Papers that use power law distributions to model/improve IR tasks. Typically these are based on observations of power laws in some dataset.

Notice that the two areas are not strictly mutually exclusive. The review is skewed towards research using Web data, as this was, by far, the most used type of data where power laws were reported or used, in the context of IR.

### 4.3.1 Power Law Observations

Power law distributions in both standard and distributed/peer-to-peer (P2P) IR, are frequently observed. In creating test collections [52, 93, 189, 281], for example, Bailey et al. [52] find that the probability of a server having $n$ pages is approximately power law-distributed. Similarly, Macdonald and Ounis [281] find that the number of outgoing links in blogs is power law-distributed. These findings were also reported by Kolari et al. [243]. PageRank scores have also been found to be power law-distributed [58, 59, 129, 337]. Craswell et al. [129], for example, find that PageRank scores in the .GOV collection are power law-distributed and Becchetti and Castillo [58] find the same for a sample of websites in the .GR domain. For distributed or P2P IR, the distribution of links [112], file sizes and participating peers [379], documents in testbeds for P2P IR [239] and the number of peer transactions on eBay [404] all follow power law distributions.

There is an extensive body of research on modelling different properties of the Web graph, with a heavy emphasis on topological or structural properties. The Web graph is a directed graph where nodes correspond to Web pages and links between nodes correspond to hyperlinks. One structural property of the Web graph that has been extensively studied is the degree distribution: the probability distribution of the degrees of all nodes in a graph.[5] The degree distribution is usually a good indicator of the graph's properties, both structurally and dynamically [177, Chap. 1] which may explain why it is so often the subject of study. Research that has studied the degree distribution and found it to be power law-distributed includes [6, 8, 13, 20, 53, 54, 75, 79, 142, 157, 173, 240, 254, 337, 401]. For example, Broder et al. [79] examine two Altavista crawls, each with more than 200 million websites and 1.5 billion links between them, and Soboroff [401] investigates if the 1.69 million WT10g[6] dataset is structurally similar to the Web. Despite the large difference in the sizes of the datasets, both Broder et al. and Soboroff find power law degree distributions for in-degrees and for most of the range of out-degrees. Other structural properties of the Web graph with power law distributions are the sizes of the connected components [46, 50, 79, 108, 142, 401] and the sizes of communities [91, 116, 187, 336]. For example, Chung et al. [108] find that (strongly) connected components–a property that assumes significance when examining the spread of e.g. computer viruses in a network [177, Chap. 1]–of the Japanese Web are power law-distributed across successive years, and Baeza-Yates and Tiberi [46] report similar findings for Web samples of many different countries. In addition, Palla et al. [336] find that community sizes–an important property when searching for e.g. single-topic websites in collections [329]–of a co-authorship network are power law-distributed.[7] Interestingly, the degree distribution of nodes in such single-topic sets of websites is also power law-distributed [100]. Finally, power law-distributed topological properties of the Web graph have been used to discover local authorities in Question-Answer communities [229] and to detect "link farms" [108]. *Temporally*, power law distributions are used to develop theories regarding growth dynamics of the Web [210] and models of graph evolution (see [337] and the references therein).

---

[5]The degree of a vertex is the number of edges incident to it.

[6]WT10g (http://ir.dcs.gla.ac.uk/test_collections/wt10g.html) is an official TREC collection used in TREC-9 and TREC-2001.

With the explosion in social data by Web applications such as Facebook, Twitter and Flickr, power law observations have followed suit and have been reported in collaborative tagging [66, 97, 171, 191, 432] and the tag vocabulary growth of these systems [96]; social bookmarking [26, 441]; Wikis [431] and their temporal evolution [86]; replies made to Web logs and Web posts [311] and for articles for ranking purposes [420]; semantic networks [412]; and Twitter networks [219] including a number of Twitter only statistics [98, 219, 256]. For example, Voss [432] compares the 25 most popular tags in Wikipedia's categorisation system with that of four other collaborative tagging systems and finds that they are all power law-distributed. Power law tag distributions are also reported in several semantic Web papers [143, 209, 335]. Similarly, Wetzker et al. [441] find that tag occurrence, bookmarks per URL and user bookmarks per month are all power law-distributed in a sample of 142 million bookmarks from the del.icio.us bookmarking service. Kwak et al. [256] observe that the number of followers on Twitter and the reach and depth of re-tweets both follow a power law. Finally, Alonso and Baeza-Yates [15], in crowdsourcing relevance assessments using Amazon Mechanical Turk, find that the number of tasks versus task completion time follows a power law.

Other areas of IR that are replete with power law observations include access statistics [5, 137, 211, 279], query or click logs [94, 130, 454], caching [47, 55, 144, 178, 265, 422, 450], NLP [232], text categorisation [455] and browsing patterns [319]. Huberman et al. [211], for example, observe that the frequency of clicks is power law-distributed, and Xue et al. [454], as well as [47, 144], find that the query frequency is power law-distributed. Finally, more peripheral areas of IR where power laws are observed include information access and database summary construction [9] and knowledge diffusion of patents [105].

## 4.3.2 Power Law Applications

Applications of power laws in IR tasks are found in such areas as classification [361], network [459] and test collection generation [93, 189]. A common use of power law distributions is to generate probabilities for unseen terms in language modelling [181, 246, 318, 421] and collaborative tagging systems [97]. For example, Koo et al. [246] construct an ontology based on hub terms in a semi-automatic fashion that can be used as an index file for document retrieval. The base ontology is constructed from hub terms - terms that are related to many other terms - using the *Wall Street Journal* collection and the assumption that term frequencies are power law-distributed. Nouns are then automatically added to the ontology using sentence extraction and relation extraction rules. Goldwater et al. [181] propose a generator/adaptor framework for specifying language models that may better learn linguistic information of text. The main idea is to use any generative model to generate text since the adaptor will transform this text into one where term frequencies are power law-distributed as this reflects natural language. Using their framework for morphological learning, their model comes close to the true distribution of suffixes of verbs in the training part of the *Penn Wall Street Journal* treebank. Power law distribution of terms has been used to model the prior probability of document relevance using citations [307] and link evidence [231]. Power laws have also been used to improve caching [49, 387]. Baeza-Yates and Saint-Jean [49], for example, show that using a Zipfian distribution for the query distribution in a search engine can improve query answering time by 5%. In addition, Serpanos et al. [387] use Zipf's law to derive the minimum number of requests made to a Web cache before cache members' popularity converges to a Zipfian distribution with a specific confidence.

---

[7]The communities are detected using their own algorithm, which is based on "quantities for the statistics of communities" [336].

Generated P2P networks should reflect the statistical properties of complex networks such as power law degree distributions and other small-world properties [439]. Examples include [222, 391, 459]. For example, Zhang et al. [459] study distributed search techniques in P2P networks where no single node has a global view of where resources are located in the network. Distributing the content of TREC-1, 2, 3 and TREC VLC over nodes in a network that exhibit both power law degree distribution and small world properties, their model demonstrates significant performance increase for retrieval. Similarly, Adamic et al. [8] demonstrate that power law-distributed networks can be searched efficiently (search costs increase sub-linearly with the size of the graph) using e.g. random walks, as such paths naturally gravitate towards high-degree nodes compared to Poisson-generated networks.

Power laws have also been used to estimate index sizes for the Semantic Web [335], to analyse the complexity of classifiers for text categorisation [455], to improve expected accuracy in unstructured P2P search using non-uniform document replication strategies [38], to provide an upper bound on the fraction of frequent queries in on-line batch query processing techniques arriving in any interval of time [144] and to design caching strategies for the Web [137].

## 4.4 RQ1: To What Extent Do Power law Approximations Hold?

This section investigates to what extent power law approximations hold for properties of IR datasets. Section 4.4.1 describes the experimental setup and Section 4.4.2 presents the experimental findings.

### 4.4.1 Experimental Setup

Table 4.1 lists the datasets we use. In total, we investigate 28 datasets of which 23 are official TREC datasets and 5 are non-TREC datasets. Each dataset was indexed with Indri 5.7[8], without stemming or stop word removal. Properties studied for each dataset are shown in Table 4.2. We restrict our investigation to English datasets here, but mention that power law-distributions of terms and *n*-grams have been reported in other languages such as Chinese [138, 190] and Hungarian [149].

Capsule descriptions of each dataset are given below.

---

[8]http://www.lemurproject.org/indri/.

| **TREC datasets** | | | |
|---|---|---|---|
| Collection | Datasets | Abbr | Size |
| | | # documents | |
| | Wall Street Journal (1987–1992) | wsj | 173,252 686M |
| | Federal Register (1988–1989) | fr | 45,820 485M |
| | Associated Press (1988–1990) | ap | 242,918 995M |
| TIPSTER | Department of Energy abstracts | doe | 226,087 299M |
| | Computer Select disks (1989–1992) | ziff | 293,121 919M |
| | San Jose Mercury News (1991) | sjm | 90,257 365M |
| | U.S. Patents (1983-1991) | patents | 6,711 216M |
| Total | | 1,078,166 | 3,965M |
| | | # documents | |
| | Financial Times Limited (1991–1994) | ft | 210,158 766M |
| | Congressional Record of the 103rd Congress (1993) | cr | 27,922 277M |
| TREC 4 & 5 | Federal Register (1994) | fr94 | 55,630 315M |
| | Foreign Broadcast Information Service (1996) | fbis | 130,471 576M |
| | Los Angeles Times (1989–1990) | latimes | 131,896 596M |
| Total | | 556,077 | 2,530M |
| | | # documents | |
| AQUAINT | New York Times (1999–2000) | nyt | 314,452 2,113M |
| | Associated Press (1999–2000) | apw | 239,576 939M |
| | Xinhua News Agency (1996–2000) | xie | 479,433 975M |
| Total | | 1,033,461 | 4,027M |
| | | # documents | |
| | Agence France Presse (2004–2006) | apf_eng | 296,967 889M |
| | Central News Agency (Taiwan) (2004–2006) | cna_eng | 19,782 50M |
| AQUAINT-2 | Xinhua News Agency (2004–2006) | xin_eng | 170,228 360M |
| | Los Angeles Times-Washington Post (2004–2006) | ltw_eng | 65,713 384M |
| | New York Times (2004–2006) | nyt_eng | 159,400 1,100M |
| | Associated Press (2004–2006) | apw_eng | 194,687 666M |
| Total | | 906,777 | 3,449G |
| | | # documents | |
| ClueWeb cat. B. | Websites in English (2009) | CW09 | 49,2M 169G |
| | Websites in English (2012) | CW12 | 52,3M 180G |
| Total | | 101,5M | 349G |
| **Non-TREC datasets** | | | |
| | | # documents | |
| | Physics articles, metadata and book recods | is_full | 453,254 4,179G |
| iSearch [280] | | # citations | |
| | Citations | is_cit | 3,768,410 102M |
| Total | | 4,221,664 | 4,281G |
| | | # queries | |
| Excite | Commercial query log (1999) | excite | 2,5M 115M |
| Microsoft | Commercial query log (2006) | MSN | 14,9M 1,1G |
| | | # $n$-gram | |
| Google books [179] | Syntactic unigrams | books | 13,6M 3,6G |

Table 4.1: Datasets used.

- ClueWeb09 cat. B. is an uncurated, domain-free crawl from 2009 of ca. 50 million Web pages in English.

- ClueWeb12 cat. B. is an uncurated, domain-free crawl from 2012 of ca. 52 million Web

pages in English.

- TIPSTER (TREC disks 1–3) comprises datasets of different style, size and domain. The Wall Street Journal (1987–1992), San Jose Mercury News (1991) and Associated Press (1988–1990) datasets are short newswire/newspaper stories. The Department of Energy dataset contains short abstracts. The Federal Register (1988–1989) dataset is whole issues of the Federal Register, a publication that serves as a reporting source for actions taken by U.S. government agencies. The Computer Select disks (1989–1992) contain articles on computer product information and the U.S. Patent (1983-1991) dataset consists of patents from the U.S. Patent Office.

- AQUAINT is a dataset of more than 1 million English news texts from the New York Times, the Associated Press and the Xinhua News Agency newswires from 1999 and 2000 (1996–2000 for Xinhua).

- AQUAINT-2 is a subset of the Linguistic Data Consortium's English Gigaword Third Edition of just under 1 million documents from Agence France Presse, Central News Agency (Taiwan), Xinhua News Agency, Los Angeles Times-Washington Post News Service, New York Times, and the Associated Press. All datasets are from 2004–2006.

- TREC disk-4 contains documents from the 103rd Congressional Record, the Federal Register from 1994 and the Financial Times (1992-1994).

- TREC disk-5 contains documents from the Foreign Broadcast Information Service and randomly selected articles from the Los Angeles Times.

- Excite and Microsoft 2006 query logs are samples from commercial search engine logs of different sizes (ca. 2.5M and 14M queries respectively).

- iSearch is an IR test collection from the domain of academic physics, with bibliographic citations of documents. The documents are scientific publications, their bibliographic and metadata records. All of these document types are relatively standardised in length, format and language.

- Google-released dataset of syntactic $n$-grams from the UK English subset of books published between 1520 and 2008. We use syntactic 1-grams (unigrams) which consist of a term and its syntactic role. E.g. `matter-NNdobj`, means that `matter` is a noun (NN) and the direct object (dobj) of a sentence. Only $n$-grams seen at least 10 times are included in the data by Google to keep its size manageable.

### 4.4.2 Experimental Findings

The results of each of the six dataset properties in Table 4.2 are presented separately in the following sections. To determine the best-fitting statistical model for each property, we applied our principled approach from Chapter 3. For each property, we report the best-fitting discrete and continuous model. All tables are found on-line.[9]

---

[9]http://www.nonplayercharacter.dk/tois-appendices.zip.

| Collections | Datasets | Term frequency | Document length | Query frequency | Query length | Citation frequency | Syn. $n$-gram frequency |
|---|---|---|---|---|---|---|---|
| TIPSTER | All datasets (see Table 4.1) | ✓ | ✓ | | | | |
| TREC 4 & 5 | All datasets (see Table 4.1) | ✓ | ✓ | | | | |
| AQUAINT | All datasets (see Table 4.1) | ✓ | ✓ | | | | |
| AQUAINT-2 | All datasets (see Table 4.1) | ✓ | ✓ | | | | |
| ClueWeb cat. B. | Websites in English (2009)<br>Websites in English (2012) | ✓<br>✓ | ✓<br>✓ | | | | |
| iSearch [280] | Full-text physics articles, metadata and book records | ✓ | ✓ | | | | |
| | Citations | | | | | ✓ | |
| MSN | Commercial query log | | | ✓ | ✓ | | |
| Excite | Commercial query log | | | ✓ | ✓ | | |
| Google books | Syntactic unigrams | | | | | | ✓ |

Table 4.2: Data properties studied for each dataset.

### 4.4.3 Term Frequency

The best-fitting discrete and continuous statistical model for each of the 24 term frequency distributions are shown in Table 4.3.

The best-fitting statistical model for all datasets is the Generalized Extreme Value (GEV). By the Fisher-Tippett-Gnedenko Theorem, the (normalised) extrema of any sequence of i.i.d. variables are GEV distributed, and hence the GEV is often used to describe extreme or rare events [394]. A rare event corresponds here to a rare term that is likely to appear in the tail of the term frequency distributions. As the number of rare terms increases, the tail of the distribution becomes sparser, which the GEV can approximate better than the other models. That the GEV is the best overall model, might be caused by the presence of stop words (which were not removed) which have much larger frequencies that non-stop words. The presence of such words may create a sufficient number of outliers which results in our approach selecting the GEV as the best overall distribution.

The best-fitting discrete model for 19/24 datasets is the Yule–Simon distribution. That a Yule–Simon model is the best-fitting discrete model suggests a preferential attachment mechanism, i.e. that "popular" terms are likely to "attach" to new terms. This seems likely as term frequencies are reportedly biassed by content or semantics [18].

The best-fitting discrete model for the Congressional Record, Federal Register, iSearch and both ClueWeb datasets is the power law model. While the results show that there is a statistically significant difference at the $p < .05$ level between the power law and Yule–Simon model, comparing the two models (both through their LR value and qualitatively by inspecting their fit to the distributions) show that the two models are very similar. For example, comparing the two models for ClueWeb09 cat. B. shows an LR difference of 84.2 indicating a nearly identical fit. Figure 4.6 shows term frequency distribution for 6 of the datasets in Table 4.1 which highlights the similarity for the power law and Yule–Simon distribution. Qualitatively, both quantify the stem of the distribution equally well, but neither can approximate the sparse tail. In contrast, the GEV appears to fit the distribution of lower frequency terms slightly worse than the Yule–Simon or power law but quantifies the high-frequency terms better.

| Collection | Dataset | Discrete | Continuous |
|---|---|---|---|
| TIPSTER | Wall Street Journal | Yule ($p = 1.506$) | Generalized Extreme Value ($k = 4.198, \sigma = 0.819, \mu = 1.195$) |
| | Federal Register | Yule ($p = 1.563$) | Generalized Extreme Value ($k = 4.318, \sigma = 0.79, \mu = 1.183$) |
| | Associated Press | Yule ($p = 1.529$) | Generalized Extreme Value ($k = 4.142, \sigma = 0.6999, \mu = 1.169$) |
| | Department of Energy | Yule ($p = 1.646$) | Generalized Extreme Value ($k = 4.221, \sigma = 0.7295, \mu = 1.173$) |
| | Computer Select disks[‡] | Yule ($p = 1.597$) | Generalized Extreme Value ($k = 3.819, \sigma = 1.853, \mu = 1.484$) |
| | San Jose Mercury News | Yule ($p = 1.515$) | Generalized Extreme Value ($k = 4.229, \sigma = 0.7439, \mu = 1.176$) |
| | U.S. Patents[‡] | Yule ($p = 1.554$) | Generalized Extreme Value ($k = 4.211, \sigma = 0.7171, \mu = 1.17$) |
| TREC 4 & 5 | Financial Times Limited[‡] | Yule ($p = 1.536$) | Generalized Extreme Value ($k = 4.139, \sigma = 0.6858, \mu = 1.166$) |
| | Congressional Record of the 103rd Congress | Power law ($\alpha = 1.635$) | Generalized Extreme Value ($k = 3.825, \sigma = 0.5372, \mu = 1.14$) |
| | Federal Register | Power law ($\alpha = 1.631$) | Generalized Extreme Value ($k = 4.117, \sigma = 0.677, \mu = 1.164$) |
| | Foreign Broadcast Information Service[‡] | Yule ($p = 1.604$) | Generalized Extreme Value ($k = 3.796, \sigma = 1.759, \mu = 1.462$) |
| | Los Angeles Times | Yule ($p = 1.509$) | Generalized Extreme Value ($k = 3.75, \sigma = 1.216, \mu = 1.323$) |
| AQUAINT | New York Times | Yule ($p = 1.498$) | Generalized Extreme Value ($k = 3.936, \sigma = 2.866, \mu = 1.727$) |
| | Associated Press | Yule ($p = 1.475$) | Generalized Extreme Value ($k = 3.991, \sigma = 3.381, \mu = 1.846$) |
| | Xinhua News Agency[‡] | Yule ($p = 1.618$) | Generalized Extreme Value ($k = 4.193, \sigma = 0.7286, \mu = 1.174$) |
| AQUAINT-2 | Agence France Presse | Yule ($p = 1.502$) | Generalized Extreme Value ($k = 3.994, \sigma = 3.296, \mu = 1.834$) |
| | Central News Agency (Taiwan) | Yule ($p = 1.496$) | Generalized Extreme Value ($k = 3.845, \sigma = 2.585, \mu = 1.671$) |
| | Xinhua News Agency | Yule ($p = 1.55$) | Generalized Extreme Value ($k = 3.788, \sigma = 1.622, \mu = 1.427$) |
| | L.A. Times-Washington Post | Yule ($p = 1.503$) | Generalized Extreme Value ($k = 3.941, \sigma = 3.119, \mu = 1.79$) |
| | New York Times[‡] | Yule ($p = 1.524$) | Generalized Extreme Value ($k = 4.215, \sigma = 0.7283, \mu = 1.173$) |
| | Associated Press | Yule ($p = 1.503$) | Generalized Extreme Value ($k = 3.832, \sigma = 2.562, \mu = 1.667$) |
| iSearch [280] | Full-text physics articles, metadata and book records | Power law ($\alpha = 1.844$) | Generalized Extreme Value ($k = 4.126, \sigma = 0.6834, \mu = 1.165$) |
| ClueWeb cat. B. | Web pages in English (2009) | Power law ($\alpha = 1.817$) | Generalized Extreme Value ($k = 3.664, \sigma = 0.4585, \mu = 1.125$) |
| | Web pages in English (2012) | Power law ($\alpha = 1.958$) | Generalized Extreme Value ($k = 3.648, \sigma = 0.442, \mu = 1.121$) |

Table 4.3: Best-fitting discrete and continuous models for term frequencies. [‡] indicates where Vuong's test and the AICc found different best-fitting models.

Vuong's LR test and AICc disagree on the best-fitting models in 5/24 datasets. Manual inspection of these cases showed that although the AICc values for all models were large, their differences were small.

Overall, the results confirm that very few terms have a very high frequency and many terms have a very low frequency. This happens because highly frequent terms belong to a *closed grammatical class* (e.g., determiners, prepositions): no new terms of that type can be produced

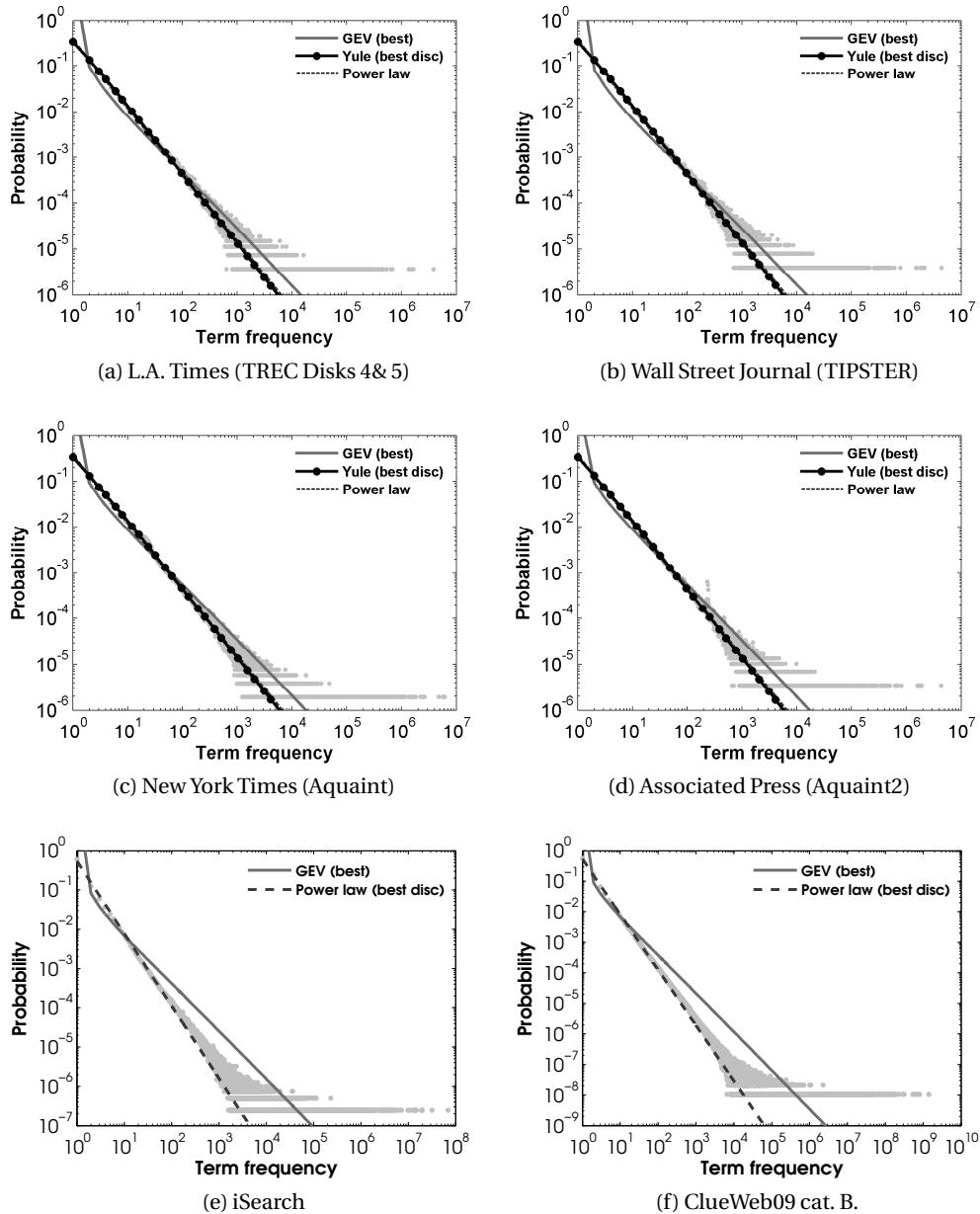

(a) L.A. Times (TREC Disks 4 & 5)

(b) Wall Street Journal (TIPSTER)

(c) New York Times (Aquaint)

(d) Associated Press (Aquaint2)

(e) iSearch

(f) ClueWeb09 cat. B.

Figure 4.6: Term frequency distributions and the best-fitting overall and discrete model. The power law is plotted as reference for those term frequency distributions where it is not the best-fitting discrete model.

in language (i.e. no new ways of producing equivalents to `the` or `of`), hence their frequency is boosted by their repeated use. Conversely, the numerous but infrequent terms belong to an *open grammatical class* (e.g. nouns, verbs): new terms of that type can be produced in language (e.g. `to tweet`, `a googler`) analogously to the new meanings that emerge. Hence, the frequency of these types of terms is "diluted" by varied usage, leading to a heavy tail. Collectively, the findings from the term frequency distribution in iSearch, ClueWeb09 cat. B., ClueWeb12 cat. B. Federal Register and the Congressional Records agree with Zipf's original findings.

#### 4.4.4 Query Frequency

Figure 4.7 shows the query frequency distributions for the Excite and Microsoft (MSN) query logs of different sizes (ca. 2.5M and 14M queries, respectively) from commercial search engines. Table 4.4F shows that the best-fitting continuous distribution was the GEV and the best-fitting discrete distribution was the power law.

| Collections | Dataset | Discrete | Continuous |
|---|---|---|---|
| Query frequency | Excite | Power law ($\alpha$=2.474) | Generalized Extreme Value ($k$=0.815,$\sigma$=0.152,$\mu$=1.076) |
| | MSN | Power law ($\alpha$=2.613) | Generalized Extreme Value ($k$=1.241,$\sigma$=0.205,$\mu$=1.106) |

Table 4.4: Best-fitting discrete and continuous models for query frequency. For the GEV, $k$ is the shape parameter, $\sigma$ is the scale parameter and $\mu$ the location parameter. For the power law, $\alpha$ is the scale parameter.

For the Excite dataset (Fig. 4.7a), the power law approximates the query frequency distribution well until the query frequency is approximately $10^2$. For queries occurring more frequently than $10^2$, the power law either underestimates or overestimates the probability of observing queries that have these frequencies as data sparsity causes the empirical distribution to fan out. The GEV underestimates the probabilities of query frequencies in the same range relative to the power law, but the "gap" between the two models narrows for queries with a frequency above $10^2$. The GEV, like the power law, also fails to approximate queries of frequencies above $10^2$. The lack of outliers in the Excite dataset may explain why the GEV does not appear to fit the dataset better than the power law. Comparatively, the MSN dataset (Fig. 4.7b) has more extreme[10] outliers. The GEV underestimates the probability of queries occurring less than $10^1$ but tends to overestimate the probability of queries with frequencies above $10^2$. Data sparsity again gives rise to a "noisy" tail, but because the tail of the GEV has extreme outliers, the GEV skews towards this phenomenon. In contrast, the power law approximates query frequencies in the interval $[10^0,10^1]$ well, but consistently underestimates the probability of query frequencies above the upper $10^1$. As the data are discrete, the power law is the best-fitting discrete model according to our framework. Consequently, our results corroborate the findings of e.g. [49], [50] and [47], and also suggest that the upper bound derived in [144] is reasonable.

#### 4.4.5 Query Length

Table 4.5 shows the results best-fitting discrete and continuous distribution of the query lengths of the Excite and Microsoft query logs. The best-fitting discrete and continuous model, in both cases, is the negative binomial and inverse Gaussian respectively. Our findings support previous results [109] that the negative binomial is useful for modelling text in documents, and has been used to derive new ranking models [119].

Figure 4.8 shows the fitted negative binomial and inverse Gaussian to the distribution of query lengths. The figure shows that, for both query logs, the power law is a poor approximation to the distribution of query lengths. In contrast, the negative binomial and inverse Gaussian both qualitatively approximate the distribution queries up to length $\approx 10$. For query lengths higher than $\approx 10$, the negative binomial drops off sharply and begins to underestimate the probability of observing a query of higher lengths. The inverse Gaussian, for the Excite query log, continues to quantify the distribution of query lengths up to $\approx 40$, but cannot capture the sparsity in the tail. In the MSN dataset, the inverse Gaussian quantifies queries up to length $\approx 20$. Our findings support those by Arampatzis and Kamps [31], who found that a mixture of a Poisson and Power law could fit the entire distribution of query lengths at the cost of calculating a transition point where one should switch between the two distributions.

---

[10]The standard deviation $\sigma^2$ of the MSN dataset is much larger than that of the Excite dataset.

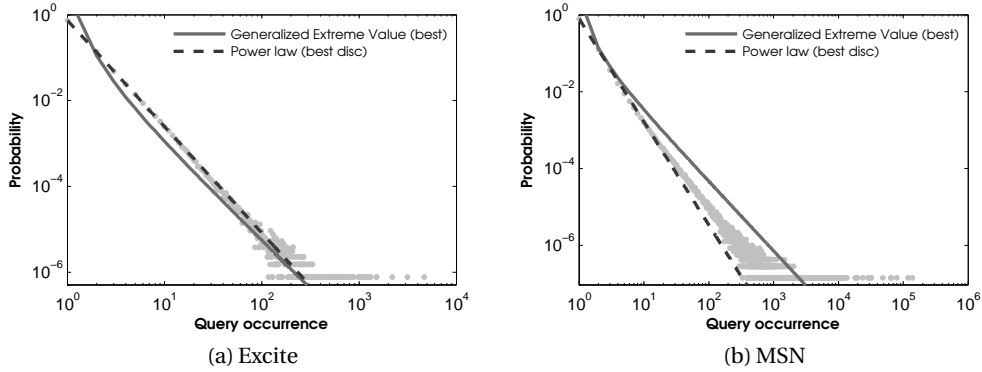

(a) Excite  (b) MSN

Figure 4.7: Query frequency distribution for the Excite and MSN query logs. We show the best-fitting (i) overall model (solid), (ii) discrete model (dot-solid) and (iii) power law (dashed). The first two models are the best-fitting according to Vuong's LR test. The power law is plotted as reference.

| Collections | Dataset | Discrete | Continuous |
|---|---|---|---|
| Query length | Excite | Negative Binomial $(r{=}4.238, p{=}0.556)$ | Inverse Gaussian $(\mu{=}3.249, \lambda{=}5.274)$ |
| | MSN | Negative Binomial $(r{=}43.671, p{=}0.948)$ | Inverse Gaussian $(\mu{=}2.401, \lambda{=}5.598)$ |

Table 4.5: Best-fitting discrete and continuous models for query lengths.

### 4.4.6 Document Length

The best-fitting discrete and continuous model for each dataset is shown in Table 4.6. The results show that the best-fitting discrete distribution in most cases is the negative binomial, and only in a few cases the geometric model. In contrast, the best-fitting continuous model changes from dataset to dataset, with the Gamma being the best-fitting continuous model in most cases. The Gamma distribution is used to model the time required for $a$ events to occur, given that the events occur randomly in a Poisson process with a mean time between events of $\beta$. Thus, if we know that a document of length $n$ occurs in an incoming stream of documents, on average, every $m^{\text{th}}$ seconds, the Gamma distribution models the number of seconds before the next document of length $n$ appears in the stream. For 8 datasets, the best-fitting models selected by Vuong's test and the AICc were different, though the disagreement is between models whose AICc values are relatively close to each other (or nested in which case we favour the most parsimonious model).

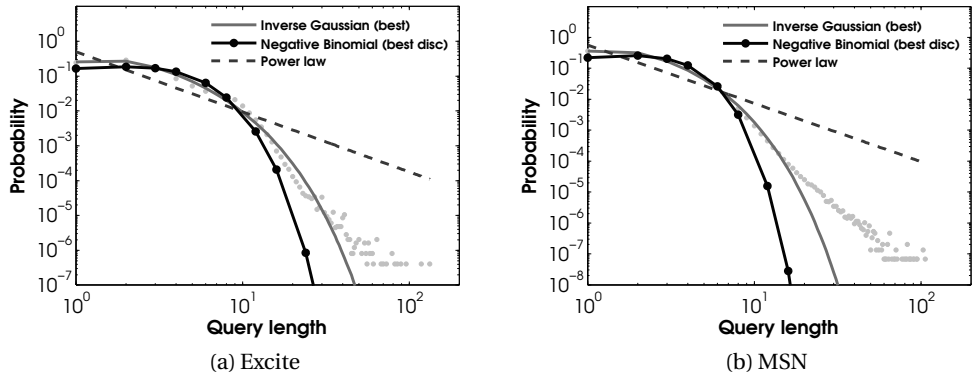

Figure 4.8: Distribution of query lengths for the Excite and MSN datasets. We show the best-fitting (i) overall model (solid), (ii) discrete model (dot-solid) and (iii) power law (dashed). The first two models are the best-fitting according to Vuong's test. The power law is plotted as reference.

Figure 4.9 shows 6 document length distributions. In all cases, the distributions deviate substantially from a straight line which suggests that the power law is a poor choice of model to quantify document lengths. The distribution of document lengths in the Xinhua dataset (Figure 4.9b) appears equally well approximated by both the Gamma and the negative binomial with small variations for documents of length $n < 10$. For long ($n \gtrsim 600$) documents, both models drop off more sharply than the document distribution.

The distribution of document lengths in ClueWeb09 cat. B. shows two distinct peaks at $n = 10$ and $100 \leq n \leq 400$ (a plateau). In addition, data sparsity gives rise to a "noisy" tail. The negative binomial distribution vastly overestimates the probability of document lengths in the interval $2 \leq n \leq 100$ but approximates the plateau and subsequent drop-off ($100 \leq n \leq 6,000$). However, for document lengths in the interval $6,000 \leq n \leq 250,000$, the negative binomial consistently underestimates their probability. The log-normal underestimates probabilities of document lengths in the interval $2 \leq n \leq 20$ and overestimates in the interval $20 \leq n \leq 100$. However, the log-normal model appears to approximate the document length distribution from $6000 \leq n \leq 20,000$ marginally better than the negative binomial does.

The document length distribution of the iSearch dataset displays three distinct local maxima. This is due to the dataset's heavily curated nature and the standardised document lengths of scientific publication environments: iSearch is a collection of full-length articles, their metadata and short bibliographic records of books in physics. These three types of data correspond to the three peaks in Figure 4.9d. The trimodality of the document length distribution makes it difficult for any of the models we consider to quantify any part of it. Indeed, while the inverse Gaussian provides the best fit, it consistently under or overestimates the probability of a document of any given length. To the best of our knowledge, no standard distribution exists that can model the oscillation of the ClueWeb09 cat. B. and the trimodality of the iSearch document length distribution. This suggests that mixture models could be a next step for quantifying document lengths.

| Collections | Dataset | Discrete | Continuous |
|---|---|---|---|
| TIPSTER | Wall Street Journal[‡] | Geometric ($p$:0.00223) | Inverse Gaussian ($\lambda$:258,$\mu$:447.5) |
| | Federal Register | Negative Binomial ($r$:0.5986,$p$:0.0004) | Generalized Extreme Value ($k$:0.7565,$\sigma$:330.6,$\mu$:324) |
| | Associated Press[‡] | Negative Binomial ($r$:3.137,$p$:0.0067) | Rayleigh ($b$:370) |
| | Department of Energy[‡] | Negative Binomial ($r$:3.989,$p$:0.03) | Rayleigh ($b$:99.24) |
| | Computer Select disks[‡] | Negative Binomial ($r$:0.7491,$p$:0.002) | Log-normal ($\mu$:5.098,$\sigma^2$:1.221) |
| | San Jose Mercury News | Geometric ($p$:0.002442) | Exponential ($\mu$:408.5) |
| | U.S. Patents[‡] | Geometric ($p$:0.0002) | — |
| TREC 4 & 5 | Financial Times Limited | Negative Binomial ($r$:1.469,$p$:0.0035) | Gamma ($a$:1.462,$b$:273.4) |
| | Congressional Record of the 103rd Congress | Negative Binomial ($r$:0.4669,$p$:0.0003) | Generalized Extreme Value ($k$:0.8658,$\sigma$:219.1,$\mu$:215.6) |
| | Federal Register | Negative Binomial ($r$:2.404,$p$:0.0035) | Log-normal ($\mu$:6.287,$\sigma^2$:0.6704) |
| | Foreign Broadcast Information Service[‡] | Geometric ($p$:0.002) | Generalized Extreme Value ($k$:0.5833,$\sigma$:164.7,$\mu$:214) |
| | Los Angeles Times[‡] | Geometric ($p$:0.002) | Gamma ($a$:1.035,$b$:485.1) |
| AQUAINT | New York Times[‡] | Negative Binomial ($r$:2.017,$p$:0.0025) | Nakagami ($\mu$:0.6925,$w$:88712) |
| | Associated Press | Negative Binomial ($r$:2.626,$p$:0.0061) | — |
| | Xinhua News Agency | Negative Binomial ($r$:4.146,$p$:0.019) | Gamma ($a$:4.055,$b$:51.09) |
| AQUAINT-2 | Agence France Presse | Negative Binomial ($r$:2.039,$p$:0.006) | Gamma ($a$:2.021,$b$:166.1) |
| | Central News Agency (Taiwan) | Negative Binomial ($r$:2.688,$p$:0.01) | Gamma ($a$:2.653,$b$:97.19) |
| | Xinhua News Agency | Negative Binomial ($r$:2.945,$p$:0.013) | Gamma ($a$:2.897,$b$:76.55) |
| | L.A. Times-Washington Post | Negative Binomial ($r$:1.715,$p$:0.0025) | Nakagami ($\mu$:0.5993,$w$:70213) |
| | New York Times | Negative Binomial ($r$:3.189,$p$:0.004) | — |
| | Associated Press | Negative Binomial ($r$:2.437,$p$:0.006) | Gamma ($a$:2.416,$b$:160.3) |
| iSearch [280] | Full-text physics articles, metadata and book records | Negative Binomial ($r$:0.3981,$p$:0.0001) | Inverse Gaussian ($\lambda$:183.2,$\mu$:2261) |
| ClueWeb cat. B. | Web pages in English (2009) | Negative Binomial ($r$:1.14,$p$:0.0014) | Log-normal ($\mu$:6.183,$\sigma^2$:0.9606) |
| | Web pages in English (2012) | Negative Binomial ($r$:0.8921,$p$:0.0011) | Generalized Pareto ($k$:0.2121,$\sigma$:558.6,$\theta$:0) |

Table 4.6: Best-fitting discrete and continuous models for document lengths. '–' in the Continuous column means the best-fitting model is the one listed in the Discrete column. [‡] indicate datasets where Vuong's test and the AICc found different best-fitting models.

### 4.4.7 Citation Frequency and Syntactic Unigrams

Table 4.7 shows the best-fitting discrete and continuous model for the iSearch citation and Google syntactic unigram dataset. The results, for both datasets, show that the Yule–Simon model is the best discrete model. Similar to the term frequency distributions (Section 4.4.3), this suggests the presence of a preferential attachment scheme which, for the iSearch citation dataset, supports previous findings from citation analysis [355, 220, 358] i.e. highly cited papers will continue to attract citations. The GEV is the best-fitting model overall as was also the case for all term frequency distributions. For the syntactic unigram distribution, the Yule–Simon distribution is the best-fitting discrete *and* overall model. Also similar to the findings from the term frequency distributions, the Yule–Simon model suggests that POS-tagged term frequencies are also biassed by content or semantics.

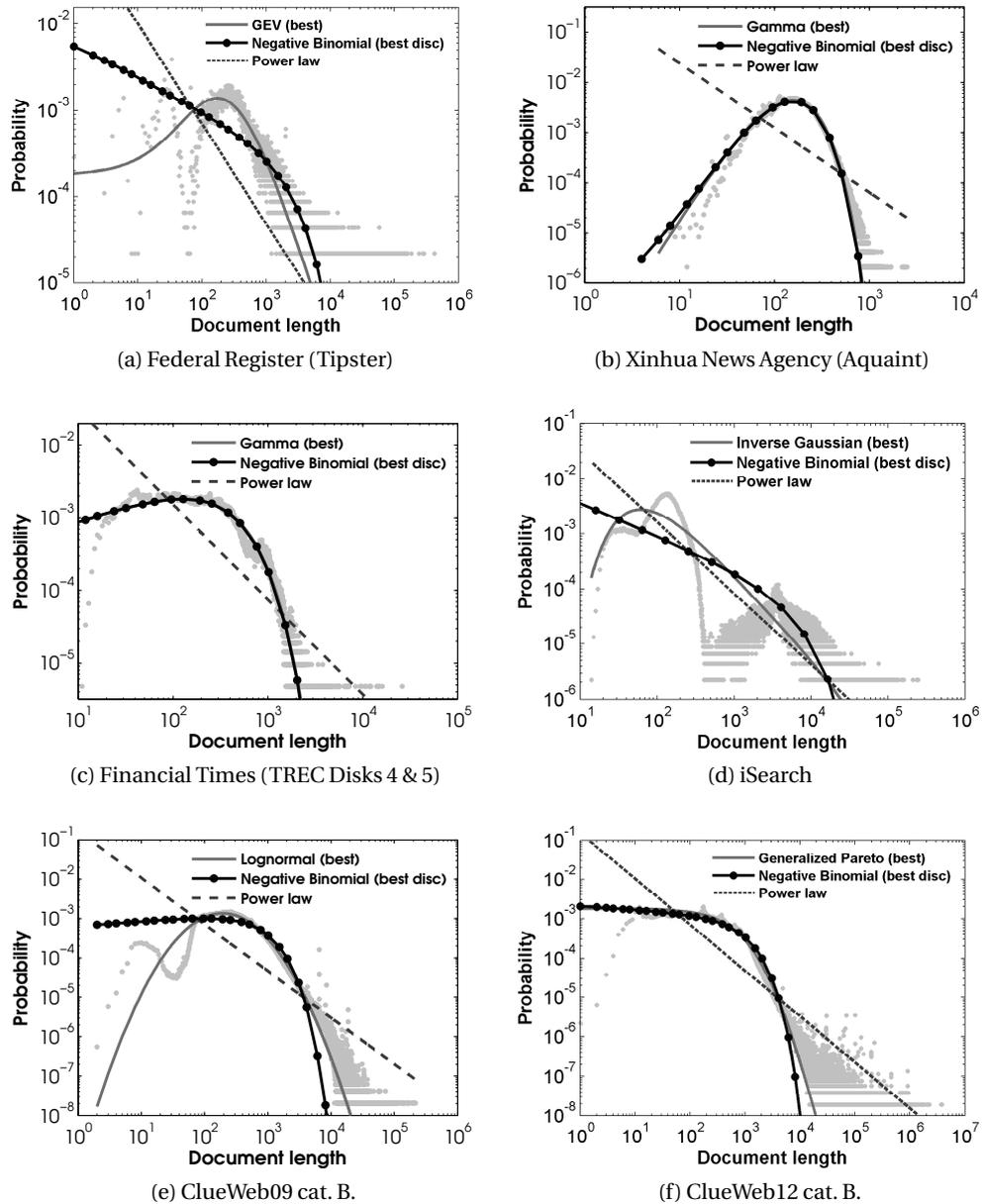

Figure 4.9: Document length distributions and the best-fitting overall and discrete model. The power law is plotted as reference for those term frequency distributions where it is not the best-fitting discrete model.

Figure 4.10 shows the distribution of both the iSearch citation dataset and the syntactic unigrams. The distribution of the iSearch citation dataset is curved, and neither the Yule–Simon nor the GEV can approximate this curvature. Both models also substantially overestimate the probability of a paper receiving $k > 100$ citations. The syntactic unigram distribution visually resembles the cone-shapes of the term frequency distributions. The Yule–Simon model visually appears to fit syntactic unigrams with frequencies $k \lesssim 5{,}000$ after which point the distribution "fans out" and the Yule–Simon fails to provide a good approximation.

| Collections | Datasets | Discrete | Continuous |
|---|---|---|---|
| iSearch | Citations | Yule ($p$=1.518) | Generalized Extreme Value ($k$=3.761,$\sigma$=2.098,$\mu$=1.556) |
| Google books | Syntactic unigrams | Yule ($p$=1.683) | — |

Table 4.7: Best-fitting discrete and continuous models for citations and syntactic unigrams. For syntactic unigrams, the Yule model was the best-fitting model overall.

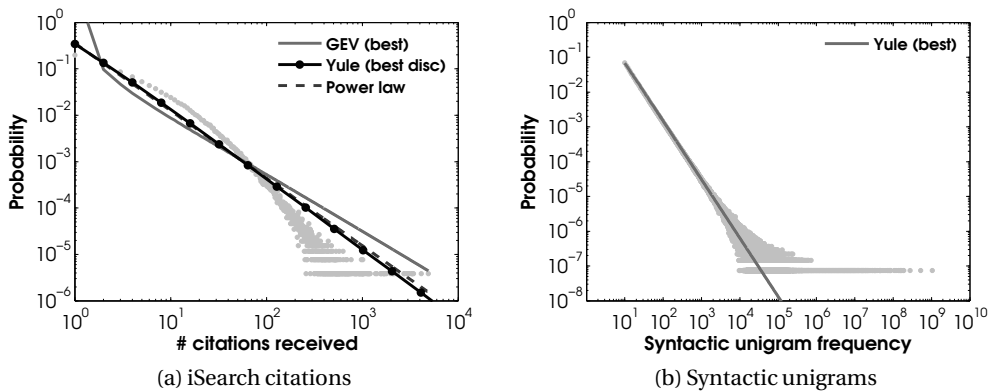

(a) iSearch citations

(b) Syntactic unigrams

Figure 4.10: Distribution of citations in iSearch and syntactic unigrams in Google books. We show the best-fitting (i) overall model (solid), (ii) discrete model (dot-solid) and (iii) power law (dashed). The first two models are the best-fitting according to Vuong's test. The power law is plotted as reference.

### 4.4.8 Summary

Our first research question was concerned with the extent to which power laws hold for properties of IR datasets. Specifically, we investigate if term frequency, document length, query frequency, query length, citation frequency and syntactic unigram frequency distributions - all of which have been associated with power laws in the past - are, indeed, (approximately) power law distributed.

Table 4.8 summarises our findings for our first research question. The table shows that the vast majority of the properties studied are *not* (approximately) power law distributed.

| Term Frequency | Document Length | Query Frequency | Query Length | Citations Received | Syntactic Unigrams |
|---|---|---|---|---|---|
| 5/24 | 0/24 | 2/2 | 0/2 | 0/1 | 0/1 |

Table 4.8: Number of approximate power laws found for each data property per dataset.

As our results show that 5 term frequency distributions are power law distributed, this suggests that computations assuming that Zipf's law, such as computation of the space requirements for an inverted index compressed with any standard encoding such as $\gamma$ encoding [289, Ch. 5], will be more accurate than if other models are used. However, in the cases where the Yule–Simon is the best-fitting model, a more accurate estimate of the size of the index might be achieved. Overall, however, the GEV is the best-fitting model which is likely caused by (very) rare terms which it is better at quantifying.

We find that distributions of document lengths tend to have one or more local maxima, which makes the power law a poor choice of model. Instead, we find that multi-modal models such as the negative binomial, Gamma, GEV or log-normal are consistently better-fitting models. These are amongst some of the models used by Zhou et al. [461] to derive a length-biased BM25 ranking model which show superior performance over a tuned BM25.

Our results show that query frequencies are approximately power law-distributed lending support to previous findings [47, 144]. Specifically, our findings indicate that the increase in expected accuracy for top-1 documents obtained by [38] is reasonable and that the upper bound of query frequency received by a search engine [144] may be correct. Similarly, our results for query length agrees with previous research [31] in that a power law is a poor choice of model for the entire range of lengths, but that a mixture model may provide superior fits.

The citation frequency distribution is best fitted by the GEV and Yule–Simon model. Our results show that the GEV is better at approximating highly cited papers which is likely why it is preferred over the Yule–Simon. The preferential attachment mechanism implied by the Yule–Simon model supports previous findings from citation analysis: highly cited articles will continue to be cited. This agrees with previous findings such as [355, 220, 307, 358]. We are not aware of any work in IR where the GEV has been used for this purpose.

For syntactic unigrams, the Yule–Simon model was the best discrete *and* overall model. Similar to our results for term frequencies, the preferential attachment mechanism implies that syntactic unigram frequencies are biassed by content or semantics [18]. Unlike the term frequency distributions, the GEV is not the best-fitting model overall. One reason might be that the tail of the syntactic unigram distribution is not sufficiently "extreme" for the GEV to be favoured.

## 4.5   What are the Computational Costs of Using Our Approach

Ideally, our approach should be an efficient (but also effective) alternative to the ad hoc graphical methods commonly used in IR to probe for approximate power laws.

To determine if our approach is efficient, we measure the elapsed time required to fit each model from Table 2.2 to each data property per dataset in Table 4.2. Each model is fitted three times and we report the median elapsed time. All experiments were done on an unladen server with a single Intel Core i7 CPU with 32GB memory running Ubuntu Linux 13.10 and MATLAB R2013B on a 256GB OCZ Agility 3 SSD.

The results are shown in Table 4.9. Most of the statistical models can be fitted to any property within at most a few seconds. This is largely caused by the closed-form solution to the MLE parameters of these models. The exception is the GEV which, for the two ClueWeb and Google datasets, takes up to $\approx 14$ hours to fit. The reason is two-fold. Firstly, we used a gradient-based optimisation routine to estimate the MLE parameters of the GEV. The reason is that the maximum likelihood estimates would otherwise converge to a boundary point and while MLE is typically robust against this [74, 107], any likelihood ratio test must be corrected [247]. Secondly, we did not parallelise the implementation. Subsequent testing has shown that the GEV can be accurately fitted to these datasets in approximately 23 minutes. This can be further improved using better hardware, or more optimised solvers for those statistical models with no closed-form solution for their MLE parameters.

Table 4.9: Elapsed time (median of 3 trials in seconds) to fit each model to each data property per dataset. For iSearch, we report the total time for term frequencies, document lengths and citations.

| | | | Exp | Gamma | Gauss | Geo | GEV | GP | IGauss | Log | Logn | Naka | NBin | Pois | Plaw | Rayl | Weib | Yule |
|---|---|---|---|---|---|---|---|---|---|---|---|---|---|---|---|---|---|---|
| **TIPSTER** | wsj | dl | .01 | .02 | .01 | .00 | 25.90 | .10 | .05 | .22 | .02 | .08 | .26 | .05 | .10 | .02 | .04 | .27 |
| | | tf | .01 | .02 | .01 | .00 | 52.16 | .38 | .05 | 1.14 | .02 | .10 | .36 | .06 | .57 | .01 | .05 | .77 |
| | fr | dl | .00 | .01 | .00 | .00 | 9.67 | .05 | .01 | .08 | .01 | .02 | .05 | .01 | .04 | .01 | .01 | .09 |
| | | tf | .01 | .02 | .02 | .00 | 57.86 | .39 | .06 | 1.08 | .02 | .11 | .42 | .07 | .66 | .02 | .06 | .88 |
| | ap | dl | .01 | .02 | .02 | .00 | 46.14 | .23 | .06 | .27 | .02 | .10 | .28 | .09 | .13 | .02 | .05 | .42 |
| | | tf | .01 | .03 | .02 | .00 | 68.43 | .50 | .08 | 1.51 | .03 | .14 | .46 | .09 | .74 | .03 | .07 | 1.01 |
| | doe | dl | .01 | .02 | .01 | .00 | 4.43 | .36 | .06 | .23 | .02 | .10 | .28 | .10 | .12 | .02 | .05 | .39 |
| | | tf | .01 | .02 | .01 | .00 | 41.67 | .30 | .04 | .65 | .02 | .08 | .25 | .05 | .43 | .01 | .04 | .59 |
| | ziff | dl | .01 | .02 | .01 | .00 | 24.98 | .26 | .07 | .39 | .03 | .11 | .45 | .08 | .14 | .02 | .06 | .43 |
| | | tf | .01 | .03 | .02 | .00 | 75.44 | .56 | .08 | 1.65 | .03 | .14 | .51 | .09 | .80 | .02 | .07 | 1.09 |
| | sjm | dl | .00 | .01 | .00 | .00 | 17.38 | .05 | .02 | .08 | .01 | .03 | .11 | .03 | .05 | .01 | .02 | .14 |
| | | tf | .01 | .02 | .01 | .00 | 42.68 | .31 | .05 | .70 | .02 | .08 | .29 | .05 | .49 | .02 | .04 | .65 |
| | patents | dl | .00 | .00 | .00 | .00 | .51 | .02 | .01 | .03 | .00 | .01 | .01 | .00 | .02 | .00 | .00 | .03 |
| | | tf | .01 | .01 | .01 | .00 | 29.86 | .21 | .03 | .53 | .01 | .06 | .18 | .04 | .36 | .01 | .03 | .47 |
| | **Total** | dl | .06 | .12 | .08 | .01 | 129.05 | 1.10 | .30 | .33 | .12 | .47 | 1.46 | .38 | .62 | .10 | .25 | 1.80 |
| | | tf | .08 | .17 | .12 | .02 | 368.14 | 2.67 | 0.42 | 7.28 | .17 | .72 | 2.49 | .47 | 4.05 | .16 | .39 | 5.48 |
| **TREC 4 & 5** | ft | dl | .01 | .02 | .01 | .00 | 31.43 | .15 | .06 | .26 | .02 | .09 | .19 | .08 | .11 | .02 | .05 | .81 |
| | | tf | .01 | .02 | .02 | .00 | 62.91 | .50 | .07 | 1.11 | .03 | .12 | .38 | .08 | .68 | .02 | .06 | .93 |
| | cr | dl | .01 | .01 | .00 | .00 | 5.28 | .04 | .01 | .19 | .00 | .02 | .05 | .01 | .03 | .01 | .01 | .06 |
| | | tf | .01 | .02 | .01 | .00 | 49.12 | .34 | .06 | .80 | .02 | .09 | .30 | .06 | .56 | .02 | .05 | .73 |
| | fr94 | dl | .00 | .01 | .00 | .00 | 11.07 | .04 | .02 | .39 | .01 | .03 | .09 | .02 | .04 | .01 | .02 | .26 |
| | | tf | .01 | .02 | .02 | .00 | 54.31 | .37 | .06 | .84 | .02 | .10 | .34 | .06 | .54 | .02 | .05 | .73 |
| | fbis | dl | .01 | .01 | .01 | .00 | 24.93 | .11 | .03 | .23 | .01 | .05 | .20 | .04 | .08 | .01 | .03 | .48 |
| | | tf | .01 | .03 | .02 | .00 | 6.03 | .44 | .07 | 1.12 | .03 | .12 | .38 | .07 | .63 | .02 | .06 | .82 |
| | latimes | dl | .01 | .01 | .01 | .00 | 22.17 | .08 | .04 | .17 | .02 | .06 | .19 | .04 | .08 | .01 | .03 | .23 |
| | | tf | .01 | .02 | .01 | .00 | 55.03 | .40 | .06 | 1.01 | .02 | .11 | .30 | .07 | .60 | .02 | .06 | .79 |
| | **Total** | dl | .04 | .07 | .05 | .01 | 94.90 | .42 | .16 | 1.25 | .07 | .25 | .72 | .20 | .35 | .06 | .14 | 1.83 |
| | | tf | .06 | .13 | .09 | .02 | 227.41 | 2.06 | .32 | 4.96 | .12 | .54 | 1.71 | .36 | 3.02 | .10 | .29 | 4.01 |
| **AQUAINT** | nyt | dl | .01 | .03 | .02 | .00 | 8.4 | .27 | .09 | .36 | .03 | .14 | .48 | .13 | .16 | .03 | .07 | .54 |
| | | tf | .02 | .05 | .03 | .00 | 106.15 | .85 | .12 | 1.81 | .04 | .20 | .67 | .14 | 1.18 | .03 | .11 | 1.65 |
| | apw | dl | .01 | .02 | .01 | .00 | 47.12 | .20 | .07 | .27 | .02 | .10 | .32 | .09 | .13 | .02 | .05 | .41 |
| | | tf | .01 | .03 | .20 | .00 | 61.28 | .48 | .07 | 1.20 | .03 | .12 | .44 | .08 | .70 | .03 | .06 | .96 |
| | xie | dl | .02 | .04 | .03 | .00 | 9.82 | .40 | .13 | .49 | .05 | .21 | .87 | .23 | .24 | .04 | .10 | 1.79 |
| | | tf | .01 | .03 | .02 | .00 | 77.92 | .52 | .09 | 1.46 | .03 | .15 | .49 | .09 | .80 | .02 | .08 | 1.10 |
| | **Total** | dl | .05 | .09 | .06 | .01 | 65.38 | .87 | .29 | 1.12 | .11 | .46 | 1.69 | .47 | .54 | .09 | .23 | 2.74 |
| | | tf | .04 | .10 | .07 | .01 | 245.35 | 1.85 | .28 | 4.49 | .11 | .48 | 1.60 | .31 | 2.68 | .12 | .26 | 3.71 |

Table 4.9: (continued)

| | | | Exp | Gamma | Gauss | Geo | GEV | GP | IGauss | Log | Logn | Naka | NBin | Pois | Plaw | Rayl | Weib | Yule |
|---|---|---|---|---|---|---|---|---|---|---|---|---|---|---|---|---|---|---|
| AQUAINT-2 | afp | dl | .02 | .03 | .01 | .00 | 7.02 | .37 | .08 | .33 | .03 | .13 | .46 | .11 | .16 | .04 | .10 | .50 |
| | | tf | .01 | .02 | .01 | .00 | 6.16 | .43 | .07 | 1.16 | .03 | .12 | .46 | .08 | .68 | .03 | .06 | .94 |
| | cna | dl | .00 | .01 | .00 | .00 | 2.8 | .03 | .01 | .03 | .01 | .01 | .02 | .01 | .03 | .01 | .01 | .13 |
| | | tf | .01 | .01 | .00 | .00 | 1.27 | .07 | .01 | .16 | .01 | .02 | .07 | .01 | .18 | .03 | .01 | .22 |
| | xie | dl | .01 | .02 | .01 | .00 | 3.73 | .15 | .05 | .20 | .02 | .08 | .24 | .07 | .10 | .02 | .04 | .29 |
| | | tf | .01 | .02 | .01 | .00 | 4.52 | .29 | .05 | .68 | .02 | .08 | .30 | .05 | .47 | .02 | .04 | .63 |
| | ltw | dl | .01 | .01 | .01 | .00 | 12.86 | .06 | .02 | .06 | .01 | .03 | .06 | .03 | .05 | .01 | .02 | .12 |
| | | tf | .01 | .02 | .01 | .00 | 45.19 | .34 | .05 | .75 | .02 | .09 | .32 | .06 | .51 | .02 | .05 | .67 |
| | nyt | dl | .01 | .02 | .01 | .00 | 4.52 | .14 | .04 | .21 | .02 | .07 | .22 | .06 | .09 | .02 | .04 | .28 |
| | | tf | .01 | .03 | .02 | .00 | 79.45 | .60 | .09 | 1.44 | .03 | .15 | .58 | .10 | .83 | .02 | .08 | 1.15 |
| | apw | dl | .01 | .02 | .01 | .00 | 38.71 | .16 | .05 | .22 | .02 | .08 | .29 | .07 | .11 | .02 | .05 | .34 |
| | | tf | .01 | .03 | .02 | .00 | 62.55 | .45 | .07 | 1.12 | .03 | .12 | .41 | .08 | .68 | .03 | .06 | .91 |
| | **Total** | dl | .05 | .10 | .07 | .01 | 69.71 | .92 | .26 | 1.05 | .10 | .41 | 1.31 | .37 | .55 | .12 | .26 | 1.67 |
| | | tf | .07 | .14 | .09 | .02 | 199.16 | 2.19 | .34 | 5.32 | .13 | 0.59 | 2.14 | .39 | 3.37 | .21 | .32 | 4.54 |
| iSearch | is_full | dl | .02 | .04 | .03 | .01 | 39.04 | .65 | .12 | .79 | .04 | .19 | .62 | .12 | .23 | .03 | .10 | 1.70 |
| | | tf | .12 | .32 | .21 | .05 | 825.28 | 9.60 | .84 | 15.46 | .31 | 1.52 | 5.64 | 1.01 | 13.00 | .24 | .82 | 16.51 |
| | is_cit | cf | .01 | .02 | .02 | .00 | 53.97 | .29 | .06 | .35 | .02 | .10 | .22 | .08 | .65 | .02 | .06 | .85 |
| | **Total** | all | .16 | .38 | .25 | .06 | 918.31 | 10.54 | 1.03 | 16.61 | .38 | 1.82 | 6.48 | 1.22 | 13.89 | .30 | .98 | 19.05 |
| Query logs | excite | qf | .06 | .10 | .08 | .02 | 8.78 | .85 | .18 | 5.03 | .10 | .33 | .98 | .49 | 1.01 | .08 | .20 | 2.28 |
| | msn | qf | .36 | .61 | .48 | .15 | 150.01 | 9.25 | 1.28 | 226.91 | 0.64 | 2.04 | 6.53 | 3.35 | 11.67 | 0.52 | 1.24 | 17.99 |
| | **Total** | ql | .08 | .21 | .14 | .03 | 24.02 | 3.65 | .56 | 3.70 | .20 | .97 | 3.74 | .90 | 1.25 | .16 | .53 | 12.69 |
| ClueWeb | CW09 | dl | 1.49 | 4.03 | 2.52 | 0.63 | 8577.09 | 91.64 | 13.20 | 77.27 | 4.46 | 20.96 | 94.24 | 18.94 | 22.60 | 3.32 | 10.39 | 184.96 |
| | | tf | 2.90 | 7.46 | 4.91 | 1.23 | 19155.91 | 266.34 | 19.62 | 384.78 | 7.27 | 35.54 | 143.14 | 24.05 | 345.19 | 5.60 | 20.06 | 585.17 |
| | CW12 | dl | 1.58 | 4.26 | 2.70 | 0.67 | 9150.52 | 118.43 | 14.04 | 191.20 | 4.74 | 22.29 | 116.01 | 21.51 | 24.07 | 3.52 | 11.53 | 161.94 |
| | | tf | 4.40 | 10.45 | 7.76 | 2.27 | 32565.05 | 575.29 | 32.58 | 619.49 | 12.58 | 55.46 | 226.17 | 51.95 | 735.27 | 10.36 | 24.28 | 971.39 |
| | **Total** | dl | 3.08 | 8.30 | 5.23 | 1.30 | 17727.62 | 210.08 | 27.24 | 268.48 | 9.20 | 43.26 | 210.25 | 40.45 | 46.67 | 6.84 | 21.93 | 346.91 |
| | | tf | 7.30 | 17.91 | 12.67 | 3.50 | 51720.96 | 841.63 | 52.20 | 1004.27 | 19.85 | 91.00 | 369.31 | 76.00 | 1080.46 | 15.96 | 44.34 | 1556.56 |
| Google | books | sn | .42 | 1.12 | .70 | .18 | 1007.37 | 37.07 | 3.62 | 62.20 | 1.23 | 5.78 | 26.26 | 4.02 | 6.28 | .90 | 3.00 | 80.41 |

## 4.6 Conclusion

Many properties of standard IR datasets are assumed to be distributed according to some statistical model. Of these, the *power law* in particular has been either assumed or observed in many of different areas of IR. These assumptions and observations, however, may not be accurate as (i) previous work have found many datasets associated with power laws are, in fact, not power laws [117] and (ii) most assumptions or observations of power laws in IR are based on ad hoc methods. Consequently, this chapter focussed on testing power law assumptions and observations frequently made in IR. Specifically, we formulate the research question: To what extent do power law approximations hold for term frequency, document length, query frequency, query length, citation frequency, and syntactic unigram frequency? These are all properties of IR datasets previously associated with power laws. To answer our research question, we systematically fit the power law and 15 candidate statistical models to the different distributions of 28 datasets (23 TREC and 5 non-TREC) of different sizes. Our results show that 5/24 term frequency distributions and 2/2 query frequency distribution are approximately power law distributed. Document length, query length, citation frequency and syntactic unigram frequency are all better approximated by an inverse Gaussian, Generalized Extreme Value, negative binomial or Yule–Simon model. As our approach is effective, we also investigated if it is efficient. To this end each statistical model was fitted three times to each dataset and the median elapsed time recorded. Our results show that most models can be fitted within at most a few seconds though depending on the size of the dataset, some model can take several hours. This effect can be mitigated using better hardware and parallelisation. Thus, we tentatively conclude that our approach is both an effective and efficient alternative to ad hoc methods such as visual inspection of distributions in double logarithmic axes.

While this chapter has shown that statistical models other than the power law should be considered, any effects this might practically have was not investigated. Previous research by Zhou et al. [461], however, indicates that using an approximate "correct" statistical model of document lengths improves ad hoc retrieval. Consequently, we are interested in whether using the best-fitting statistical model to *other* properties of IR datasets can be used to improve the effectiveness of ad hoc retrieval. Chapter 5 in this thesis will revisit this problem.

# 5 Adaptive Distributional Ranking Models for IR

This chapter derives new ranking models that adapt to the distribution of the term frequencies in a collection.

This chapter is organised as follows. Section 5.1 introduces the idea of adapting the ranking computations to the statistics of the documents being retrieved. This idea is applied to the Divergence From Randomness (DFR) ranking models, which are presented in Section 5.2. Section 5.3 reviews related work most closely related to ours. Section 5.4 presents our novel cascading methodology and Section 5.5 derives *adaptive distributional ranking* (ADR) models from the DFR framework. Section 5.6 experimentally evaluates our ADR models against several strong baselines. Finally, Section 5.7 concludes the chapter. This chapter makes use of the notation introduced in Section 2.1, though some notational overload is introduced to be consistent with the DFR notation.

## 5.1  Introduction

A central issue in IR is *ranking* [175, 370]: given a finite set of documents and queries, compute a score for each document with respect to each query, sort the documents in decreasing order of this score, and return a list of, typically, the top-$k$ ranked documents. Depending on the application, the scores may represent estimated degrees of relevance, preference or "importance". Typically, each score is taken to be an estimate of a document's relevance to the query. Estimating relevance scores for each document is done using a "ranking model", and developing effective ranking models is a fundamental problem in IR [175, 370, 451]. [1]

In this chapter, we present new ranking models that can adapt their computations to the statistics of the documents being retrieved. While this, in principle, could be done to any ranking model that relies on distributional assumptions about some property of the collection (i.e. that document lengths are distributed by a Gaussian), we choose to use DFR as it is a flexible framework that facilitates the derivation of new ranking models based on assumptions about collection statistics.

The family of ranking models from DFR [17, 19], are inspired by Harter's two hypotheses [197, 198]: (i) Given a collection of documents there exist a set of *elite* and a set of *non-elite* documents. In the elite set of documents, *informative* terms occur much more frequently than in documents outside the elite set. (ii) Terms that do not occur in the elite set of documents, so-called *non-informative* terms, are distributed differently than informative terms. Specifically, the central idea in DFR is that non-informative terms are assumed to be distributed "randomly" in documents [19]. However, because documents are not constructed randomly (i.e. by picking terms at random from a collection and placing them in documents), they cannot be considered random; indeed, term frequencies are biassed by content or semantics [18] which makes them less "random". Thus, the more a term diverges from "randomness", the more informative the term is. Documents are then ranked by considering the probability that the document term distribution would occur by chance [89, p. 295]. In DFR, the notion of "randomness" is captured using a *statistical model* (also known as a "model of randomness" [19]) (introduced in Section 5.2.2) which quantifies the distribution of non-informative terms. More specifically, let $f_{t,d}$ and $f_{t,C}$ denote the within-document and collection frequency of a term $t$ in a document $d$ in collection $C$. The divergence between $f_{t,d}$ and $f_{t,C}$ according to the statistical model, is then used as a measure of term informativeness: the smaller the probability of observing $f_{t,d}$ occurrences of $t$ under the chosen statistical model, the more informative the term is.

---

[1]Trotman [424] reports that the number of published ranking models is several thousands.

Each choice of statistical model instantiates a new DFR ranking model which quantifies the distribution of non-informative terms in a different manner. However, as the choice of statistical model can substantially affect the effectiveness of a DFR ranking model [19, 121], it becomes important to select the statistical model that best quantifies the distribution of non-informative terms. Previous work [197, 198], for example, used Poisson distributions with different means based on empirical evidence of the distribution of collection term frequencies in a curated dataset of 650 abstracts, but those findings may not hold true for current IR datasets which are usually more noisy, orders of magnitude larger and often domain-specific. Indeed, recent findings [344] suggest that collection term frequency distributions of multiple TREC collections are better quantified using the Yule-Simon, power law or Generalized Extreme Value distributions. Furthermore, considering the scale and heterogeneity of available data, the distribution of non-informative terms may vary considerably across collections. Consequently, a DFR ranking model should adapt its scoring to the distribution of term frequencies of non-informative terms on a per collection basis, instead of assuming a single statistical model. By adapting to the "true" distribution, the resulting ranking model better conforms to the theory of DFR and could, therefore, lead to increased retrieval performance.

Motivated by the above, we ask the following research question:

1. **Research Question 2 (RQ2):** Can the use of the correct statistical model improve ranking?

To answer our research question, we propose a novel cascaded methodology which first identifies the set of non-informative terms in a collection. We then use our principled approach from Chapter 3 to determine the best-fitting statistical model to the collection frequencies of non-informative terms. We finally use the best-fitting statistical model to develop ranking models and evaluate them in an ad hoc retrieval task.

Throughout this chapter, the notation in Table 5.1 is used. Documents and queries are treated as bags of terms and, with no loss of generality, set-based definitions for $f_{t,d}$ and $f_{t,q}$ are used.

| Symbol | Interpretation | Definition |
|---|---|---|
| $C$ | A collection of documents | $C = \{d\}$ |
| $N$ | # documents in $C$ | $N = |C|$ |
| $d$ | A document in $C$ | $d \in C$ |
| $t$ | A term in $C$ or $d$ | |
| $|d|$ | Length of $d$ | $|d| = |\{t \mid t \in d\}|$ |
| $f_{t,d}$ | Frequency of $t$ in $d$ | $f_{t,d} = |\{t' \mid t' = t \wedge t' \in d\}|$ |
| $f_{t,C}$ | Frequency of $t$ in $C$ | $f_{t,C} = \sum_{d \in C} f_{t,d}$ |
| $f_{t,q}$ | Frequency of $t$ in query $q$ | $f_{t,q} = |\{t' \mid t' = t \wedge t' \in q\}|$ |
| $n_t$ | Number of documents in $C$ where $t$ occurs | $n_t = |\{d \mid d \in C \wedge t \in d\}|$ |

Table 5.1: Notation used.

## 5.2 Divergence From Randomness

DFR, introduced by Amati [17], is a framework for constructing ranking models derived in a theoretical manner as a combination of different probability distributions [19]. Ranking models derived from DFR have yielded retrieval effectiveness comparable to BM25 [19] and the language modelling approach [89, p. 295].

### 5.2.1 Basics

A DFR ranking model is the product of two information content functions:

$$inf_1 \cdot inf_2 = \left(-\log_2 P_1\right) \cdot (1 - P_2) \tag{5.1}$$

where $P_1$ is the probability of having, by pure chance, $f_{t,d}$ occurrences of term $t$ in document $d$ in some collection $C$ [19].[2] $P_2$ is the probability of occurrence of $t$ in a document with respect to its "elite set". A document is elite for a term if it is "about" or on-the-topic associated with the term. $(1 - P_2)$ is the risk of accepting $t$ as a term that describes the topic of the document, which decreases as $f_{t,d}$ increases.

To rank documents with respect to a query $q$, the DFR ranking formula is:

$$R(q,d) = \sum_{t \in q \cap d} \left(-\log_2 P_1\right) \cdot (1 - P_2) \tag{5.2}$$

The remainder of this section focusses on how to estimate $P_1$ (Section 5.2.2), how to estimate $P_2$ (Section 5.2.3) and how to normalise term frequencies by document length (Section 5.2.4).

### 5.2.2 Model of Randomness

A model of randomness quantifies how non-informative terms (such as, but not only, stop words) are distributed throughout a collection. A model of randomness is a function of $f_{t,d}$. If $P_1$ is low, then $f_{t,d}$ occurrences of $t$ in $d$ are unlikely to have occurred by chance according to the chosen model of randomness, and $-\log_2 P_1$ will be large to reflect the "informativeness" of term $t$ in $d$. In contrast, if $P_1$ is high, then (i) the $f_{t,d}$ occurrences of $t$ "conform" or are expected according to the chosen model of randomness, and (ii) $-\log_2 P_1$ is low, suggesting the term is less informative.

---

[2]We use $P_1$ and $P_2$ here to conform with the original DFR notation.

Amati and Rijsbergen [19] propose five basic models of randomness shown in Table 5.2. These are the Poisson (P), geometric (G), tf-idf ($I_n$), tf-itf ($I_F$) and a tf-expected-idf ($I_n^e$). Each of these models makes a different assumption about the distribution of terms in the collection.

| Name | Abbr. | Definition |
|---|---|---|
| Poisson | $P$ | $P_1 = f_{t,d} \cdot \log_2\left(\frac{f_{t,d}}{\lambda}\right) + \left(\lambda + \frac{1}{12 \cdot f_{t,d}} - f_{t,d}\right) \cdot \log_2 e + 0.5 \cdot \log_2(2\pi \cdot f_{t,d})$ |
| Geometric | $G$ | $P_1 = -\log_2\left(\frac{1}{1+\lambda}\right) - f_{t,d} \cdot \log_2\left(\frac{\lambda}{1+\lambda}\right)$ |
| tf-idf | $I_n$ | $P_1 = f_{t,d} \cdot \log_2\left(\frac{N+1}{n_t + 0.5}\right)$ |
| tf-itf | $I_F$ | $P_1 = f_{t,d} \cdot \log_2\left(\frac{N+1}{f_{t,C} + 0.5}\right) + \log_2\left(\frac{f_{t,C}}{N}\right)$ |
| tf-expected-idf | $I_n^e$ | $P_1 = f_{t,d} \cdot \log_2\left(\frac{N+1}{\left(N \cdot \left(1 - \left(\frac{N-1}{N}\right)^{f_{t,C}}\right) + 0.5\right)}\right)$ |

Table 5.2: The five basic models of randomness [19]. $\lambda$ is a free parameter set to either $\lambda = \left\{\frac{n_t}{N}, \frac{f_{t,C}}{N}\right\}$.

As an example of a model of randomness, consider randomly placing $k$ occurrences of $t$ in $N$ documents. The number of ways that $k$ can be distributed over $N$ is expressed using the binomial coefficient:

$$\binom{N+k-1}{k} = \frac{(N+k-1)!}{(N-1)!k!} \tag{5.3}$$

Assume now that a document $d$ with $f_{t,d}$ occurrences of $t$ is found. The remaining $k - f_{t,d}$ occurrences can be distributed over the remaining $N-1$ documents as:

$$\binom{(N-1)+\left(k-f_{t,d}\right)-1}{k-f_{t,d}} = \frac{\left((N-1)+(k-f_{t,d})-1\right)!}{(N-2)!\left(k-f_{t,d}\right)!} \tag{5.4}$$

and the probability that a random document contains $f_{t,d}$ occurrences of $t$ is then the ratio of Eqn. 5.3 to Eqn. 5.4:

$$P_1 = \frac{\left((N-1)+(k-f_{t,d})-1\right)!(N-1)!k!}{(N-2)!,\left(k-f_{t,d}\right)!(N+k-1)!} \tag{5.5}$$

Assuming $N \gg f_{t,d}$, Eqn. 5.5 reduces to a geometric distribution [19, p. 366]:

$$P_1 = \left(\frac{1}{1+k/N}\right)\left(\frac{k/N}{1+k/N}\right)^{f_{t,d}} \tag{5.6}$$

which is "parameterised" by $f_{t,d}$. Eqn. 5.6, then, returns the probability of observing $f_{t,d}$ occurrences of $t$ in $d$ under the assumptions that non-informative terms are distributed according to a geometric distribution. Eqn. 5.6 is therefore referred to as the geometric model of randomness.

### 5.2.3 First Normalisation Principle

The notion of "term eliteness" is used in the DFR framework to estimate $P_2$, the probability of occurrence of term $t$ in $d$ with respect to $t$'s elite set. The higher $f_{t,d}$ is, the more likely $t$ is expected to be informative for the document's topic(s) [19]. Thus, $P_2$ is the conditional probability that $t \in d$ if $d$ is in $t$'s elite set:

$$(1 - P_2) = (1 - \Pr(t \in d | d \in \text{Elite set of } t)) \tag{5.7}$$

The more frequently $t$ occurs in the elite set, the less its frequency is due to randomness which is estimated using either Laplace's law of succession or a ratio of Bernoulli trials [19]. Both normalisations "resize" the information gained by selecting $t$ as a good descriptor of a potentially relevant document. A higher risk means a higher gain, as $t$ occurs relatively infrequently in its elite set. In contrast, a low risk means a low gain as $t$ occurs relatively frequently in its elite set. Heuristic alternatives, such as Lidstone's law of succession [99, p. 151], could also be used.

#### 5.2.3.1 Laplace Normalisation

Let $Z$ be a binary random variable and $z = \{z_1, ..., z_n\}$ a sample of $n$ i.i.d. realisations of $Z$. Assuming $s$ observations are successes ($z_i = 1$) and $n - s$ observations are failures ($z_j = 0$), Laplace's law of succession gives the conditional probability that observation $n + 1$ is a success:

$$\Pr(Z_{n+1} = 1 | z_1, ..., z_n = s) = \frac{s+1}{n+2} \tag{5.8}$$

which increases with the number of successes. Eqn. 5.8 can be adopted to term frequencies [19] as the conditional probability of observing $f_{t,d} + 1$ occurrence of $t$ having observed $f_{t,d}$ as:

$$\Pr(t \in d | d \in \text{Elite set of } t) = \frac{f_{t,d}}{f_{t,d} + 1} \tag{5.9}$$

The assumption behind Eqn. 5.9 is that a high probability of seeing $f_{t,d} + 1$ occurrences is not accidental, but due to some underlying semantic cause, and that "the probability that the observed term contributes to select a relevant document is high, if the probability of encountering one more token of the same term in a relevant document is similarly high" [19, p. 369]. Using Laplace's law of succession, the resulting geometric DFR ranking model is obtained by substituting Eqn. 5.9 into Eqn. 5.2:

$$R(q,d) = \sum_{t \in q} \left( -\log_2 \left( \left( \frac{1}{1 + k/N} \right) \left( \frac{k/N}{1 + k/N} \right)^{f_{t,d}} \right) \right) \cdot \left( 1 - \frac{f_{t,d}}{f_{t,d} + 1} \right) \tag{5.10}$$

where the geometric model of randomness (Eqn. 5.6) has been substituted for $P_1$.

#### 5.2.3.2 Bernoulli Process Normalisation

A Bernoulli process is a finite sequence of i.i.d. Bernoulli trials (a binary random variable). Estimating $\Pr(t \in d | d \in \text{Elite set of } t)$ using a Bernoulli process is done using the following urn model [19]:

$$\alpha = \frac{B(n_t, f_{t,C}, f_{t,d}) - B(n_t, f_{t,C} + 1, f_{t,d} + 1)}{B(n_t, f_{t,C}, f_{t,d})} = 1 - \frac{B(n_t, f_{t,C} + 1, f_{t,d} + 1)}{B(n_t, f_{t,C}, f_{t,d})} \tag{5.11}$$

where $B(n_t, f_{t,C}+1, f_{t,d}+1)$ is the probability of obtaining one more occurrence of $t$ in $d$ out of all $n_t$ documents using a Bernoulli process. If the ratio:

$$\frac{B(n_t, f_{t,C}+1, f_{t,d}+1)}{B(n_t, f_{t,C}, f_{t,d})} \tag{5.12}$$

is smaller than 1, the probability of document $d$ receiving, at random, the new occurrence of $t$ increases. In other words, the higher $f_{t,d}$, the more likely it is to observe one more token of $t$. Eqn. 5.12 can be rewritten as [19]:

$$\Pr(t \in d | d \in \text{Elite set of } t) = \frac{f_{t,C}+1}{n_t(f_{t,d}+1)} \tag{5.13}$$

where the intuition is that a high probability of seeing $f_{t,d}+1$ occurrences given $f_{t,d}$ is not accidental but due to some underlying semantic cause [19]. Using the ratio of Bernoulli trials, the geometric DFR ranking model is obtained by substituting Eqn. 5.13 into Eqn. 5.2:

$$R(q,d) = \sum_{t \in q} \left( -\log_2 \left( \left( \frac{1}{1+k/N} \right) \left( \frac{k/N}{1+k/N} \right)^{f_{t,d}} \right) \right) \cdot \left( 1 - \frac{f_{t,C}+1}{n_t(f_{t,d}+1)} \right) \tag{5.14}$$

where the geometric model of randomness (Eqn. 5.6) have been substituted for $P_1$.

### 5.2.4 Second Normalisation Principle

The first normalisation principle assumes that documents are of equal length [89, p.301], which would bias a DFR ranking model towards longer documents where $f_{t,d}$ is expected to be higher. The second normalisation principle defines a probability density function $\Pr(|d|)$ and estimates the density assuming either that (i) $\Pr(|d|)$ of a term is uniform, or (ii) $\Pr(|d|)$ is a decreasing function of $|d|$ [19].

The *uniform* term frequency normalisation assumes that the distribution of $t$ in $d$ is uniform:

$$\hat{f}_{t,d} = f_{t,d} \cdot \frac{\text{avg\_l}}{|d|} \tag{5.15}$$

where avg_l is the average document length in the collection. Notice that Eqn. 5.15 will "reward" terms occurring in short documents (relative to the average document length).

The *logarithmic* term frequency normalisation:

$$\hat{f}_{t,d} = f_{t,d} \cdot \log_2 \left( 1 + c \cdot \frac{\text{avg\_l}}{|d|} \right) \tag{5.16}$$

where $c \in \mathbb{R}^+$ is a free parameter, is derived under the assumption that the density of $f_{t,d}$ in a document is a decreasing function of $|d|$. The logarithmic normalisation ensures that the difference in $\hat{f}_{t,d}$ for a term with the same number of tokens in $d_1, d_2$, with $|d_2| \ll |d_1|$ documents, is substantially smaller than using Eqn. 5.15. Eqn. 5.15 or 5.16 is substituted for $f_{t,d}$ in Eqns. 5.10, 5.14 and all other DFR ranking models.

### 5.2.5 Summary

A DFR ranking model is the product of two information content functions $inf_1 \cdot inf_2 = -\log_2(P_1) \cdot (1-P_2)$. $P_1$ is the probability of having, by chance, $f_{t,d}$ occurrences of term $t$ in document $d$, and $P_2$ resizes the informative content of $t$ using either Laplace's law of succession (Eqn. 5.9) or a ratio of Bernoulli trials (Eqn. 5.13). To ensure that DFR models are not biassed towards longer documents where term tokens are expected to be higher, normalised term frequencies (Section 5.2.4) $\hat{f}_{t,d}$ substitute for all $f_{t,d}$.

## 5.3 Related Work

A number of extensions to the original DFR models of Amati and Rijsbergen [19] have been proposed in the literature. We review previous work concerning DFR here as it relates specifically to this chapter, and the concepts introduced are referred to later in the chapter.

### 5.3.1 Beta Negative Binomial Model

Clinchant and Gaussier [119] propose to use the discrete Beta Negative Binomial (BNB) model for text modelling. The BNB distribution is proposed on the basis of (i) integrating text burstiness (once a term appears in a text it is much more likely to appear again), and (ii) previous findings showing the "good behaviour" of the negative binomial (NB) distribution for text modelling [10, 109, 363]. The negative binomial is a discrete statistical model with PMF:

$$\begin{aligned}
\text{Nbin}(x|\theta) &= \left\{ f\big(x|r,\beta\big) \right\} \\
&= \left\{ \frac{\Gamma(r+x)}{x!\Gamma(r)} \big(1-\beta\big)^r \beta^x : r>0, \beta \in (0;1) \right\}
\end{aligned} \tag{5.17}$$

where $\Gamma$ is the gamma function, ! is the factorial function and $x \in \mathbb{Z}$. Because the negative binomial is an infinite mixture of Poisson distributions, it can be considered a generalisation of the 2-Poisson [197, 198] and $N$-Poisson [292]. By imposing a prior Beta distribution over $\beta$, the resulting distribution becomes:

$$\begin{aligned}
\text{BNBGen}(x|\theta) &= \left\{ f(x|r,a,b) \right\} \\
&= \left\{ \frac{\Gamma(x+r)\Gamma(a+x)}{x!\Gamma(r)\Gamma(a)\Gamma(b)} \cdot \frac{\Gamma(a+b)\Gamma(r+b)}{\Gamma(a+b+r+x)} : r>0, a>0, b>0 \right\}
\end{aligned} \tag{5.18}$$

where $a,b$ are the Beta distribution's parameters. Assuming the Beta prior is uniform ($a=b=1$) and substituting $x$ with the normalised within-document term frequency $f_{t,d}$, the BNB model is defined as:

$$\begin{aligned}
\text{BNB}\big(\hat{f}_{t,d}|\theta\big) &= \left\{ f\big(\hat{f}_{t,d}|r_t\big) \right\} \\
&= \left\{ \frac{r_t}{(r_t+\hat{f}_{t,d}+1)(r_t+\hat{f}_{t,d})} : r_t>0 \right\}
\end{aligned} \tag{5.19}$$

where $r_t$ is a free parameter. The probability of a term under the BNB model is [119]:

$$\text{Pr}_{\text{BNB}}\big(\hat{f}_{t,d} \geq 1|r_t\big) = \frac{r_t}{r_t+1} \tag{5.20}$$

which, when $r_t = \hat{f}_{t,d}$, equals Laplace's law of succession. Referring to Eqn. 5.2, $P_1$ is set to either $P_{\text{BNB}}\big(\hat{f}_{t,d} \geq 1 | r_t\big)^{\hat{f}_{t,d}}$ or $P_{\text{BNB}}\big(\hat{f}_{t,d} | r_t\big)$ where the subscript refers to the statistical model used. In the former case, $r_t$ is either set to $f_{t,C}/N$ or its maximum likelihood estimate (MLE). In the latter case, $r_t$ is set to its MLE estimate or an estimate that takes into account the effect of the document length on the number of occurrences of a term. For $P_2$, another BNB models is used where $r_t$ is set to either $\hat{f}_{t,d}$ or its MLE value. Five versions of the BNB ranking model are considered, including two versions which do *not* renormalise the informative content of terms i.e. do not use Laplace's law of succession or the ratio of Bernoulli trials to resize the informative gain of a term. These ranking models take the form:

$$R(q,d) = \sum_{t \in q \cap d} -\log\big(\Pr_{\text{BNB}}\big(\hat{f}_{t,d} \geq 1 | r_t\big)\big) \tag{5.21}$$

All versions are evaluated on the English part of the CLEF-2003 corpus containing $\approx 160{,}000$ documents, and the associated 60 queries using both short (title) and long (title and description) queries using Laplace normalisation and logarithmic term frequency normalisation.

Using short queries, all BNB ranking models with Laplace normalisation, perform on par with mean average precision (MAP), R-precision (R-PREC) and precision at $k$ (P@$k$ for $k = 5, 10$) compared to a geometric and $I_F$ DFR baseline. For long queries, the observations are the same, though performance is higher compared to using the short queries. The results for the BNB models without Laplace normalisation are on par with both baselines using short queries, but perform substantially worse for long queries on MAP and R-PREC. On this basis, the authors suggest that effective DFR models can be derived using only a single distribution and without the first normalisation principle.

### 5.3.2 Log-logistic Model

Clinchant and Gaussier [120] set out to simplify the DFR framework by formulating four heuristic retrieval constraints similar to Fang et al. [158]:

$\mathbf{C}_1$: $\forall (|d|, z_t, \theta), \dfrac{\partial h(\hat{f}_{t,d}, |d|, z_t, \theta)}{\partial \hat{f}_{t,d}} > 0$ $\qquad$ $\mathbf{C}_2$: $\forall (|d|, z_t, \theta), \dfrac{\partial^2 h(\hat{f}_{t,d}, |d|, z_t, \theta)}{\partial \hat{f}_{t,d}^2} < 0$

$\mathbf{C}_3$: $\forall (\hat{f}_{t,d}, z_t, \theta), \dfrac{\partial h(\hat{f}_{t,d}, |d|, z_t, \theta)}{\partial |d|} < 0$ $\qquad$ $\mathbf{C}_4$: $\forall (\hat{f}_{t,d}, |d|, \theta), \dfrac{\partial^2 h(\hat{f}_{t,d}, |d|, z_t, \theta)}{\partial z_t} < 0$

Table 5.3: Heuristic retrieval constraints [120].

where $z_t = \left\{ \dfrac{f_{t,C}}{N}, \dfrac{n_t}{N} \right\}$, $\theta$ is a set of parameters and $h$ is the generic ranking function:

$$R(q,d) = \sum_{t \in q \cap d} h\big(\hat{f}_{t,d}, |d|, z_t, \theta\big) \tag{5.22}$$

The formulated constraints collectively state that $h$ should be an increasing and concave function of $\hat{f}_{t,d}$ and decrease with $|d|$ and the document and collection frequency.

Under these constraints, the authors show that their BNB ranking model [119] is not a valid IR ranking model. In addition, the BNB ranking model should only be used with unnormalised within-document term frequencies as the BNB distribution is discrete. As Clinchant and Gaussier point out, this is also a limitation of the Poisson and Geometric DFR model (see Table 5.2) which they call theoretically flawed [121]. To facilitate the use of normalised term frequencies, Clinchant and Gaussier propose the log-logistic (LL) distribution, which can be seen as a continuous approximation to the BNB distribution:

$$\Pr_{\text{LL}}\big(X < x | \alpha, \beta\big) = \frac{x^{\beta}}{x^{\beta} + \alpha^{\beta}} \tag{5.23}$$

where $\Pr_{\text{LL}}(\cdot)$ is the CDF of the log-logistic model and $\alpha, \beta$ are its parameters. Setting $r = \alpha$ and $\beta = 1$, the resulting log-logistic distribution becomes:

$$\forall x \in \mathbb{R}^{+}, P_{\text{LL}}(x \le X < x + 1 | r) = \frac{x+1}{r+x+1} - \frac{x}{r-x} = \frac{r}{(r+x+1)(r+x)} \tag{5.24}$$

A simplified DFR ranking model, called the LL ranking model, based solely on $inf_{1}$ and the log-logistic distribution is then defined as:

$$R(q,d) = \sum_{t \in q \cap d} -\log\big(\Pr_{\text{LL}}\big(X \ge \hat{f}_{t,d} | r\big)\big) \tag{5.25}$$

where $r = \left\{ \frac{f_{t,C}}{N}, \frac{n_{t}}{N} \right\}$. Like the BNB ranking model, the LL ranking model does not renormalise using $P_{2}$ [119], but is shown to verify all heuristic retrieval constraints in Table 5.3.

The LL ranking model is evaluated on TREC-3 (queries 151-200), ROBUST 2004 (queries 301 - 450, 600 - 700), CLEF 2003 (60 queries in total), CLEF domain-specific 2004-2006 (75 queries in total) and CLEF 2008 ad hoc (50 queries). The baselines are the BNB and the standard $P$ and $I_{n}$ DFR models. All ranking models use Laplace's law of succession and logarithmic term normalisation (with $c = 0.5, 1, 5, 10$), and performance is measured using MAP and P@10. The results for all datasets show that using $r = \frac{n_{t}}{N}$ gives the best performance for the LL model. For MAP, the LL ranking model mainly improves performance for low values of $c$ but tends to do so consistently. For higher values of $c$, the two standard DFR models tend to perform better. For P@10, the standard DFR models tend to achieve the best performance regardless of the value of $c$. The authors also show that a Dirichlet-smoothed language model performs equal to or lower than the standard DFR ranking models and both $r = \left\{ \frac{f_{t,C}}{N}, \frac{n_{t}}{N} \right\}$ versions of the LL ranking model.

### 5.3.3  Information Models

Based on their previous work [119, 120], Clinchant and Gaussier [121] introduce a family of information-based retrieval models for ad hoc IR. By information, the authors refer to Shannon's information when observing a stochastic event. The intuition behind these ranking models is to use statistical models that can quantify text "burstiness" and also adhere to their heuristic retrieval constraints (see Table 5.3). According to [121], a statistical model $h$ is bursty for all $\epsilon > 0$, if:

$$h_{\epsilon}(x) = P(X \ge x + \epsilon | X \ge x) \tag{5.26}$$

is a strictly increasing function of $x$. The notion of burstiness is similar to the first normalisation principle, and, informally, means it is easier to generate or observe higher values of $f_{t,d}$ once lower values have been observed. Finally, all information-based retrieval models are of the form given by Eqn. 5.21. From the family of information-based models, two power law ranking models are instantiated. The first information model is the LL ranking model [120] (see Eqn. 5.25) though it is not explained why the LL is considered a power law. The second information model is coined a "smoothed power law" (SPL), defined as:

$$R(q,d) = \sum_{t \in q \cap d} -\log\left(\frac{\lambda_t^{\frac{\hat{f}_{t,d}}{\hat{f}_{t,d}+1}} - \lambda_t}{1 - \lambda_t}\right) \tag{5.27}$$

where $\lambda_t$ is a free parameter. Notice that the SPL ranking model integrates the first normalisation principle, as the exponent $\frac{\hat{f}_{t,d}}{\hat{f}_{t,d}+1}$ is identical to Laplace's law of succession (see Eqn. 5.9). Essentially, both the LL and SPL models assign higher weights to terms with lower frequencies than the DFR models. Practically, this means that as soon as one token of a term appears, the probability of observing one more token increases more when using an information ranking model than a standard DFR ranking model.

Both the LL and SPL information models are evaluated on the TREC-3, ROBUST 2004, CLEF 2003 and CLEF domain specific 2004 - 2006 with associated queries and qrels, using logarithmic term normalisation and setting $c = \{0.5, 0.75, 1, 2, 3, 4, 5, 6, 7, 8, 9\}$. Performance is measured using MAP and P@10. Compared to tuned language models with both Dirichlet and Jelinek-Mercer smoothing, the LL and SPL models are almost consistently the best-performing regardless of using short (title) or long (description) queries. Both information models also perform on par or better compared to a tuned BM25 baseline, though mostly for short queries. Finally, the information models yield results similar to the $P$ and $I_n$ DFR ranking models.

### 5.3.4 Multinomial Model

DFR ranking models have also been used for fielded retrieval i.e. where the structure of documents can be integrated into the scoring, by weighting different parts of a document differently. Plachouras and Ounis [346] propose to use a multinomial DFR ranking model for Web retrieval which combines evidence from multiple fields to produce a final ranking of documents. Suppose document $d$ has $k$ fields. The multinomial statistical model is then given by:

$$\begin{aligned}
\mathrm{MN}(x|\theta) &= \left\{f(t|f_{t,C}, p)\right\} \\
&= \left\{\binom{f_{t,C}}{tf_1 \cdots tf_k \cdot tf'} p_1^{tf_1} \cdots p_k^{tf_k} \cdot p'^{tf'} : 0 < p_i \leq 1, \sum_i p_i = 1\right\}
\end{aligned} \tag{5.28}$$

where $tf_i : i = 1, \ldots, k$ is the frequency of $t$ in field $i$, $tf' = f_{t,C} - \sum_{i=1}^{k} tf_i$ (frequency of $t$ in documents other than $d$), $p_i = \frac{1}{k \cdot N}$ is the prior probability that $t$ occurs in field $i$ of $d$ and $p' = \frac{N-1}{N}$ is the prior probability that $t$ does not appear in any field of $d$. Eqn. 5.28 denotes the probability of a term occurring $tf_i$ times in the $i^{\text{th}}$ field of $d$. The intuition behind the use of the multinomial model is that term frequencies are no longer assumed to be drawn from the same distribution, thus capturing the observation that term occurrences in different document fields have different levels of importance.

The multinomial DFR ranking model formulated using Eqn. 5.28 with Laplace and logarithmic term normalisation is given by:

$$R(q,d) = \sum_{t \in q \cap d} f_{t,q} \cdot \left(-\log_2\left(\Pr_{\text{MN}}\left(t|f_{t,C},p\right)\right)\right) \cdot (1 - P_2) \tag{5.29}$$

$$= \sum_{t \in q \cap d} \left(-\log_2(f_{t,C}!) + \sum_{i=1}^{k} w_i + y\right) \cdot \frac{f_{t,q}}{\sum_{i=1}^{k} tfn_i} \tag{5.30}$$

where $\Pr_{\text{MN}}\left(t|f_{t,C},p\right)$ refers to Eqn. 5.28, $tfn_i = tf_i \cdot \log_2\left(1 + c_i \cdot \frac{l_i^{avg}}{l_i}\right)$ is the normalised frequency of $t$ in field $i$, $c_i$ is a tunable hyperparameter for field $i$, $l_i^{avg}$ is the average length of the $i$th field in $C$, $w_i = \log_2(tfn_i!) - tfn_i \cdot \log_2(p_i)$ and $y = \log_2(tfn'!) - tfn'\log_2(p')$. The authors evaluate the multinomial DFR ranking model in Eqn. 5.30 (and a limiting form based on "an information theoretic approximation") on the .GOV collection using the body, anchor text and title document fields. All fields are given a uniform prior weight and only the $c_i$ parameter of each field is optimised for w.r.t. MAP using the training tasks for topic distillation, named page finding and home page finding from the TREC 2003 Web Track. The tuned multinomial DFR ranking model is then evaluated on the corresponding TREC 2004 Web Track tasks of topic distillation, named page finding and home page finding. The results show that both multinomial DFR ranking models perform on par with a tuned PL2F ranking model [282] (a Poisson DFR ranking model adapted for fielded retrieval) for all three tasks using MAP and MRR.

### 5.3.5   NG Models

The work closest to ours is by Hui et al. [212]. In their work, new DFR ranking models, called NG models, are derived by replacing $P_1$ with either the Poisson, gamma, exponential, Weibull, Rayleigh or $\chi^2$ distribution. Each statistical model is fitted to the distribution of $f_{t,C}$ of several datasets and their fit is evaluated using Pearson's correlation coefficient. The optimal parameters of each statistical model are determined using gradient search on half of the queries of each collection. Furthermore, their NG models do not rely on collection-wide statistics such as document frequency and term frequencies. The NG models are evaluated on TREC disks 1 & 2, TREC disks 4 & 5, WT10g and .GOV2 with associated queries and relevance assessments, against an untuned Dirichlet-smoothed query language model, BM25 and the standard Poisson DFR ranking model with Laplace normalisation. The results show that all baselines perform statistically significantly better than any of the proposed models. The authors also propose simplified versions of their NG models. These are defined as:

$$R(q,d) = \sum_{t \in q \cap d} f_{t,q} \cdot \log_2(1 - P_1) \tag{5.31}$$

which, identically to [121], do not make use of the first normalisation principle. Evaluating these simplified NG models shows statistically significant improvement over the standard NG models, but these models do not outperform any of the baselines for any dataset.

### 5.3.6   Criticism of Reviewed Work

An important caveat in the reviewed work is the absence of empirical evidence or testing (or reporting of either) that supports the use of the selected statistical models to quantify the distribution of $f_{t,C}$. Clinchant and Gaussier [119, 120] base their use of the beta negative binomial and log-logistic distribution on previous findings [10, 109], but these findings (i) result from using small, curated and domain-specific datasets (e.g. the million term Brown corpus and an IMDB dataset[3]) and (ii) consider (or report) only the Poisson, binomial or negative binomial distribution as candidates for the distribution of term frequencies. Similarly, the use of the SPL distribution [121] is not supported by any empirical evidence, though recent findings [344] suggest that this distribution is useful for quantifying the distribution of $f_{t,C}$ in some collections.

Similar issues exist in the work by Hui et al [212]. Firstly, it is not clear why the authors chose to use only the Poisson, gamma, exponential, Weibull, Rayleigh and $\chi^2$ model when off-the-shelf commercial (such as MATLAB) and open-source (such as R) statistical software includes many more. Personal communication with the first author did not clarify this choice further. Secondly, all aforementioned statistical models are used for experimentation instead of the single best one. While this may be feasible for small collections, the authors give no indication that their approach is scalable to large-scale datasets. Thirdly, the use of Pearson's correlation coefficient to assess each distribution's fit to the distribution of $f_{t,C}$ is sub-optimal as (i) it assumes that both variables $x$ and $y$ are normally distributed and (ii) the correlation coefficient is significantly affected by extreme values [389, p. 455] which are not uncommon for term frequencies in TREC collections [344]. Both of these assumptions are likely incorrect based on recent findings [344]. Fourthly, their NG models intentionally disregard collection-wide statistics such as a term's collection frequency on the basis that these are difficult to accurately maintain for large collections. By doing so, their NG models assume that the most frequent terms in a document reflect its main topic(s). Thus, their models do not discriminate between such terms and those frequent in most documents which do not reflect the document's topic. This may also explain why their simplified NG models achieve higher performance than the standard NG models. Recall that $-\log_2(P_1)$ in DFR, is meant to quantify the probability of observing $k$ tokens of a term $t$ in some document $d$ relative to some distribution of "randomness". A low probability suggests that $k$ tokens of $t$ are highly unlikely and thus not random. The $\log_2(1 - P_1)$ reverses this idea: a low probability suggests that $k$ tokens of $t$ are highly likely and thus random. This, essentially, rewards terms which are frequent in *all* documents. Hence, the performance improvements for the simplified NG models may be caused by clusters of non-informative terms in some relevant documents.

That most current statistical IR models do not seem to be based on strong empirical evidence is the weakness addressed in this chapter. We propose to rectify this using the novel cascading methodology presented in the next section.

## 5.4 Cascading Methodology

The goal of this chapter is to derive ranking models that adapt to the distribution of the data being processed. We refer to such a ranking model as an *adaptive distributional ranking* (ADR) model. An ADR model explicitly uses the statistical model that best quantifies the distribution of the data and is derived using the following novel cascading methodology:

    **Step 1:** Identify the set $T$ of non-informative terms of $C$ and extract their collection term

---

[3]The Internet Movie DataBase is a large collection of movie information with information on actors, filming locations, budget etc.

frequencies $f_{t,C}$.

**Step 2:** If $T$ is large, subsample $T$.

**Step 3:** Select a set of candidate statistical models that may quantify the distribution of $\forall t \in T : f_{t,C}$ .

**Step 4:** Determine, for each candidate statistical model, the values of its parameters (if any) that make the distribution fit the distribution of $f_{t,C}$.

**Step 5:** Select, among all the fitted statistical models, the one that best quantifies the distribution of $f_{t,C}$

**Step 6:** Integrate the best-fitting statistical model into the DFR framework.

We next briefly describe the motivation underpinning each step:

Step 1:   Because DFR postulates that non-informative terms are distributed randomly across a collection and that a term's divergence from the distribution of these non-informative terms is a measure of its informativeness, Step 1 is concerned with identifying the non-informative terms.

Step 2:   Because the original set of non-informative terms may be too large to handle with available resources, Step 2 examines how to extract a representative sample from the original set of non-informative terms which will act as a replacement.

Step 3:   Identify a set of parametric statistical models that may reasonably well quantify the distribution of all collection term frequencies, based, in part, on statistical distributions used elsewhere in IR research. The set comprises 16 different parametric statistical models widely available in commercial and open-source software and is the same as used in Chapter 3 (see Section 2.3 for a full list of statistical models used).

Step 4:   To determine which parametric statistical model to integrate into our ADR model, Step 4 fits each parametric statistical model to the distribution of collection term frequencies using maximum likelihood estimation.

Step 5:   This step uses Vuong's likelihood ratio test to determine which of the fitted models provide the best fit.

Step 6:   Lastly, Step 6 integrates the best-fitting statistical model into DFR.

The output of Step 6 is what will be referred to as an ADR model. All of the above steps (apart from Step 1) are performed automatically. Steps 3 through 5 are addressed using our principled approach from Chapter 3. Thus, the remainder of this section focusses on Steps 1,2 and 6. We emphasize that while this approach may be used for any framework of ranking models that relies on the use of a statistical model, we focus here on the DFR framework as it facilitates a straight-forward application of our proposal.

### 5.4.1   Identifying Non-informative Terms

DFR postulates that non-informative terms are distributed randomly across a collection $C$, and that a term's divergence from the distribution of these non-informative terms (according to some "model of randomness" - see Section 5.2.2) is a measure of its informativeness [19]. An important first step is thus to identify the non-informative terms. Previous work on DFR ranking models, however, has disregarded this step and instead used all terms regardless of their informativeness. Because the distribution of non-informative terms in $C$ may be different from the distribution of all terms in $C$, the chosen "model of randomness" may not be suitable, thus risking to significantly reduce the performance of the DFR ranking model.

Estimating term informativeness is fundamental to many NLP tasks [449] and is widely used to discriminate between informative and non-informative terms [360]. Such *term weighting* approaches can be categorised as either unsupervised or supervised [259, 447]. Traditionally, term weighting methods from IR belong to unsupervised term weighting methods as the calculation of these weighting methods do not involve the information on any membership of documents to specific categories. Examples of unsupervised term weighting methods from IR include TF-IDF [374], IDF [226], $x^I$ [73], lnu.ltc [82], residual-IDF (RIDF) [110], variance and burstiness [110], gain [339], BM25 [369], statistical dispersion [305] and mixture models [360]. In contrast, supervised term weighting methods make use of prior information about document membership in predefined categories to produce a weighting of each term. Examples of supervised term weighting methods include odds ratio [317], information gain [140], relevance frequency [448] and, more recently, evolutionary computation techniques [136]. One disadvantage of using supervised term weighting approaches for discriminating non-informative from informative terms is that supervised term weighting approaches typically require terms labelled as both non-informative and informative. However, given two lists - one containing informative terms and one containing non-informative terms - a supervised learning approach could learn to tell non-informative from informative terms. To facilitate this learning, each term should be represented as a feature vector, which are used as input to train the actual supervised learning approach. One way to obtain these features is by using *unsupervised* term weighting approaches, which output, for a given term, a score without requiring any training. This score can then be used as a feature in the feature vector for that term. Using $k$ term weighting approaches would then, for each term, produce a feature vector of size $k$. We will detail the supervised learning approach in Section 5.5.2. The unsupervised term weighting methods we consider for the purpose of producing a feature vector for each term are the inverse document frequency (IDF), $x^I$, lnu.ltc, residual-IDF (RIDF), variance and burstiness, gain and mixture models. We focus on these as they (i) appear to be widely used, and (ii) have been found to be effective in finding keywords in large collections [449].

The IDF measure is defined as:

$$\text{IDF}(t) = -\log\left(\frac{n_t}{N}\right) \tag{5.32}$$

A high IDF signals a rare term that is useful to discriminate between documents. The IDF has long been used to weight terms in IR and has also been used in text classification [360]. Papineni [339] argues that IDF is not synonymous with term informativeness and proposes the gain measure:

$$g(t) = \frac{n_t}{N}\left(\frac{n_t}{N} - 1 - \log\frac{n_t}{N}\right) \tag{5.33}$$

where very frequent and infrequent terms will have low gain. Papineni shows that the gain measure favours medium-frequency terms.

Bookstein and Swanson [73] propose $x^I$:

$$x^I(t) = f_{t,C} - n_t \tag{5.34}$$

which measures the "clusteredness" of the tokens of $t$ relative to their "elite" set. A high positive value of $x^I$ indicates that terms are clustered in a few documents. This measure, however, is biassed towards frequent, and often less informative, terms and generally does a poor job of finding informative terms [360]. Another early approach by Buckley et al. [82] is "lnu.ltc" where terms in documents (lnu) are weighted by a logarithmically smoothed term frequency and a pivoted length normalisation. Terms in queries (ltc) are computed by a logarithmically smoothed query term frequency in combination with idf and cosine normalisation. Church and Gale [110] introduce the burstiness score:

$$burstiness(t) = \frac{f_{t,C}}{n_t} \tag{5.35}$$

which also measures the relative clusteredness of $t$. Church and Gale also introduce the RIDF measure:

$$\text{RIDF}(t) = \text{IDF}(t) - \widehat{\text{IDF}}(t) \tag{5.36}$$

where $\widehat{\text{IDF}} = -\log\left(1 - e^{-F_t/N}\right)$ is the expected IDF of $t$. The RIDF is based on the assumption that nearly all terms have IDF scores larger than what one would expect according to an independence-based statistical model (such as the Poisson). Church and Gale also note that informative terms tend to have the largest deviations from what would be expected. Similar to gain, the RIDF measure favours medium-frequency terms [110, 339]. A weakness of both RIDF and gain is that they rely solely on document frequency and do not consider the heavy-tailed frequency distribution of informative terms [360]. Harter suggested that informative terms may be identified by their fit to a mixture of two Poisson distributions, and defined the $z$-measure (previously introduced by [80]) as:

$$z = \frac{\mu_1 - \mu_2}{\sqrt{\sigma_1^2 + \sigma_2^2}} = \frac{\lambda_1 - \lambda_2}{\sqrt{\lambda_1 + \lambda_2}} \tag{5.37}$$

where $\lambda_1, \lambda_2$ are the means (and variances) of the Poisson distributions. Eqn. 5.37 was reportedly useful for identifying informative terms [198], but has been found to be a poor discriminator of term informativeness in contemporary collections [360].

Orăsan [334] argues that term weighting approaches should exploit the information provided by a small set of user specified documents, and propose the relative document frequency (RelDF) measure:

$$RelDF(t) = \frac{r}{R} - \frac{n_t}{N} \tag{5.38}$$

where $r$ is the number of user-specified documents containing $t$, $R$ is the total number of user-specified documents. The intuition behind Eqn. 5.38, is that the first term favours terms that exhaustively describe the user-specified documents and, by extension, the topic of interest. The second term biasses the calculation towards terms that are discriminative in $C$.

### 5.4.2   Subsampling Non-informative Terms

Because large document collections contain a very large number of of terms, it may not be possible to use all non-informative terms as advocated in our cascading methodology. In this scenario, a sample of non-informative terms can be extracted and used instead provided the sample obtains the *same* results - the same best-fitting statistical model - as if *all* non-informative terms are used. The smaller the size of the sample required to meet this requirement, the better our cascading methodology can scale to large collections.

For our sample to accurately reflect the characteristics of the full dataset, it should be representative. Let $T$ denote the list of all non-informative terms of some collection $C$. A representative sample is then a subset $z \subseteq T$ which provides an unbiassed indication of the members of $T$. To create a representative sample, each member of the population should have an equal probability of being selected. Consequently, *probability sampling* reflects the characteristics of the population better than other types of sampling, because sample bias (an error in the selection process that systematically favours some outcomes over others) is avoided [372, Chap. 9]. A probability sampling method uses some form of randomisation to select members of the sample and is therefore also known as random sampling. Three common random sampling methods are (i) simple random sampling, (ii) stratified random sampling and (iii) systematic sampling [372, Chap. 9]. In the following, assume a random sample of size $k$ is to be extracted from $T$. Then, using:

1. **Simple random sampling:** Each member $t \in T : w = |T|$ is assigned a unique ID from $v = \{1,...,w\}$. $v$ is then permuted according to some random process or generator creating $\hat{v}$. Finally, the $k$ terms from $T$ having an ID that matches one of the first $k$ entries in $\hat{v}$ are extracted.

2. **Stratified random sampling:** $T$ is divided into $m$ non-overlapping homogeneous subgroups $T_1 \subset T,...,T_m \subset T$ such that $T_1 + ... + T_m = T$. Within each subgroup, extract a simple random sampling of size $k/m$. The homogeneous subgroups may be divided based on e.g. the range of the frequencies of non-informative terms in $T$ or some other aspect.

3. **Systematic sampling:** Every $k^{\text{th}}$ entry in $T$ is sampled. The only two restrictions are that (i) the first member (i.e. after which every $k$ member is sampled) is chosen at random, and (ii) $T$ must be randomised to avoid periodicity: a cyclic pattern that coincides with the sampling interval (every $k^{\text{th}}$ member) producing a potentially biassed sample.

While all the above methods can produce representative samples, stratified random sampling can achieve a higher degree of representativeness [372, Chap. 9]. Simple random sampling, however, appears to be the most widely used approach in practice.

To generate a "random" permutation of $T$ or $v$, the Fisher–Yates shuffling method [151, 167] seems a popular choice. Given a list of $n$ elements, the Fisher–Yates swaps two elements (one selected at random) $n$ times. A single random number, such as the index of the first member in systematic sampling or the random element in Fisher-Yates shuffling, can be generated using e.g. the Mersenne twister [298].

To show that sampling of collection term frequencies of non-informative terms is a viable strategy for our cascading approach, simple random sampling was used to extract samples of size 5%,10%,...,95% of collection term frequencies of non-informative terms from ClueWeb09 cat. B (the exact method used to determine non-informative terms is given in Section 5.5.2). The best-fitting distribution of each sample is then determined [344] and compared to the best-fitting distribution found when using the collection term frequencies for all non-informative terms. This was repeated $n = 5$ times and sample size averaged. The results show that at minimum 10% of all collection term frequencies for non-informative terms is required to accurately determine the best-fitting distribution. Repeating the experiment for ClueWeb12 cat. B. shows that also here a minimum of 10% of all term frequencies is required. Practically, each iteration (depending on the sample size) took between 8 seconds and 60 seconds.

Taken together, the observations suggest that our cascading methodology can be applied to large datasets with negligible computational overhead required to fit and compare all statistical models. These results, however, are dependent on the method used to identify the non-informative terms, as other methods might create much larger (or smaller) sets of non-informative terms which could substantially inflate the results.

### 5.4.3  Integrating Best-Fitting Distribution into Ranking

The final step of our cascading methodology derives new ADR models. At this step, the collection term frequencies of non-informative terms have been extracted and potentially subsampled, and the best-fitting statistical model to the distribution of these collection term frequencies determined. Let $M$ denote this best-fitting statistical model. An ADR model is then obtained by substituting $M$ for $P_1$ in Eqn. 5.2. The resulting ADR model is given by:

$$R(q,d) = \sum_{t \in q \cap d} \left( -\log_2 M \right) \cdot (1 - P_2) \tag{5.39}$$

An ADR model is very similar to a DFR ranking model, but we see it as having several desirable properties compared to a DFR ranking model:

- An ADR model captures the same assumptions as a DFR ranking model. The only difference is that $P_1$ is substituted for the best-fitting statistical model.

- An ADR model is formulated on the basis of a principled approach to statistical model selection, rather than assumptions regarding said distributions. The practical effect is that it "adapts" to the distribution of collection term frequencies of non-informative terms on a per dataset basis.

- An ADR model conforms more closely to the theory of DFR since the best-fitting statistical model is determined using only non-informative terms, rather than all terms.

Eqn. 5.39 is the generic ADR model from which specific instantiations for individual collections are derived based on the best-fitting statistical model.

## 5.5    Deriving ADR Models

This section derives ADR models for several TREC datasets. Section 5.5.1 describes the datasets and queries used.  Section 5.5.2 presents how non-informative terms are identified and extracted, and Section 5.5.3 presents the results from applying our principled approach to statistical model selection from Chapter 4 to determine the best-fitting statistical model to the collection term frequencies of these non-informative terms. Finally, Section 5.5.4 presents the derived ADR models.

### 5.5.1    Datasets

All datasets used are shown in Table 5.4. ClueWeb09 cat. B. consists of approximately 50 million Web sites in English and is representative of a large, heterogeneous and noisy collections. For ClueWeb09 cat. B.  we use the queries from TREC WebTrack 2009 - 2012 (queries 1 - 200). For queries 1-50, we use the relevance judgements for the category B runs only and remove query 20 for which no documents are judged relevant.  TREC disks 1 and 2 consist of news articles from e.g. the Associated Press and Wall Street Journal. We use the queries from TREC-1, TREC-2 and TREC-3 (queries 51 - 200). TREC disks 4 and 5 consist of news articles from e.g. L.A. Times and Financial Times, and we use queries 301-450,601-700 from the TREC ROBUST 2004 track. iSearch [280] is an IR test collection from the domain of physics, containing scientific publications, their bibliographic and metadata records. All of these document types are relatively standardised in length, format and language. The iSearch collection consists of approximately 453,000 documents and $n = 66$ queries, from which we use the keywords as queries.

All datasets are indexed without stop word removal and without stemming using Indri 5.10. Dataset characteristics are shown in Table 5.4.

| Dataset | Size (GB) | Number of docs | Average doc length | Number of unique terms | Percentage non-informative |
|---------|-----------|----------------|--------------------|------------------------|----------------------------|
| CWEB09    | 150 | 50,220,423 | 807.3  | 94,889,361 | 5.9%  |
| TREC-d12  | 2,4 | 741,856    | 412,9  | 805,262    | 90.3% |
| TREC-d45  | 2,0 | 528,155    | 479,7  | 783,212    | 90.1% |
| iSearch   | 4,1 | 453,254    | 2252.8 | 4,032,383  | 94.8% |

Table 5.4:  Datasets used.  Average document length is the total number of terms divided by the number of documents. Last column shows the percentage of unique terms in the collections that are non-informative.

### 5.5.2    Identifying Non-Informative Terms

The first step in the proposed cascading methodology is to separate the non-informative from the informative terms.  The method adopted here combines human assessors and machine learning to classify terms as being either informative or non-informative based on term weighting methods.

Our choice of combining human assessors with machine learning is based on two observations.

Firstly, qualitative inspection of the term weights for both non-informative and informative terms (how these were determined is addressed below) suggest that no single approach can separate the two. Examples are shown in Figure 5.1 for six of the term weighting approaches in Section 5.4.1. Thus, thresholding - i.e. specifying a fixed value above which terms are considered informative and noninformative otherwise - is likely an insufficient approach. Secondly, term weighting methods are frequently automatic and provide quantitative estimates of informativeness, but are based on term and collection-wide statistics which may be a poor indicator of informativeness. In contrast, a human assessor can carefully evaluate a term's informativeness, but this method is subjective and slow. Thus, by combining the scrutiny of human assessors with the speed of term-weighting methods, the goal is to create an effective yet efficient compromise.

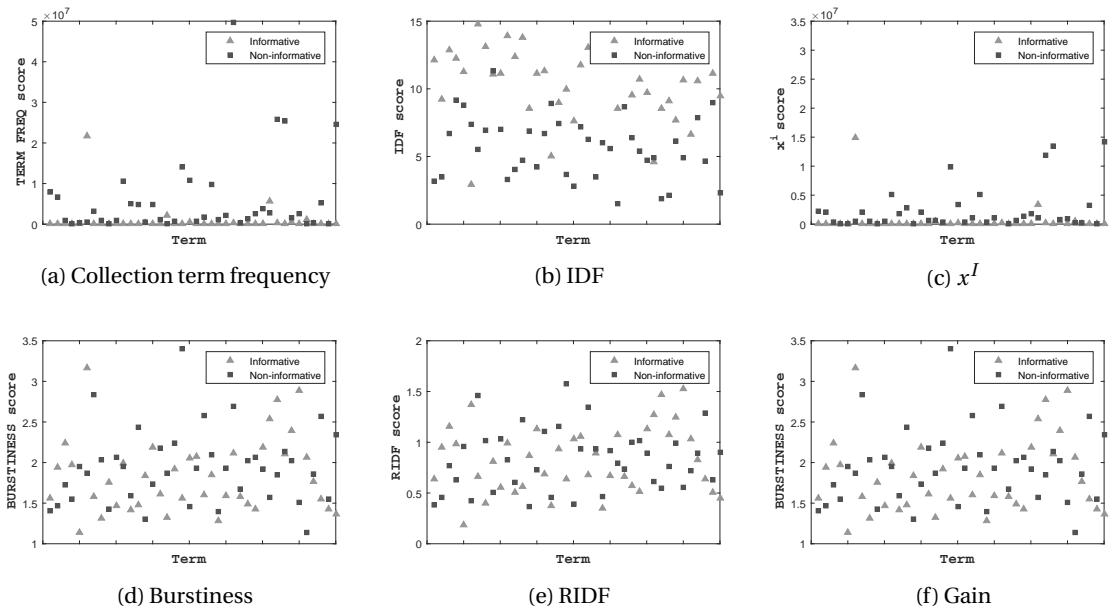

(a) Collection term frequency  (b) IDF  (c) $x^I$

(d) Burstiness  (e) RIDF  (f) Gain

Figure 5.1: Weights of the informative and non-informative terms identified in the first step of our approach according to five of the term weighting approaches in Section 5.4.1. Collection term frequency is included on the assumption that highly frequent terms are considered stop, or function, words which may be less likely to be informative.

The proposed method consists of three steps:

**Step 1: Human Assessment** Lists of non-informative and informative terms are compiled from ClueWeb12 cat. B. (though any collection can be used) by a human assessor. ClueWeb12 cat. B. is used as it contains a large number of unique terms ($\approx 168$M). The human assessor may apply any technique, knowledge and heuristic(s) when selecting these terms.

**Step 2: Term weights** For each term in the two lists, their weight is calculated using the IDF, $x^I$, burstiness, RIDF and gain. The $z$-measure and RelIDF measure are not considered as (i) the $z$-measure relies on the distributions of recall (the probability of retrieving a document given it is relevant) and fall-out (the probability of retrieving a document given it is not relevant) for each term [80], and (ii) RelIDF requires a set of user-specified documents. Each term's collection frequency was also included as a separate weight under the assumption that highly frequent terms are considered stop, or function, terms which may be less likely to be informative.

**Step 3: Classification** Given the lists of informative and non-informative terms and their weights according to the IDF, $x^I$, burstiness, RIDF, gain and collection frequency weights, the problem of separating informative from non-informative terms is cast as a binary classification task. In this task, the weights are "features" (in this case each term has 6 features corresponding to a weight according to one of the term weighting methods) and the goal is to classify each term as either informative or non-informative by learning a classifier that minimises the number of incorrectly classified terms. For each combination of features, a binary support vector machine (SVM) classifier using a Gaussian radial basis function kernel is trained and validated using 10-fold cross-validation.[4] This was repeated 10 times for each classifier and the classification errors averaged. The best results were obtained when classifying terms using their RIDF and gain scores, which incurred a mean classification accuracy of $\approx 86\%$ ($\sigma^2 = 2\%$).

To compare our trained classifier from Step 3 to simple thresholding, the optimal threshold, i.e. the value (or term weight) above which the *largest* number of non-informative terms and the *smallest* number of informative terms exist, was determined. The best results were obtained using IDF term weights which incurred a classification accuracy of $\approx 70\%$. The worst results came when using RIDF which resulted in a $\approx 12.5\%$ classification accuracy. This is interesting as previous research [360] has found the RIDF to perform on par with a high-performing term weighting method based on mixture models. One reason is likely the difference in task (named entity recognition) and domain (restaurants), but also that our training set is substantially smaller and taken from a collection with characteristics different from those used in previous research. Because our method achieves higher accuracy compared to thresholding, we use it to classify each term in each collection in Table 5.4 as being either informative or non-informative. Table 5.4 also lists the percentage of non-informative terms found in each collection using this approach. The differences in the percentage of non-informative terms identified in the ClueWeb09 cat. B. and TREC disks are profound. There are at least two possible explanations for this. First, this might be an artefact of the sample selected. Given that the classifier is trained using 40 samples of informative and non-informative words, there is a risk that these are poor representatives of either or both classes. This implies that, regardless of the classifier used, the outcome will be skewed. A second explanation is that the classifier is over-fitted. Indeed, the trained classifier uses about 73% of the samples as support vectors which imply an absence of regularity (or presence of high variance) in the input data which, in turn, means that removing a single point could drastically alter the decision boundary. Such over-fitting means the trained classifier does not generalise well to unseen data and that the resulting classification of terms as informative or non-informative may be quite poor. While the classifier was trained using a regularisation term and cross-validation, a larger sample constructed using multiple assessors may produce very different classifications and, consequently, different ADR models. It is also worth noting that the over-fitted classifier is trained on terms from a noisy and heterogeneous dataset, but also used to classify terms from domains having different characteristics. This may help explain the difference in the number of terms classified as informative and non-informative for the different datasets.

---

[4]A total of 126 classifiers was trained.

For iSearch, we also see a high percentage of non-informative terms. This is not surprising as the lists of informative and non-informative terms come from an ad hoc domain, but are used to classify terms in a collection using a highly specialised vocabulary which is different from both the TREC Disks and ClueWeb collections. For those terms labelled non-informative, their collection term frequencies are extracted and stored. Each set of collection term frequencies is next used to determine the best-fitting model that will be integrated into DFR to produce an ADR model.

### 5.5.3 Fitting Distributions to Non-informative Terms

The next step in our cascading methodology is to determine the statistical model that best fit the distribution of collection term frequencies of non-informative terms. We do this using our approach from Chapter 3. The results are shown in Table 5.5.

For all datasets, the best-fitting discrete and continuous distribution is the Yule–Simon, power law and Generalized Extreme Value respectively. These are the same statistical models (although with different parameter values) as reported in [344] to quantify the distribution of *all* term frequencies. Practically, this means that the non-informative terms can be used instead of all terms to determine the correct distribution, but also that the number of non-informative terms may be too small (in the case of ClueWeb09 cat. B.) or too large (for both TREC datasets) to capture finer nuances of terms that further help separate informative from non-informative terms. For iSearch, the reason is likely that the articles are from a domain very different from the TREC collections. Consequently, the "features" of informative and non-informative terms may identify only a small group of terms which cannot change the empirical distribution of the term frequencies and, subsequently, what statistical model is the best-fitting. An important caveat in this context are the terms themselves. In a follow-up study using $n = 2$ additional assessors, an inter-rater agreement of $0.211, p < .05$ using Fleiss' $\kappa$.[5] was obtained for the non-informative terms. Thus, the bias introduced by the single human assessor is non-negligible and the informative and non-informative terms reflects this. Another factor is the term weighting methods used. These methods were selected as they are well-known term weighting approaches that have previously been used to discriminate between informative and non-informative terms. Recent research, however, has shown that using semantic methods [273] or mixture models [360] to weigh terms give (i) increased retrieval performance, and (ii) better separation than any single term weighting approach reviewed in Section 5.4.1. While it is unknown if these methods will be beneficial for the purpose of deriving effective ranking models using our proposed cascading methodology, comparing Rennie and Jaakkola's [360] top-10 most informative terms produced by their mixture model to our classification approach shows only a 20% agreement. This suggests that using other term-weighting approaches may result in a markedly different set of non-informative terms being selected, whose distribution may be different than the Yule–Simon, power law or Generalized Extreme Value model.

As the outcome of this step is the *single* best-fitting statistical model, either the Yule–Simon, power law or Generalized Extreme Value must be discarded. Because the fit of all models were determined on the basis of discrete data, the Yule–Simon and power law models are kept. The reason why we report the best-fitting continuous model, is because it provides a starting point as to what continuous model we could emphasise discretising first in future work.

---

[5]The two assessors were a mathematics graduate and an actuary undergraduate. Each was given a single (scrambled) list containing both the informative and non-informative terms and asked to mark those terms they considered informative. No help was given regarding what might constitute an informative term.

| Dataset | Best-fitting discrete distribution | Best-fitting continuous distribution |
|---------|-----------------------------------|--------------------------------------|
| CWEB09 | Yule–Simon ($p = 1.627$) | Generalized Extreme Value ($k = 1.322, \sigma = 30.28, \mu = 36.18$) |
| TREC-d12 | Yule–Simon ($p = 1.738$) | Generalized Extreme Value ($k = 3.843, \sigma = 0.5491, \mu = 1.143$) |
| TREC-d45 | Yule–Simon ($p = 1.804$) | Generalized Extreme Value ($k = 4.198, \sigma = 0.7295, \mu = 1.174$) |
| iSearch | Power law ($p = 1.853$) | Generalized Extreme Value ($k = 2.254, \sigma = 0.1598, \mu = 1.066$) |

Table 5.5: Best-fitting discrete and continuous statistical models for each dataset. Below each statistical model are given their parameters and values. For the Yule–Simon and power law mode, both $p$ and $\alpha$ is called the scale parameter [168]. For the Generalized Extreme Value, $k$, $\sigma$ and $\mu$ are called the scale, shape and location parameter, respectively [168].

The distributions of collection term frequencies for non-informative terms for all datasets are shown in Figure 5.2, with the best-fitting Yule–Simon or power law model superimposed. Both the Yule–Simon and the power law model qualitatively fit terms with low collection term frequencies better than terms with higher collection term frequencies. Compared to the empirical distributions in [344] for the same datasets, *informative* terms are clustered around the mean collection frequencies. This suggests that our approach, because it is based on RIDF and gain which favours medium frequency terms, corroborates Luhn's observation that such terms have optimal resolution power [278]. As we keep the Yule–Simon and power law models based on the above, we next consider how to integrate them into the DFR framework for ranking.

### 5.5.4 Integrating the Yule–Simon Model into Ranking

The Yule–Simon (YS) model is a generalisation of Zipf's law which states that the frequency of any term is inversely proportional to its rank. Hence, in a Yule–Simon model, a small set of terms will have very high frequency, and many terms will have frequencies that are orders of magnitude smaller. The YS model has been used for e.g. text generation [397], citation analysis [354] and for describing the emergence of complex networks [53]. The differences between the YS and power law are two-fold. Firstly, only for large $x$ does the YS model give rise to a power law [328]. Secondly, a YS model decreases faster than a power law [344] and is therefore said to capture stardom, but not super stardom [406].

Formally, the discrete YS model is defined as:

$$\begin{aligned} \text{YS}(x|\theta) &= \left\{ f(x|p) \right\} \\ &= \left\{ \frac{p \cdot \Gamma(x) \cdot \Gamma(p+1)}{\Gamma(x+p+1)} : p > 0 \right\} \end{aligned} \tag{5.40}$$

where $\Gamma$ is the gamma function and $x \in \mathbb{Z}^+$. The emergence of a YS model can be explained by the following modified Polya urn process [174]: consider finitely many documents, each containing a single term, and a stream of terms. For each new term that arrives, with probability $p$, a new document is created, and the term is placed in this document. With probability $1 - p$ the term is placed in one of the existing documents, such that the probability of placing the term in document $i$ is proportional to the number of terms already in this document. The YS model for different values of $p$ is shown in Figure 5.3.

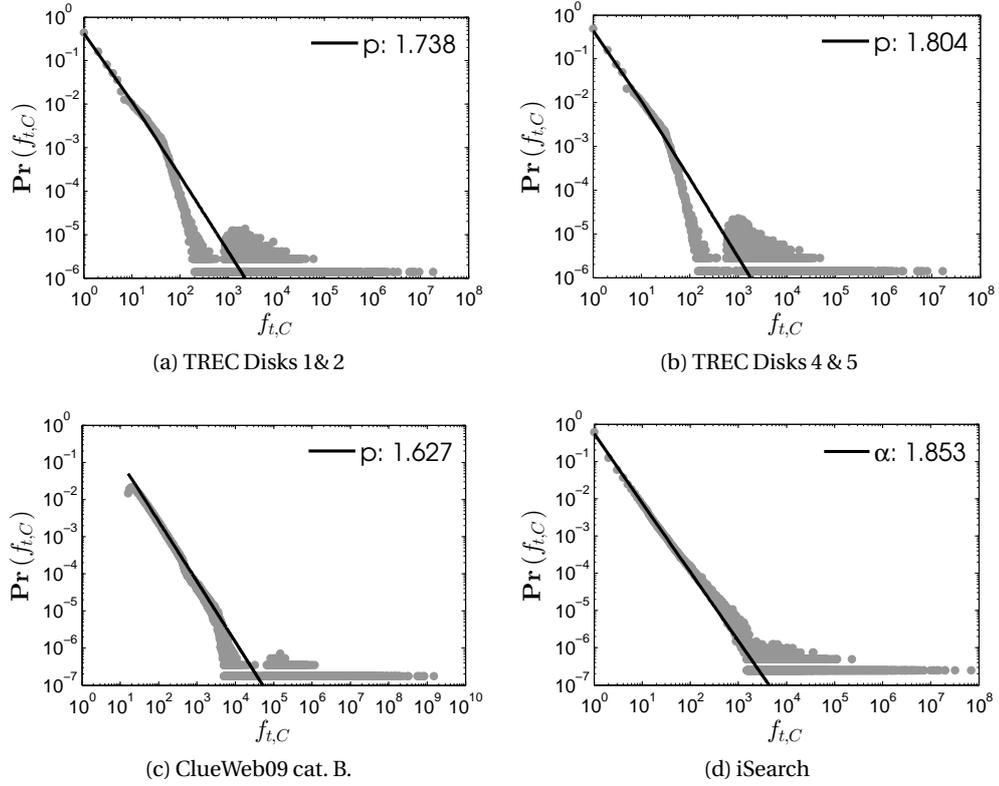

(a) TREC Disks 1 & 2

(b) TREC Disks 4 & 5

(c) ClueWeb09 cat. B.

(d) iSearch

Figure 5.2: Distributions of collection term frequencies for TREC Disks 1 & 2 (5.2a), TREC Disks 4 & 5 (5.2b), ClueWeb09 cat. B. (5.2c) and iSearch (5.2d). Superimposed on each empirical distribution is the best-fitting Yule–Simon distribution (for both TREC and ClueWeb09 cat. B.) and power law for iSearch. `cf` refers to collection term frequency.

We use Eqn. 5.40 to estimate $P_1$ in the DFR ranking model (Eqn. 5.2), by replacing $x$ with the normalised term frequencies from Eqns. 5.8 or 5.13, yielding:

$$\text{Pr}^{\text{YS}}(\hat{f}_{t,d} | p) = \left\{ p \cdot \frac{\Gamma(\hat{f}_{t,d}) \cdot \Gamma(p+1)}{\Gamma(\hat{f}_{t,d} + p + 1)} : p > 0 \right\} \tag{5.41}$$

which, for a term $t$, returns the probability of having $\hat{f}_{t,d}$ occurrences of $t$.

Observe that Eqn. 5.41 is an approximation to Eqn. 5.40 as the gamma function in Eqn. 5.40 is only defined for non-zero integers. Two widely used approaches to approximating the gamma function for real-valued input are Stirling's formula and Lanzco's approximation [353, p. 213]. Of these, Lanzco's approximation is used as it has a lower approximation error than Stirling's formula [346]. Thus, the computation of $\Gamma$ in Eqn. 5.41 is done using Lanzco's approximation. Because the YS model is a discrete model and we apply to it continuous data, it is a theoretically flawed model (similar to the BNB ranking model [119]).[6]

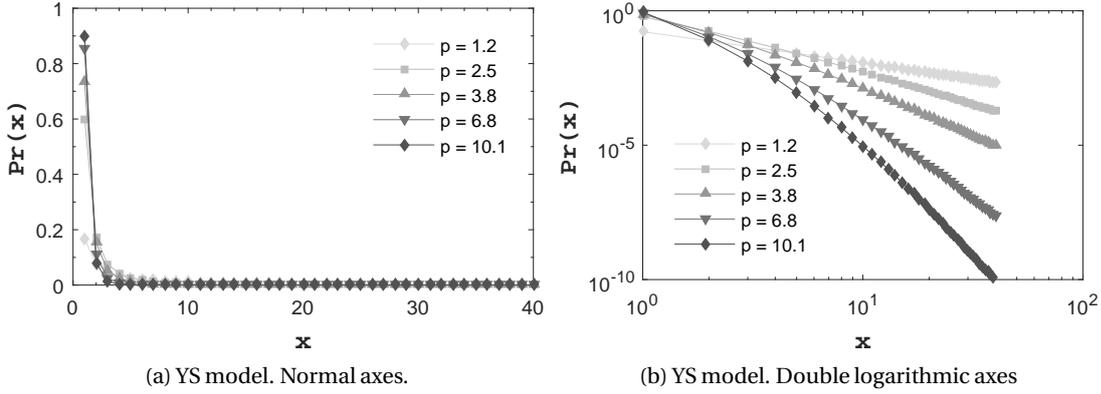

(a) YS model. Normal axes.                    (b) YS model. Double logarithmic axes

Figure 5.3: YS model (Eqn. 5.40) for different values of $p$ on normal (5.3a) and double-logarithmic axes (5.3b).

Inserting Eqn. 5.41 into Eqn. 5.2 gives:

$$R(q,d) = \sum_{t \in q \cap d} \left( -\log_2 \Pr^{\mathrm{YS}}\!\left(\hat{f}_{t,d}|p\right)\right) \cdot (1 - P_2) \tag{5.42}$$

where the superscript to $\Pr^{\mathrm{YS}}\!\left(\hat{f}_{t,d}|p\right)$ denotes the best-fitting statistical model to the collection term frequencies for the non-informative terms in some collection. Because Eqn. 5.42 is a DFR ranking model where $\Pr^{\mathrm{YS}}\!\left(\hat{f}_{t,d}|p\right)$ has been "adapted" to the collection term frequencies, we refer to Eqn. 5.42 as the YS ADR model.

### 5.5.4.1 Varying the Yule–Simon Parameter

Previous research [19, 120] has used two possible values for the parameter of DFR models: $f_{t,C}/N$ and $n_t/N$. To compare our ADR models to previous DFR ranking models, the same parameter values are used.

In addition to $f_{t,C}/N$ and $n_t/N$, we propose a variation of these based on the following observation. The implication of using the YS model for estimating term informativeness may be seen by Figure 5.3b. Let the values on the $x$-axis in Figure 5.3b denote $\hat{f}_{t,d}$. According to the YS model, terms with high $\hat{f}_{t,d}$ will thus receive a low probability according to Eqn. 5.2 and thus a high score from $-\log_2(P_1)$. However, the difference in probability (and subsequent score) between terms with high $\hat{f}_{t,d}$ and low $\hat{f}_{t,d}$ is "smoothed" by the value of $p$ as shown in Figure 5.3b. For example, for $p = 1.2$ a term with $\hat{f}_{t,d} = 2.0$ and $\hat{f}_{t,d} = 10.0$ has probabilities $P(2|p=1.2) = .04$ and $P(10|p=1.2) = .0013$. In contrast, setting $p = 18.9$, the probabilities are $P(2|p=18.9) = .0452$ and $P(10|p=18.9) \approx 1.58e-08$. Thus, a higher value of $p$ will generate a more skewed distribution which results in low probabilities to terms with relatively low $\hat{f}_{t,d}$, whereas a low value for $p$ will generate more uniform probabilities even for terms with very different $\hat{f}_{t,d}$.

---

[6] Both Clinchant and Gaussier [119] and Amati and Rijsbergen [19] apply discrete models to continuous data.

Consequently, we are interested in how the value of $p$ affects a term's informativeness under the YS ADR model and, subsequently, the effectiveness of the YS ADR model. The proposed variation squares the parameter values i.e. $\left(f_{t,C}/N\right)^2$ and $(n_t/N)^2$ respectively. The reasoning is as follows: Consider $n_t/N$ which always lies in the interval $0 < n_t/N \le 1$. According to the observation above, squaring $n_t/N$ will "smooth" out the difference in probabilities of terms with different $\hat{f}_{t,d}$, so only terms with high $\hat{f}_{t,d}$ are given a low probability as $n_t/N$ goes towards 0. Practically, this means that documents where $\hat{f}_{t,d}$ is very high (i.e. documents where $t$ is clustered) are scored highly.

In contrast, $f_{t,C}/N$ is not restricted to the interval $(0,1]$. For $f_{t,C}/N < 1$, the intuition is the same as for $n_t/N$. For $f_{t,C}/N > 1$, increasingly fewer occurrences of $t$ in $d$ are needed to give a term a high score. Practically, this means that given some multi-term query $q$, the risk of single query term with high $f_{q_i,d} : q_i \in q$ dominating the scoring of $d$ is mitigated.

All parameters and variations used for experimentation are given in Table 5.6.

| Abbreviation | Definition | Explanation |
|---|---|---|
| $T_{tc}$ | $f_{t,C}/N$ | Mean number of tokens of $t$ in $C$ |
| $T_{dc}$ | $n_t/N$ | Mean number of documents in $C$ where $t$ occurs |
| $T_{tc}^2$ | $\left(f_{t,C}/N\right)^2$ | Variation of $T_{tc}$ |
| $T_{dc}^2$ | $(n_t/N)^2$ | Variation of $T_{dc}$ |

Table 5.6: YS ADR model parameter values and variations.

All ranking models are post-fixed with either $T_{tc}$ or $T_{dc}$ to denote what parameter value is used. The two variations $T_{tc}^2$ and $T_{dc}^2$ are used only for the YS ADR model because the YS model was found to be the best-fitting model. Such variations could not have been reasoned without knowing the best-fitting statistical model was the Yule–Simon.

### 5.5.5 Integrating the Power Law Model into Ranking

The power law (PL) model is nearly identical to the YS model. In a power law model, a small set of terms will have very high frequency but most terms have frequencies that are orders of magnitude smaller. The power law has been used for text classification [361], to improve caching [49, 387] and test collection generation [93, 189].

Formally, the discrete PL model, repeated here from Section 4.1.1, is defined as:

$$\begin{aligned}
\text{PL}(x|\theta) &= \left\{ f(x|\alpha, x_{\min}) \right\} \\
&= \left\{ \frac{x^{-\alpha}}{\zeta(\alpha, x_{\min})} : \alpha > 1 \right\}
\end{aligned} \tag{5.43}$$

where $\alpha > 1$ is the power law exponent, $x \in \mathbb{Z}^+$ and $\zeta$ is the generalised or Hurwitz zeta function [117]:

$$\zeta(\alpha, x_{\min}) = \sum_{n=0}^{\infty} (n + x_{\min})^{-\alpha} \tag{5.44}$$

and $x_{min}$ is a threshold above which the power law "holds"; that is, given some sample of i.i.d. data points $x = \{x_1, ..., x_n\}$, $x_{min} = x_i \in [1, n]$ above which the power law "optimally" fits the data. Possible mechanisms for generating power law distributions are abundant. Newman [328] provides a comprehensive review of many such methods, including a combination of exponentials, inverses of quantities, random walks, self-organised criticality, highly optimised tolerance and coherent noise. Simon [397] has conducted what is arguably the most rigorous and extensive treatment of power law models. See e.g. [61, 286, 314, 338] for further information. Figure 5.4 shows the PL model for different values of $\alpha$.

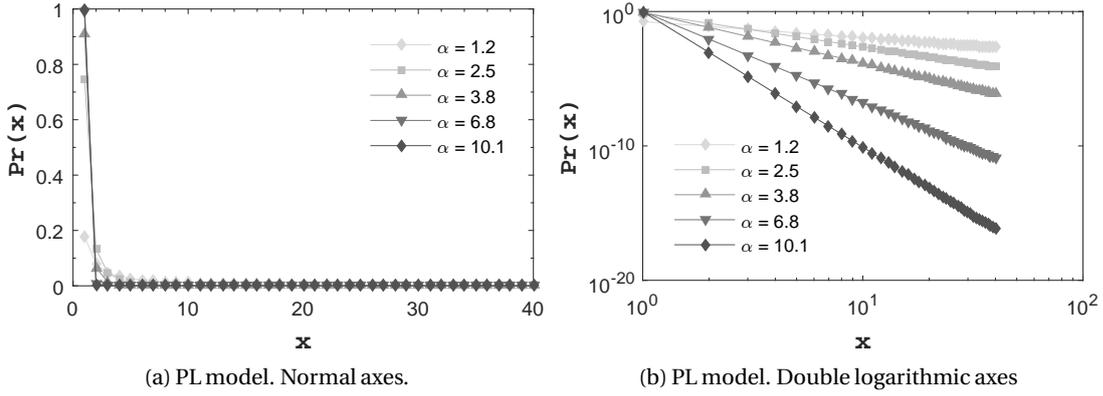

(a) PL model. Normal axes.  (b) PL model. Double logarithmic axes

Figure 5.4: PL model (Eqn. 5.43) for different values of $\alpha$ on normal (5.4a) and double-logarithmic axes (5.4b).

By replacing $x$ with the normalised term frequencies from Eqns. 5.8 or 5.13, and $x_{min}$ with $\hat{f}_{t,d_{min}}$, the power law model becomes:

$$\Pr^{\text{PL}}(\hat{f}_{t,d}|\alpha, \hat{f}_{t,d_{min}}) = \left\{ \frac{\hat{f}_{t,d}^{-\alpha}}{\zeta(\alpha, \hat{f}_{t,d_{min}})} : \alpha > 1 \right\} \tag{5.45}$$

which, for a term $t$, returns the probability of having $\hat{f}_{t,d}$ occurrences of $t$. Also here, $\hat{f}_{t,d}$ refers to logarithmic term normalisation. Identical to the YS model, the power law is a discrete model and thus Eqn. 5.43 is only defined for positive integers. Consequently, substituting $\hat{f}_{t,d}$ for $x$ produces a theoretically flawed model. As we are unaware of a version of Hurwitz's zeta function defined for continuous variables, we use the *continuous* power law model [117]:

$$\Pr^{\text{PL}}(\hat{f}_{t,d}|\alpha, \hat{f}_{t,d_{min}}) = \left\{ (\alpha - 1)\hat{f}_{t,d_{min}}^{\alpha-1} \cdot \hat{f}_{t,d}^{-\alpha} : \alpha > 1 \right\} \tag{5.46}$$

which is defined for all non-zero real numbers. We use Eqn. 5.46 to estimate $P_1$ in the DFR ranking model (Eqn. 5.2).

Inserting Eqn. 5.46 into Eqn. 5.2 gives:

$$R(q, d) = \sum_{t \in q \cap d} \left( -\log_2 \Pr^{\text{PL}}(\hat{f}_{t,d}|\alpha, \hat{f}_{t,d_{min}}) \right) \cdot (1 - P_2) \tag{5.47}$$

where the superscript to $\Pr^{\text{PL}}(\cdot|\cdot)$ denotes the best-fitting statistical model to the collection term frequencies for the non-informative terms. Because Eqn. 5.47 is a DFR ranking model where $\Pr^{\text{PL}}(\cdot|\cdot)$ has been "adapted" to the collection term frequencies, Eqn. 5.42 is referred to as the PL ADR model. We emphasize again (see Section 3.5.1) that we do *not* estimate a lower cutoff when experimenting with the power law, but include it in the definition given here to adhere to the notation.

### 5.5.5.1 Varying the Power Law Parameter

Similar to the YS model, we are interested in varying the power law's $\alpha$ parameter. The intuition behind these variations is the same as for the YS model, but because $\alpha$ is required to be larger than 1, the $T_{dc}$ variation (see Table 5.6) cannot be used, and only if $f_{t,C} > N$ can the $T_{tc}$ variation be used. Thus, instead of variations to $T_{tc}$ and $T_{dc}$ we are interested in approaches that ensure $T_{tc}$ and $T_{dc}$ are strictly greater than 1.

A number of options exist by which $T_{tc}$ and $T_{dc}$ can be made strictly greater than 1. Denote by $\hat{T}$ either $T_{tc}$ or $T_{dc}$. The first option is to add 1 to all $\hat{T}$. While this uniformly scales all parameter values, it skews the distribution which, returning to the discussion in Section 5.5.4.1, results in lower probabilities, and subsequently higher $-\log_2(P_1)$ values, for terms with relatively low $\hat{f}_{t,d}$. A second option is to determine, for each value of $\hat{T}$ smaller than 1, the smallest value of $\epsilon$ such that $\forall \hat{T} : \hat{T} + \epsilon > 1$. This, however, only scales some parameter values and means that all $\hat{T} < 1$ gets the same value, as $\epsilon$ is determined on a per-term basis. A variation of this approach is, amongst *all* terms with $\hat{T} < 1$, to determine the largest $\epsilon$ such that $\hat{T} + \epsilon > 1$ and add this to all $\hat{T}$ values regardless of whether $\hat{T} < 1$. A third option is to, for each value of $T_{tc} < 1$, take the reciprocal of $T_{tc}$ i.e. $1/T_{tc}$. Similarly to the approach involving the calculation of $\epsilon$, however, only some parameter values are scaled.

Here, we choose to experiment with the first option as it is simple and uniformly scales all values regardless of the value of $\hat{T}$ thus not, perhaps unintentionally, favouring some terms over others. Following the convention from Section 5.5.4.1, all PL ADR models using this approach is post-fixed with $T_{tc}^{+1}$ or $T_{dc}^{+1}$. Table 5.7 lists the parameter values we consider for all PL ADR models.

| Abbreviation | Definition | Explanation |
|---|---|---|
| $T_{tc}^{+1}$ | $1 + (f_{t,C}/N)$ | Mean number of tokens of $t$ in $C$ plus one |
| $T_{dc}^{+1}$ | $1 + (n_t/N)$ | Mean number of documents in $C$ where $t$ occurs plus one |

Table 5.7: PL ADR model parameter values.

Similarly, to the YS model, this variation could not have been reasoned without knowing that the best-fitting statistical model was the power law.

## 5.6 Retrieval Experiments

This section describes the baselines (Section 5.6.1) and the results of evaluating our ADR models against these baselines (Section 5.6.2).

### 5.6.1 Baselines

We experimentally evaluate our ADR models to the following baselines: P, $I_n$, LL and SPL. P, $I_n$ are original DFR models (see Table 5.2); LL and SPL are DFR extensions introduced in Section 5.3. For all models we use L2 normalisation: Laplace's law of succession (Section 5.2.3.1) and logarithmic term frequency normalisation (Section 5.2.4) as these are reported to give the best results [19]. We also compare our ADR models to a query likelihood language model baseline with Dirichlet smoothing (LMDir). All model names (except LMDir) are post-fixed with either $T_{tc}$ or $T_{dc}$ (see Section 5.5.4.1) which denotes the parameter value for each model. For example, YSL2-$T_{tc}$ is the Yule–Simon ADR model with Laplace's law of succession and logarithmic term frequency normalisation, using the $f_{t,C}/N$ parameter value. For each query, we retrieve the top-1000 documents according to each ranking model using short (title) queries, and measure performance using mean average precision (MAP), precision at 10 (P@10), Bpref, expected reciprocal rank [102] at 20 (ERR@20) and normalised discounted cumulative gain at 10 (nDCG@10) and nDCG [218]. Statistical significant differences are detected using a two-tailed $t$-test at the standard .05 level. We tune $\mu$ in LMDir in the range $\mu = \{100,500,800,1000,2000,3000,4000,5000,8000,10000\}$ [458] using 3-fold cross-validation. Similarly to Clinchant and Gaussier [121], we vary $c$ in the range $\{0.5,1,2,4,6,8\}$. $c$ is tuned for all DFR, information and ADR models using 3-fold cross-validation. For all ranking models and baselines, we report the average performance over all three test folds.

### 5.6.2 Findings

Tables 5.8, 5.9 and 5.10 show the results for all ranking models across all datasets. The tuned LMDir baseline is used as reference for statistical significance testing as it a well-known and strong baseline.

All results in Tables 5.8, 5.9 and 5.10 are somewhat lower compared to previous findings [19, 120]. The main difference is that we did not filter out stop words, use stemming nor did custom pruning of documents. We did not consider stop word removal as, historically, IR systems has moved from using quite large stop word lists (200+ terms) to using no stop word list [289, Chap. 2], and Web search engines are reported to generally not use stop word lists as contemporary IR systems has focused on how to exploit the statistics of language so as to be able to cope with common words in better ways [289, Chap. 2]. Furthermore, while stop word lists are generally beneficial to retrieval performance, evidence suggests that such improvement is model-dependent [146]. Consequently, we choose to not remove stop word in these experiments. Similarly, while the role of stemming in IR is well explored for English, results are controversial [77, 248, 452]. This is a recurring issue in several studies [196, 213, 255] with many sources of variation such as e.g. linguistic vs non-linguistic stemmers, language, query and document length and evaluation measures [248]. Hull [213] found that the average absolute improvement ranged from 1-3% and Harman [196] found no statistical significant difference between stemming and non-stemming. In contrast, Krovetz [251] finds an increase of 15-35% in retrieval performance when stemming collection comprising of very short documents and queries and much more modest improvements (similar to [213]) for collections comprising longer documents. Consequently, we also chose to not apply any stemming in these experiments.

The results for all collections and performance measures show that the standard DFR models

| Model | TREC Disks 1 & 2 | | | TREC Disks 4 & 5 | | | ClueWeb09 cat. B. | | |
|---|---|---|---|---|---|---|---|---|---|
| | MAP | nDCG | Bpref | MAP | nDCG | Bpref | MAP | nDCG | Bpref |
| LMDir | .1737 | .4095 | .2487 | .2107 | .4643 | .2239 | .1112 | .2973 | .2209 |
| PL2-$T_{tc}$ | .0551* | .2032* | .1384* | .0613* | .2524* | .1009* | .0396* | .1448* | .1258* |
| PL2-$T_{dc}$ | .0448* | .1521* | .1095* | .0604* | .2487* | .0960* | .0396* | .1444* | .1252* |
| $l_n$L2-$T_{tc}$ | .0865* | .2844* | .1761* | .0777* | .2917* | .1114* | .0430* | .1596* | .1405* |
| $l_n$L2-$T_{dc}$ | .0865* | .2846* | .1763* | .0781* | .2818* | .1088* | .0431* | .1596* | .1407* |
| LLL2-$T_{tc}$ | .1773 | .4195 | .2587 | .2225 | .4812 | .2341 | .1223 | .3184 | .2349 |
| LLL2-$T_{dc}$ | .1757 | .4176 | .2580 | .2210 | .4810 | .2329 | .1226 | .3180 | .2349 |
| SPLL2-$T_{tc}$ | .1763 | .4136 | .2548 | .2257 | .4863 | .2375 | .1231 | .3207 | .2372 |
| SPLL2-$T_{dc}$ | .1762 | .4145 | .2551 | .2271 | .4876 | .2387 | .1251 | .3224 | .2370 |
| YSL2-$T_{tc}$ | .1084* | .2593* | .1704* | .2102 | .4644 | .2280 | .1232 | .3197 | .2359 |
| YSL2-$T_{dc}$ | .1696 | .4004 | .2476 | .2257 | .4860 | .2381 | .1273 | .3240 | .2376 |
| YSL2-$T_{tc}^2$ | .1013* | .2297* | .1502* | .2140 | .4638 | .2262 | .1282 | .3249 | .2385 |
| YSL2-$T_{dc}^2$ | **.1822** | **.4227** | **.2594** | **.2281** | **.4885** | **.2394** | **.1294** | **.3265** | **.2391** |

Table 5.8: Performance of all ranking models on TREC Disks 1 &2, TREC Disks 4 & 5 and ClueWeb09 cat. B. for MAP, nDCG and Bpref. All ranking models are tuned and use Laplace's law of succession (L) and logarithmic term frequency normalisation (2). Grey cells denote entries strictly larger than the LMDir baseline. Bold entries denote the best results. $^*$ denotes statistically significant difference from the LMDir baseline using a two-tailed $t$-test at the .05% level. All models are post-fixed with either $T_{tc}$ or $T_{dc}$ (see Section 5.5.4.1) (or their squared version) which denotes the parameter estimate for each model.

| Model | TREC Disks 1 & 2 | | | TREC Disks 4 & 5 | | | ClueWeb09 cat. B. | | |
|---|---|---|---|---|---|---|---|---|---|
| | P@10 | ERR@20 | nDCG@10 | P@10 | ERR@20 | nDCG@10 | P@10 | ERR@20 | nDCG@10 |
| LMDir | .4580 | **.0922** | **.4769** | .3845 | .1043 | .3968 | .2586 | .0973 | .1769 |
| PL2-$T_{tc}$ | .1766* | .0316* | .1681* | .1273* | .0359* | .1332* | .0712* | .0211* | .0472* |
| PL2-$T_{dc}$ | .1246* | .0523* | .1180* | .1217* | .0347* | .1273* | .0709* | .0314* | .0471* |
| $l_n$L2-$T_{tc}$ | .2660* | .0488* | .2578* | .1627* | .0478* | .1742* | .0782* | .0352* | .0511* |
| $l_n$L2-$T_{dc}$ | .2660* | .0488* | .2578* | .1626* | .0481* | .1745* | .0783* | .0352* | .0512* |
| LLL2-$T_{tc}$ | .4440 | .0849 | .4475 | .4049 | .1072 | .4142 | .2542 | .0926 | .1706 |
| LLL2-$T_{dc}$ | .4300 | .0830 | .4339 | .3982 | .1069 | .4097 | .2542 | .0928 | .1707 |
| SPLL2-$T_{tc}$ | .4586 | .0881 | .4643 | .4144 | .1103 | .4276 | .2529 | .0945 | .1720 |
| SPLL2-$T_{dc}$ | .4573 | .0878 | .4620 | .4176 | .1107 | .4299 | .2586 | .0958 | .1752 |
| YSL2-$T_{tc}$ | .2826* | .0552* | .2876* | .3982 | .1048 | .4069 | .2601 | .0951 | .1752 |
| YSL2-$T_{dc}$ | .4546 | .0877 | .4624 | **.4182** | **.1113** | **.4312** | .2666 | .0985 | .1810 |
| YSL2-$T_{tc}^2$ | .2506* | .0484* | .2543* | .3961 | .1039 | .4024 | .2660 | .0990 | .1815 |
| YSL2-$T_{dc}^2$ | **.4640** | .0893 | .4697 | .4148 | .1100 | .4266 | **.2678** | **.0998** | **.1836** |

Table 5.9: Performance of all ranking models on TREC Disks 1 &2, TREC Disks 4 & 5 and ClueWeb09 cat. B. for early precision measures. All ranking models are tuned and use Laplace's law of succession (L) and logarithmic term frequency normalisation (2). Grey cells denote entries strictly larger than the LMDir baseline. Bold entries denote the best results. $^*$ denotes statistically significant difference from the LMDir baseline using a two-tailed $t$-test at the .05% level. All models are post-fixed with either $T_{tc}$ or $T_{dc}$ (see Section 5.5.4.1) (or their squared version) which denotes the parameter estimate for each model.

PL2 and $I_n$L2 are consistently statistically significantly different from the LMDir baseline. Comparatively, the information models outperform the LMDir baseline on most datasets. For TREC Disks 1 & 2 both information models outperform the LMDir baseline on MAP, nDCG and

| | iSearch | | | | | |
|---|---|---|---|---|---|---|
| Model | MAP | nDCG | Bpref | P@10 | ERR@20 | nDCG@10 |
| LMDir | **.1064** | .2379 | .1920 | **.4041** | **.1306** | **.2996** |
| PL2-$T_{tc}$ | .0398* | .1735* | .2295* | .0833* | .0470* | .0618* |
| PL2-$T_{dc}$ | .0370* | .1612* | .2170* | .0833* | .0471* | .0619* |
| $l_n$L2-$T_{tc}$ | .0326* | .1498* | .2074* | .0742* | .0578* | .0641* |
| $l_n$L2-$T_{dc}$ | .0325* | .1498* | .2074* | .0742* | .0576* | .0639* |
| LLL2-$T_{tc}$ | .1023 | **.3090** | .2901 | .2318 | .1295 | .1816 |
| LLL2-$T_{dc}$ | .0905 | .2937 | .2709 | .2106 | .1228 | .1648 |
| SPLL2-$T_{tc}$ | .1058 | .3083 | .3012 | .2076 | .1200 | .1667 |
| SPLL2-$T_{dc}$ | .1028 | .3053 | .2952 | .2151 | .1225 | .1719 |
| PL2-$T_{tc}^{+1}$ | .1012 | .2680 | .2796 | .1818 | .1052 | .1469 |
| PL2-$T_{dc}^{+1}$ | .1043 | .3079 | **.3014** | .2012 | .1194 | .1596 |

Table 5.10: Performance of all ranking models on iSearch for MAP, nDCG, Bpref, P@10, ERR@20 and nDCG@10. All ranking models are tuned and use Laplace's law of succession (L) and logarithmic term frequency normalisation (2). Grey cells denote entries strictly larger than the LMDir baseline. Bold entries denote the best results. * denotes statistically significant difference from the LMDir baseline using a two-tailed $t$-test at the .05% level. All models are post-fixed with either $T_{tc}$ or $T_{dc}$ (see Section 5.5.4.1) which denotes the parameter estimate for each model.

Bpref, though these results are not statistically significant. Similarly, the LMDir baseline has better performance for ERR@20 and nDCG@10 but these results are not statistically significant either. The YSL2-$T_{tc}$, however, performs statistically significantly different than the LMDir baseline, whereas the YSL2-$T_{dc}$ performs on par with the LMDir. For TREC Disks 4 & 5, the information models outperform the LMDir baseline on all performance measures, though no results are statistically significant. Unlike TREC Disks 1 & 2, the YSL2-$T_{dc}$ model outperforms the LMDir baseline on all performance measures. The YSL2-$T_{tc}$ performs similar, but slightly worse for MAP. None of the results are statistically significant, however. On ClueWeb09 cat. B., the information models are outperformed by the LMDir baseline on all early precision measures, but perform better than LMDir on both nDCG and Bpref. This supports the findings of Clinchant and Gaussier, even though no stop word removal or stemming is used in our experiments. This is interesting as TREC Disks 4 & 5 and ClueWeb09 cat. B. differ in their domain, size and heterogeneity but suggests they may share some characteristics that are being exploited by the YSL2 model. Both the YSL2-$T_{tc}$ and YSL2-$T_{dc}$ ranking models perform better than the LMDir baseline on MAP, nDCG, P@10 and Bpref, but only the YSL2-$T_{dc}$ outperform the LMDir baseline on ERR@20 and nDCG@10. None of the improvements are statistically significant, however.

Using the YSL2-$T_{tc}^2$ and YSL2-$T_{dc}^2$ ADR models show mixed results. The YSL2-$T_{tc}^2$ ranking model tends to decrease performance for both TREC datasets (TREC Disks 1 & 2 in particular) compared to the YSL2-$T_{tc}$, but increases performance for ClueWeb09 cat. B. In contrast, the YSL2-$T_{dc}^2$ ranking model consistently improves performance on all datasets and performance measures, and is overall the best performing ranking model. Taken together, these results hint at the importance of selecting the right parameter value to improve retrieval performance.

That the YSL2-$T_{dc}^2$ ranking model performs on par or better than the information models is perhaps not surprising as the YS model is asymptotically a power law through the relation [267,

Chap. 6]:

$$\Pr(x|p) \propto 1/x^{p+1} \tag{5.48}$$

for large $x$, and should thus perform on par with the SPL model. Following [121], we also initially experimented with a YSL2 information model defined by:

$$R(q,d) = \sum_{t \in q} -\log_2\left[1 - \left(1 - \hat{f}_{t,d}B(\hat{f}_{t,d}, p+1)\right)\right] \tag{5.49}$$

where $B$ denotes the Beta function and $1 - \hat{f}_{t,d}B(\hat{f}_{t,d}, p+1)$ is the CCDF for the YS distribution. Evaluation of this model on ClueWeb09 cat. B. showed performance similar to both the LLL2 and SPLL2. The performance difference may, therefore, be attributed to the use of the first normalisation principle (Section 5.2.3), which is not used by the information models. Because Laplace's law of succession increases as a function of $\hat{f}_{t,d}$ and is always in the interval $(0;1)$, multiplying it with $-\log_2(P_1)$ in Eqn. 5.2 practically "dampens" the informativeness of each term. This indicates that the "raw" text burstiness phenomenon quantified by the information models can benefit from normalisation in some cases, but the results using the information models are not statistically significantly different than any of our ADR models.

Compared to the TREC collections, the results for the iSearch collection in Table 5.10 are quiet different. For MAP, the LMDir baseline is the best overall though the results are only statistically significantly different compared to the standard DFR models. Similar observations are made for nDCG, except all information and PLL2 ranking models outperform the baseline, though the results are not statistically significant. For bpref, the two-tailed $t$-test finds the standard DFR models to be statistically significantly different from the LMDir baseline.[7] Using Wilcoxon's signed-rank test suggests no statistical significant difference. The PLL2 models perform on par with the information models, with the $T_{dc}^{+1}$ variation giving the best overall performance for bpref. Similarly to the TREC collections, the $T_{tc}$ variation gives the lowest performance of the PLL2 models. This suggests that, also for iSearch, a more uniform distribution of normalised collection term frequencies are beneficial for retrieval.

Early precision performance on iSearch sees all ranking models consistently outperformed by the LMDir baseline. For example, the LMDir places, on average, four relevant documents in the top-10 results, whereas the best information and PLL2 models place just over two relevant documents in the top-10. For ERR@20 and nDCG@10, only the information models perform on par with the LMDir baseline though the results are not statistically significantly different. The results for the PLL2 models are somewhat lower compared to the information models on all performance measures, but the results are not statistically significantly different. A likely reason why the results for iSearch are so different from the TREC collections, is the collection characteristics. Specifically, iSearch is heavily curated and consists of documents from a narrow domain using a specialised vocabulary. Furthermore, because our experiments use both the publications, their bibliographic content and metadata which all fluctuate in length and the content they describe, our ADR models may perform better when applied to e.g. just the publications. Indeed, post hoc investigation showed, when averaged over all queries for iSearch, that over 70% of the documents ranked in the top$-10$ are metadata articles, with the remaining documents being publications.

---

[7]This is caused by a large difference between the mean values of the standard DFR models and the LMDir baseline, which forces the value of $t$-statistic down.

To understand the behaviour of the proposed ADR models relative to the LMDir baseline, Figure 5.5 shows the sorted per-query difference in nDCG for all models using all 199 queries for ClueWeb09 cat. B. relative to the LMDir baseline.[8] A positive difference in nDCG favours the ADR model over the LMDir. Figure 5.5 shows that the standard DFR models perform worse than the LMDir baseline for most queries (167 queries for PL2 and 163 queries for $I_n$L2). In contrast, both the information models and the YSL2-$T_{dc}^2$ outperform the LMDir baseline on 53% to 64% of the queries though the individual differences between e.g. our ADR models and the LMDir are mostly small.

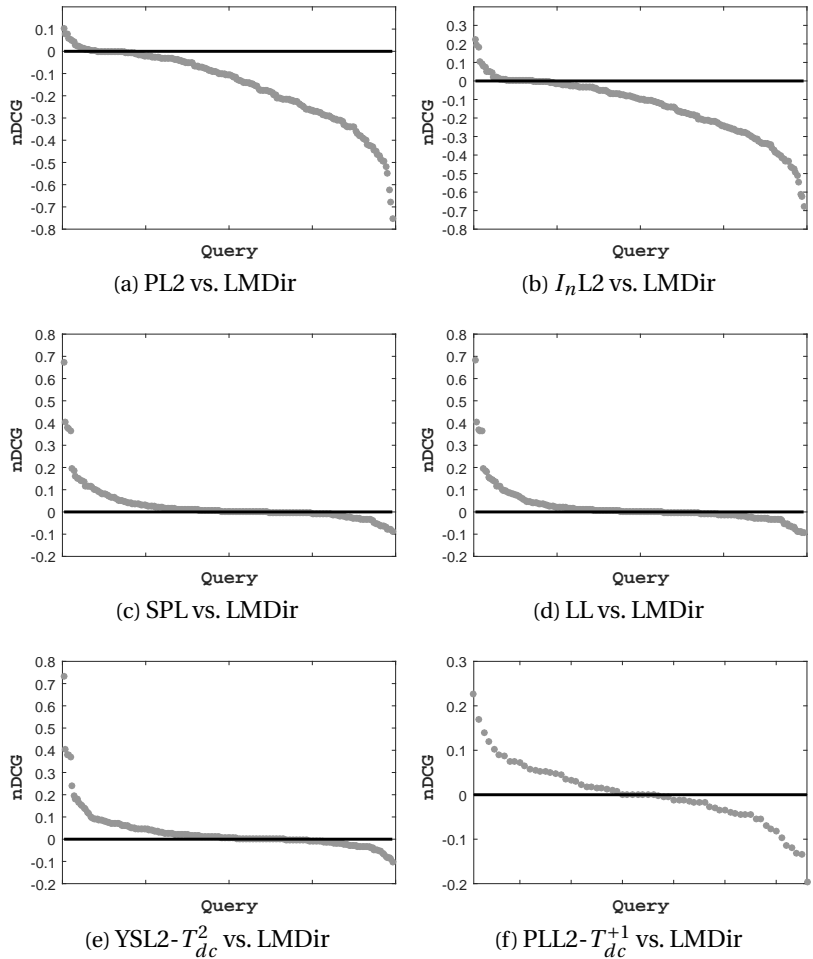

(a) PL2 vs. LMDir    (b) $I_n$L2 vs. LMDir

(c) SPL vs. LMDir    (d) LL vs. LMDir

(e) YSL2-$T_{dc}^2$ vs. LMDir    (f) PLL2-$T_{dc}^{+1}$ vs. LMDir

Figure 5.5: Sorted per-query difference in nDCG between the LMDir baseline and each ranking model on ClueWeb09 cat. B (199 queries) and iSearch (Figure 5.5f only - 66 queries). Only the YSL2-$T_{d,c}^2$ ADR model is shown as it was the best performing of our ADR models.

Comparing YSL2-$T_{d,c}^2$ to the information models shows that retrieval performance improves for a minimum of 93 (46.7%) of the queries, and decreases retrieval performance for a maximum of 41 (20.6)% of the queries when measuring performance using nDCG. Similar results are observed for the remaining performance measures. For iSearch, we find a much more even split with the LMDir outperforming the PLL2-$T_{dc}^{+1}$ model on 30 queries, whereas PLL2-$T_{dc}^{+1}$ is the better ranking model on 29 queries. The difference in nDCG can be attributed to a few queries which score substantially higher than the LMDir baseline. Furthermore, because the number of queries for iSearch is at least $\approx 33\%$ smaller than all other query sets used, the performance of individual queries can quickly skew the results.

[8]Query 20 was removed as stated in Section 5.5.1.

| | Top-5 worst-performing queries | | |
|---|---|---|---|
| Position | **PL2** | **$I_nL2$** | |
| 1 | alexian brothers hospital (46) | french lick resort and casino (2) | |
| 2 | french lick resort and casino (2) | alexian brothers hospital (46) | |
| 3 | ralph owen brewster (108) | ralph owen brewster (108) | |
| 4 | wedding budget calculator (18) | wedding budget calculator (18) | |
| 5 | continental plates (84) | continental plates (84) | |
| Position | **LLL2** | **SPLL2** | **YSL2-$T_{d,c}^2$** |
| 1 | wedding budget calculator (18) | dinosaurs (14) | fact on uranus (130) |
| 2 | bowflex power pro (124) | bowflex power pro (124) | dinosaurs (14) |
| 3 | gmat prep classes | barbados (167) | wedding budget calculator (18) |
| 4 | dinosaurs (14) | wedding budget calculator (18) | bowflex power pro (124) |
| 5 | fact on uranus (130) | fact on uranus (130) | barbados (167) |

Table 5.11: Top-5 worst performing queries for ClueWeb09 cat. B. for each ranking model. Number in parenthesis is the query number.

Next, we extract the top-5 worst-performing queries (as per nDCG) for ClueWeb09 cat. B. for each model. These are shown in Table 5.11 for all ranking models. Table 5.11 shows that only 1 in 5 queries (`wedding budget calculator`) is "hard" for all ranking models. The same five queries (although in different positions) are found for standard DFR models. Similar observations are made for both the information models and the YSL2-$T_{d,c}^2$. This suggests that these queries and the collections exhibit some general, but unknown characteristics that reduce retrieval performance when using DFR and ADR models. For example, while TREC Disks 1 & 2 and TREC Disks 4 & 5 come from the same domain, the proposed ADR models perform substantially (and consistently) better on the latter dataset. To determine if this observation is isolated to this particular domain, all experiments were repeated on the ClueWeb12 cat. B. collection comprising $\approx 52M$ Web documents. We verified that the distribution of collection term frequencies of non-informative terms for ClueWeb12 cat. B. was also best approximated by the YS model (and contained roughly the same percentage of non-informative terms). The results are shown in Table 5.12.

Compared to ClueWeb09 cat. B., the results show that YSL2 ADR models are largely outperformed by all baselines on all measures except for P@10. Only two results (for the PL2) are statistically significantly different from the LMDir. Indeed, Table 5.12 suggests that ClueWeb12 cat. B. and corresponding queries exhibit characteristics that are different from ClueWeb09 cat. B., even though both datasets are from the same domain. The information-based models perform substantially better than our proposed ADR models on the ClueWeb12 cat. B. dataset and on par on ClueWeb09 cat. B. despite coming from the same domain. A key difference between our ranking models and the information-based is the use of the first normalisation principle (see Section 5.2.3), which, for our ranking models, was Laplace's law of succession. As Laplace's law of succession is a monotonically increasing function of the number of tokens of $t$, this normalisation "rewards" terms in some document $d$ depending on the number of tokens; the more tokens of $t$ in $d$, the higher the reward. This means that even if a term $t$ gets an initial high score in $d$ because of the chosen "model of randomness", the subsequent normalisation can substantially reduce this score depending on the term's number of tokens. Seeing as the "model of randomness" already captures a term's informativeness/importance by its divergence estimated from the whole collection, the use of the first normalisation principle may (involuntarily) diminish performance in our proposed ranking models. The use of the first normalisation principled was also questioned by Clinchant and Gaussier [120]. However, the first normalisation principle does not account for the observation that the standard DFR models also outperform our proposed models on ClueWeb12 cat. B. As our experiments showed that a variation of the model parameters which produced a more uniform distribution gave the best results overall, the use of a statistical model (the Yule–Simon) which is used to quantify heavily skewed data might explain why our ranking models do not perform as well for ClueWeb12 cat. B. More fundamental, however, is the question of using a statistical model based on raw term frequencies, when scoring is based on normalised term frequencies which may give rise to a very different distribution. For ClueWeb09 cat. B., TREC Disks 1 & 2 and TREC Disks 4 & 5, this may not be the case as ADR models substantially outperform the standard DFR models and perform on par with the information models. For ClueWeb12 cat. B., however, this appears to not be the case. Indeed, post hoc qualitative examination of the distribution of normalised term frequencies for several query terms from TREC WebTrack 2013 suggests that a unimodal statistical model (such as the Poisson, negative binomial, inverse Gaussian or Rayleigh) may be a better choice.

| | ClueWeb12 cat. B. | | | | | |
|---|---|---|---|---|---|---|
| Model | MAP | nDCG | Bpref | P@10 | ERR @20 | nDCG @10 |
| LMDir | .0318 | .1087 | .0844 | .2359 | .0982 | .1640 |
| PL2-$T_{tc}$ | .0272* | .1004 | .0831 | .2095* | .0831 | .1329* |
| PL2-$T_{dc}$ | .0315 | .1095 | .0845 | .2406 | .0924 | .1606 |
| $l_n$L2-$T_{tc}$ | .0322 | .1089 | .0843 | **.2415** | .0980 | .1637 |
| $l_n$L2-$T_{dc}$ | .0325 | .1007 | .0844 | .2405 | **.1085** | .1647 |
| LLL2-$T_{tc}$ | .0319 | .1100 | .0840 | .2401 | .1026 | **.1670** |
| LLL2-$T_{dc}$ | .0323 | .1101 | .0840 | .2377 | .0997 | .1646 |
| SPLL2-$T_{tc}$ | .0317 | **.1104** | **.0850** | .2348 | .0986 | .1629 |
| SPLL2-$T_{dc}$ | .0317 | .1101 | .0846 | .2349 | .0976 | .1620 |
| YSL2-$T_{tc}$ | .0301 | .1043 | .0831 | .2339 | .0866 | .1563 |
| YSL2-$T_{dc}$ | .0304 | .1052 | .0832 | .2350 | .0878 | .1583 |
| YSL2-$T_{tc}^2$ | .0309 | .1045 | .0823 | .2397 | .0887 | .1596 |
| YSL2-$T_{dc}^2$ | .0311 | .1044 | .0818 | .2402 | .0898 | .1604 |

Table 5.12: Performance of all ranking models on ClueWeb12 cat. B. All ranking models are tuned and use Laplace's law of succession (L) and logarithmic term frequency normalisation (2). Grey cells denote entries strictly larger than the LMDir baseline. Bold entries denote the best results. * denotes statistically significant difference from the LMDir baseline using a $t$-test at the .05% level. All models are post-fixed with either $T_{tc}$ or $T_{dc}$ (see Section 5.5.4.1) which denotes the parameter estimate for each model.

## 5.6.3 Performance Analysis for ClueWeb09 and ClueWeb12

The difference in performance of our ADR models between ClueWeb09 cat. B. and ClueWeb12 cat. B. is interesting as both collections come from the same domain, and were crawled in a similar fashion. In this section we investigate this issue further, and find that the presence of stop words can significantly affect the performance of our ADR models.

Recall the definition of an ADR model (see Eqn. 5.39) is:

$$R(q,d) = \sum_{t \in q \cap d} \left(-\log_2 M\right) \cdot (1 - P_2) \tag{5.50}$$

which we see as comprising three basic components: (i) the probability model ($M$) which was found to be the Yule–Simon model for both collections, (ii) the parameter of the probability model (e.g. $T_{dc}$ or $T_{tc}$), (iii) the informative normalisation resizing $(1 - P_2)$ and (iv) the term normalisation. We will look at each of the latter three in an effort to understand how they can affect the final ranking. We do not consider the choice of model $M$ as it was determined to be the same on both collections. To facilitate our analysis we look more closely at the five highest-performing queries (according to nDCG) for both ClueWeb09 cat. B. and ClueWeb12 cat. B. We focus here on the $T_{dc}$ parameterisation as this was found to be the best-performing for ClueWeb09 cat. B.

### 5.6.3.1 Effect of Parameterisation on Scoring

The results for ClueWeb09 cat. B. showed that the $T_{dc}$ and the $T_{dc}^2$ variation produced the best overall results. Because $T_{dc}$ is always between 0 and 1, this suggested that a more uniform distribution is beneficial to retrieval, and thus $T_{dc}$ can be seen as the reciprocal of the inverse document frequency (idf): the lower $T_{dc}$, the higher the idf. In contrast, when using either $T_{dc}$ or $T_{dc}^2$, the increase in results for ClueWeb12 cat. B. is, at best, small and inconsistent relative to the baseline and the information models. As the value of $T_{dc}$ controls the uniformity of a term's distribution, we begin by looking at how the value of $T_{dc}$ varies for the five highest-performing queries for each collection. To do this, we calculate the $T_{dc}$ for each query term of each query of the top-5 highest-performing queries. Tables 5.13 and 5.14 show descriptive statistics for these two sets of queries.

| Query Nr. | Min | Max | Mean | Std Dev. | # terms |
|---|---|---|---|---|---|
| 22 | 0.0012 | 0.0078 | 0.0047 | 0.0033 | 3 |
| 31 | 0.0100 | 0.0105 | 0.0102 | 0.0004 | 2 |
| 35 | 0.0020 | 0.0020 | 0.0020 | 0 | 1 |
| 46 | 0.0007 | 0.0007 | 0.0007 | 0 | 1 |
| 108 | 0.0000 | 0.0267 | 0.0161 | 0.0142 | 3 |

Table 5.13: $T_{dc}$ statistics aggregated over query terms for each query for the highest-performing queries for ClueWeb09 cat. B.

| Query Nr. | Min | Max | Mean | Std Dev. | # terms |
|---|---|---|---|---|---|
| 211 | 0.0069 | 0.7432 | 0.3474 | 0.3292 | 5 |
| 214 | 0.0050 | 0.0510 | 0.0296 | 0.0198 | 4 |
| 217 | 0.0349 | 0.2264 | 0.0920 | 0.0901 | 4 |
| 232 | 0.0001 | 0.7954 | 0.1361 | 0.3230 | 6 |
| 236 | 0.0048 | 0.8179 | 0.2419 | 0.3860 | 7 |

Table 5.14: $T_{dc}$ statistics aggregated over query terms for each query for the highest-performing queries for ClueWeb12 cat. B.

Comparing the mean $T_{dc}$ in the two tables, it can be seen that the mean $T_{dc}$ for the best-performing queries in ClueWeb09 cat. B. is substantially lower compared to the best-performing queries for ClueWeb12 cat. B. One reason for this, is that several of the best-performing queries for ClueWeb12 cat. B. contain function words, such as stop words, which, because stop words occur in most documents, have a (relatively) much higher $T_{dc}$ value. This difference is reflected in the standard deviations. For example, query 211 asks "`what is madagascar known for`" where the stop words *what, is* and *for* give rise to larger $T_{dc}$ values and a more skewed distribution. More generally, the queries of ClueWeb12 cat. B. contain more stop words than ClueWeb09 cat. B. and are substantially longer. For example, the TREC WebTrack 2014 queries for ClueWeb12 cat. B. contain four times the number of stop words than e.g. TREC WebTrack 2009 for ClueWeb09 cat. B.

### 5.6.3.2 The Effect of the First Normalisation Principle on Scoring

Because a low value of $T_{dc}$ (and, by extension, $T_{dc}^2$) produces a more uniform distribution, query terms having either low or high $\hat{f}_{t,d}$ will be scored similarly. For example, assume the $\hat{f}_{t,d}$ of query term $q_1$ and $q_2$ in document $d$ is 2 and 1,000 respectively. If both have a low $T_{dc}$, the less the difference in value by $-\log_2(M)$ despite the high difference in $\hat{f}_{t,d}$. Thus, for queries containing query terms having low $T_{dc}$, we can approximate Eqn. 5.39 by:

$$\begin{aligned} R(q,d) &= \sum_{t \in q \cap d} \left(-\log_2 M\right) \cdot (1 - P_2) \\ &\approx \sum_{t \in q \cap d} K \cdot (1 - P_2) \end{aligned} \quad (5.51)$$

where $K$ is a constant. This effectively means that a document's score relative to the query term is determined by $(1 - P_2)$. Recall, we used Laplace's law of succession in our experiments:

$$1 - P_2 = 1 - \frac{\hat{f}_{t,d}}{1.0 + \hat{f}_{t,d}} \quad (5.52)$$

meaning that as $\hat{f}_{t,d}$ increases, its informative content is resized at an almost exponential rate. Referring to Eqn. 5.51, this means that terms with low $\hat{f}_{t,d}$ will be resized less than terms with high $\hat{f}_{t,d}$. Figure 5.6 depicts Eqn. 5.52 and shows that the decrease in informative content is steepest for smaller values of $\hat{f}_{t,d}$. Referring back to Eqn. 5.51, this means that a document's score is based on the clusteredness or density of the query terms alone. This is problematic for at least two reasons: Firstly, consider the query $q = \{$`magnesium,rich,foods`$\}$ and two documents $d_1, d_2$ where the $\hat{f}_{t,d}$ for `magnesium`, `rich` and `foods` is 2, 1 and 1 respectively, and the $\hat{f}_{t,d}$ of `foods` in $d_2$ is 5. Because of the clusteredness, $d_2$ will be scored higher than $d_1$ using Eqn. 5.51 despite $d_1$ contains all query terms. Secondly, because this clusteredness is similar to tf in the tf-idf term weighting scheme, without the idf component discriminating between documents is based solely on the magnitude of a term's $\hat{f}_{t,d}$ meaning that spam documents and/or documents which do not contain any discriminative terms (discriminative according to idf for example) can heavily impact a document's score.

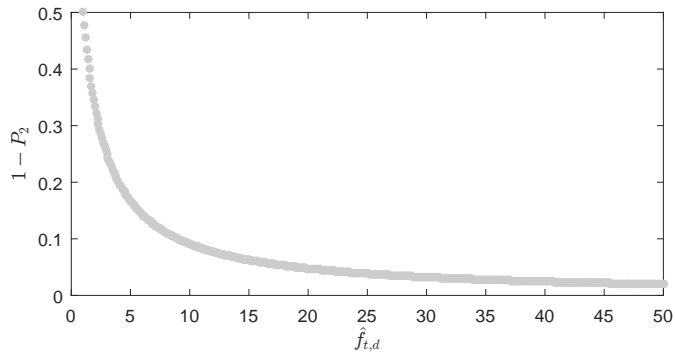

Figure 5.6: Behaviour of $1 - P_2$.

Because the $T_{dc}$ component for the best-performing queries of ClueWeb12 cat. B. is, on average, more skewed than the similar component for ClueWeb09 cat. B., the impact of $-\log_2(M)$ on the final scoring is larger. Especially stop words become problematic in this sense as (i) they occur in most documents, (ii) they tend to occur somewhat frequently in documents and (iii) their $T_{dc}$ is larger than non-function words. Combined with the skewed document distribution of the query terms, the $\hat{f}_{t,d}$ of stop words ends up contributing substantially to the overall score of the document. Looking at the best-performing queries for ClueWeb12 cat. B., we see that three out five queries contain one or more stop words. In contrast, none of the queries for ClueWeb09 cat. B. contains any stop words. Taking query 211 as an example, Table 5.15 shows the counts of the different query terms in the top-10 documents ranked by our Yule model using the $T_{dc}$ parameter value. The results are similar for the remaining queries.

| Pos | what | is | madagascar | known | for | doc. length |
|-----|------|-----|------------|-------|-----|-------------|
| 1 | 0 | 0 | 94 | 4 | 1 | 599 |
| 2 | 0 | 12 | 137 | 2 | 5 | 1,674 |
| 3 | 1 | 5 | 19 | 4 | 7 | 711 |
| 4 | 0 | 56 | 41 | 6 | 25 | 2,872 |
| 5 | 1 | 39 | 31 | 4 | 4 | 1,551 |
| 6 | 0 | 18 | 22 | 4 | 3 | 1,159 |
| 7 | 0 | 6 | 31 | 4 | 3 | 1,633 |
| 8 | 0 | 17 | 153 | 2 | 2 | 2,724 |
| 9 | 1 | 25 | 20 | 7 | 18 | 2,256 |
| 10 | 0 | 5 | 29 | 2 | 2 | 828 |

Table 5.15: Top-10 documents for query 211 and the frequency of the query terms in those documents.

Table 5.15 shows that short documents where term frequencies are high are placed at higher positions modulated by the document length. For example, the document ranked in position 3 contains a total of 36 query term occurrences, whereas the document in position 8 contains nearly five times that (174). However, because the latter document is nearly four times longer, each query term's $\hat{f}_{t,d}$ is substantially more dampened despite the ratio of query term to the document length is larger in the latter case. The impact of stop words can also be seen. For example, the documents in positions four and eight are of similar length, but the high frequencies for stop words in the former boosts it considerably in the ranking. The problem with the presence and frequency of stop words increases as we move down the ranking, where documents that are not judged relevant, are scored predominantly on the frequency of stop words. The higher up the ranking these documents occur the more detrimental the effect on the performance. Indeed, post-hoc inspection of the number of occurrences of discriminative words in the ranked documents of the queries for ClueWeb12 cat. B. shows a decrease in the number of informative terms as we move down the ranked list.

### 5.6.3.3   The Effect of Term Normalisation

An important distinction between $\hat{f}_{t,d}$ and $f_{t,d}$ is that the former is also based on the length of the document in which a term occurs. Recall that the $\hat{f}_{t,d}$ we use in our experiments is:

$$\hat{f}_{t,d} = f_{t,d} \cdot \log_2\left(1.0 + c\frac{\text{avg\_l}}{|d|}\right) \tag{5.53}$$

showing that a high $\hat{f}_{t,d}$ can occur if (i) both $f_{t,d}$ and $\log_2\left(1.0 + c\frac{\text{avg\_l}}{|d|}\right)$ is high, or if either of the two is high. Analysis of the distributions of $f_{t,C}$ of the query terms in the queries used for both ClueWeb09 and ClueWeb12 show that these are heavily skewed, meaning that most terms will have a low frequency. More importantly, the distributions of $|d|$ for the query terms are also skewed. However, because the average document length is calculated as the arithmetic mean, this means that most document (at least 73% of all documents according to our experiments) are below the average document length and, consequently, the $\hat{f}_{t,d}$ in most documents increases proportionally to $\log_2\left(1.0 + c\frac{\text{avg\_l}}{|d|}\right)$. This is further complicated by the value of $c$ which amplifies $\hat{f}_{t,d}$ the smaller the current document becomes relative to the average document length.

Figure 5.7 shows the optimal $c$ value for each of the three test-folds for each ranking model for ClueWeb09 cat. B. and ClueWeb12 cat. B (both containing stop words - the results for ClueWeb12 cat. B. without stop words is very similar). The figure shows that for ClueWeb09 cat. B., the optimal $c$ value is fairly uniform across all models. A high $c$ value boosts the impact of finding $f_{t,d}$ occurrences of $t$ in a short document relative to the average document length (see Eqn. 5.16). In contrast, the results for ClueWeb12 cat. B. show the optimal $c$ value to fluctuate more both between models and from test fold to test fold. For example, the optimal $c$ value for the $I_nL2$ model for the first two test folds is .5 whereas it was 6 for ClueWeb09 cat. B. These results indicate that for ClueWeb12 cat. B. the baseline ranking models are less reliant on boosting $\hat{f}_{t,d}$ through the logarithmic transformation, and instead consider documents closer to the average document length (see Eqn. 5.16). Thus, our YS ADR models are remarkably stable to variations of $c$, suggesting that $c$ does not need to be tuned for these ranking models on other collections in order to obtain the best performance. In contrast, for iSearch we observe (regardless of performance metric), that low values of $c$ gave almost consistently the best performance for all ranking models. This means that the impact of the current document length relative to the average document length, is less emphasised and instead scoring is biassed towards the raw term frequencies. This may be caused by the three distinct local maxima observed in the distribution of document lengths in iSearch (see Figure 4.9 in Chapter 4).

### 5.6.3.4   Retrieval Results

Our analysis shows that stop words may affect the performance of our ADR models. To test if the presence of stop words can help account for the differences in retrieval effectiveness between ClueWeb12 cat. B. and ClueWeb09 cat. B., we re-run the experiments on ClueWeb12 cat. B. where stop words have been removed. The results are shown in Table 5.16.

The removal of stop words seems to corroborate our analysis to some extent. Compared to the result when stop words are not removed for ClueWeb12 cat. B. (see Table 5.12), the perfor-mance of all metrics for all models decreases. This is expected as stop words have previously been reported to have a negligible effect on retrieval effectiveness. Identical to the results with stop words, none of the results are statistically significant. The standard DFR models outperforms the LMDir baseline more consistently compared to when stop words were not removed. The ERR@20 suggests that, on average, users stop inspecting the list of ranked documents at marginally higher positions when using the $I_n$L2-$T_{dc}$ compared to the other retrieval models. The biggest performance gain for the standard DFR models compared to the LMDir is for P@10, which increases by 0.01. For the information models, both Bpref and nDCG (both are deep precision measures) rarely beats the LMDir baseline. For Bpref, this suggests that the relative ranks of the relevant documents are higher for the LMDir baseline, and for nDCG that the ranking of the documents relative to the optimal ranking is slightly worse compared to the baseline. For all remaining performance measures, the information-based models outperform the baseline. Notice that because the information-based models do not rely on the first normalisation principle, a query that contains one or more stop words will automatically inflate the score of a document (roughly) proportionally to the number of tokens of these stop words in the document. This may account for why the information-based models had high retrieval effectiveness when using the ClueWeb12 cat. B. collection with stop words. However, as Table 5.12 indicates that including stop words tend to result in marginally higher retrieval effectiveness, considering stop words when scoring documents using the information-based models may be meaningful. Whether this increase in performance is sufficient compared to the increase in storage and query-processing costs must be determined on a case-by-case study. The YSL2 models tend to outperform the LMDir baseline on most performance measures regardless of the parameterisation used. Specifically, the YSL2-$T_{dc}^2$ model achieves the best performance for MAP, nDCG, P@10 and nDCG@10. While the improvements are nearly consistent, they remain, however, modest expect for P@10 and nDCG@10 where the improvement over the LMDir baseline is .017 and 0.0092 for the YSL2-$T_{dc}^2$, respectively. The use of stemming may further improve these results. Interestingly, the YSL2-$T_{dc}$ is also consistently better than the information-based models. Because, apart from the choice of statistical model, the use of the first normalisation principle is the only difference between YSL2-$T_{dc}$ and the information-based models, the use of the first normalisation principle warrants further research to gain a better understanding of how it works.

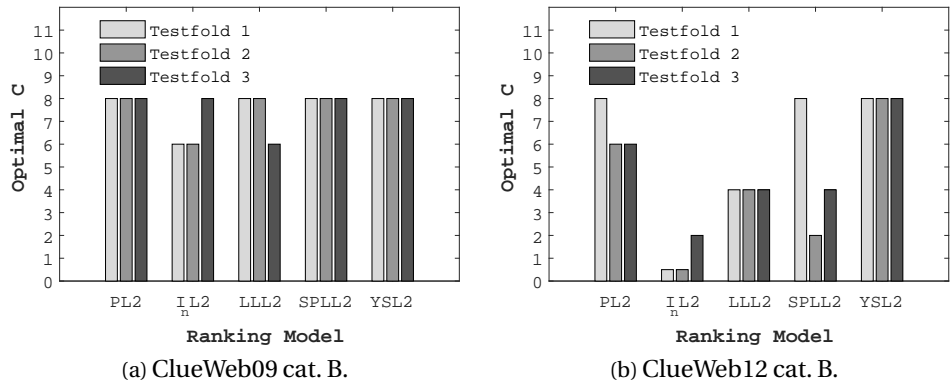

(a) ClueWeb09 cat. B.    (b) ClueWeb12 cat. B.

Figure 5.7: Optimal $c$ values for each of test fold and each ranking model. The PL2, $I_n$L2, LLL2 and SPLL2 all use the $T_{dc}$ parameterisation. The YSL2 uses the $T_{dc}^2$ parameter variation. The results were obtained by optimising for MAP.

| Model | ClueWeb12 cat. B. | | | | | |
|---|---|---|---|---|---|---|
|  | MAP | nDCG | Bpref | P@10 | ERR@20 | nDCG@10 |
| LMDir | .0294 | .1057 | .0842 | .2230 | .0826 | .1511 |
| PL2-$T_{tc}$ | .0292 | .1054 | .0844 | .2270 | .0806 | .1487 |
| PL2-$T_{dc}$ | .0299 | .1063 | .0844 | .2340 | .0826 | .1537 |
| $I_n$L2-$T_{tc}$ | .0304 | .1060 | .0832 | .2310 | .0907 | .1556 |
| $I_n$L2-$T_{dc}$ | .0307 | .1062 | .0842 | .2330 | **.0907** | .1559 |
| LLL2-$T_{tc}$ | .0299 | .1055 | .0837 | .2290 | .0847 | .1531 |
| LLL2-$T_{dc}$ | .0299 | .1055 | .0834 | .2260 | .0859 | .1531 |
| SPLL2-$T_{tc}$ | .0298 | .1054 | .0842 | .2320 | .0850 | .1554 |
| SPLL2-$T_{dc}$ | .0300 | .1067 | .0842 | .2330 | .0855 | .1561 |
| YSL2-$T_{tc}$ | .0296 | .1063 | .0839 | .2330 | .0857 | .1553 |
| YSL2-$T_{dc}$ | .0308 | .1077 | **.0845** | .2350 | .0879 | .1583 |
| YSL2-$T_{tc}^2$ | .0310 | .1077 | .0844 | .2400 | .0887 | .1596 |
| YSL2-$T_{dc}^2$ | **.0312** | **.1079** | .0843 | **.2400** | .0898 | **.1603** |

Table 5.16: Performance of all ranking models on ClueWeb12 cat. B. with stop words removed for all performance measures. All ranking models are tuned and use Laplace's law of succession (L) and logarithmic term frequency normalisation (2). Grey cells denote entries strictly larger than the LMDir baseline. Bold entries denote the best results. $^*$ denotes statistically significant difference from the LMDir baseline using a two-tailed $t$-test at the .05% level. All models are post-fixed with either $T_{tc}$ or $T_{dc}$ (see Section 5.5.4.1) (or their squared version) which denotes the parameter estimate for each model.

In summary, the YSL2 ADR models tend to outperform standard DFR models on several datasets and most performance measures. This suggests that using the best-fitting distribution to the collection term frequencies of non-informative terms is beneficial to retrieval using ranking models derived under the DFR framework. Several instantiated YSL2 ADR models perform on par or better than a tuned LMDir baseline and information-based ranking models though these results are not statistically significant, and collection dependent. Furthermore, the results must be viewed in light of the cascading methodology and its limitations.

## 5.7 Conclusion

In this chapter, we derived new ranking models that can adapt their computations to the statistics of the documents being retrieved. We presented a cascading methodology for doing so, specifically for the DFR framework, as it a powerful and flexible framework. These new ranking models, called *adaptive distributional ranking* models, or ADR, models, adapt to the distribution of non-informative terms, by determining the best-fitting statistical model of these terms on a per collection basis. Our cascading methodology first uses a simple approach to identify the non-informative terms from a collection, before deriving ADR models by integrating into DFR, the best-fitting statistical model to collection term frequencies of non-informative terms. The proposed ADR models are evaluated in an ad hoc retrieval task and compared to standard DFR ranking models, a recent extension of these called information models and a tuned query likelihood Dirichlet-smoothed language model on TREC disks 1 & 2, TREC disks 4 & 5 and ClueWeb09 cat. B and iSearch. The results show that the ADR models perform better than the standard DFR models, and on par with the information models baseline on most collections according to several performance measures. However, none of the ADR models statistically significantly outperform the language model. We next proposed a simple variation to parameter values of the ADR models and find this variation generally achieves the highest performance on the datasets and performance measures, though these new results are again not statistically significantly different from the language model baseline. We also find that our ADR models tend to be outperformed by all baselines on the ClueWeb12 cat. B. and iSearch collections. An analysis of this problem on ClueWeb12 cat. B. revealed that an attributing factor was the presence of stop words in the queries. By removing these stop words and repeating the retrieval experiment on ClueWeb12 cat. B., our ADR models were overall found to be the best performing.

# 6 Discussion

This chapter reviews the extent to which the research questions have been answered and then discusses the potential impact of the work to other areas in IR. This impact must be interpreted in light of the limitations in our work also discussed.

## 6.1 Review of Research Questions

This thesis is concerned with the following research objective:

*How can a principled approach to statistical model selection be used in IR?*

for which this thesis investigated these specific research questions:

1. **Research Question 1 (RQ1):** To what extent does commonly assumed statistical models fit common properties of IR datasets?

2. **Research Question 2 (RQ2):** Can the use of the correct statistical model improve ranking?

The first research question was addressed in Chapter 4, and the second in Chapter 5.

For RQ1, our results show that only 5/24 term frequency and 2/2 query frequency distributions can be said to be approximately power law-distributed. All remaining term frequency, document length, query length, citation frequency and syntactic unigram frequency distributions are better approximated by the inverse Gaussian, Generalized Extreme Value, negative binomial or Yule–Simon model. Hence, for RQ1, our results suggest that common properties of IR datasets are *not* always approximately power law-distributed. Our results also show that most statistical models can be fitted to a given dataset in a few seconds. These results can be further improved by using better hardware and optimisers.

For RQ2, we found that integrating the best-fitting statistical model - the Yule–Simon and power law - to collection term frequencies of non-informative terms into DFR ranking models, tended to perform better than standard (tuned) DFR ranking models on several collections and all performance measures. A small variation of the parameter of the Yule–Simon model produced ranking models that gave further performance improvements. However, compared to a strong baseline (tuned query likelihood language model with Dirichlet smoothing) and a recent class of information models, none of our proposed model performed statistically significantly better on any collection regardless of the performance measure. Furthermore, for ClueWeb12 cat. B, integrating the best-fitting statistical model, produced ranking models that did not outperform the standard DFR ranking models. An analysis of this problem on ClueWeb12 cat. B. revealed that an attributing factor was the presence of stop words in the queries. By removing these stop words and repeating the retrieval experiment on ClueWeb12 cat. B., our ADR models were overall found to be the best performing. Consequently, for RQ2, our results indicate that ranking models based on the best-fitting statistical model can improve ranking on a per collection basis, albeit not in a statistically significant way.

Taken together, we tentatively conclude that a principled approach to statistical model selection can be useful for IR, as it removes the need for making distributional assumptions and may offer moderate performance improvements to ranking for IR collections.

## 6.2 Impact

We see our principled approach to statistical model selection (and results) as a first step towards introducing statistical model selection as a common tool in IR. While some previous researchers in IR [31, 41] have used the original approach by Clauset et al. [117], the approach does not seem to have been widely adopted. In most cases (based on our review of the literature), this is because knowing the exact distribution is not used for any modelling or prediction task, but is rather seen as a perhaps interesting, but not essential part of the subject of investigation. However, this thesis has sought to argue that this task should be done as accurately as possible to avoid incorrect claims and, if used for modelling, to obtain as accurate a model as possible. Furthermore, some of our results demonstrated that using the correct statistical model in a specific retrieval framework can result in increased retrieval performance compared to a strong baseline.

In light of this, a number of areas in IR offer interesting avenues for investigating the use of principled statistical model selection. Due to the time constraints, these were not investigated in this thesis. For example, correct modelling of the distributions of scores of relevant and non-relevant documents can be highly useful [233] in e.g. information filtering and topic detection to set the threshold which separates relevant from non-relevant documents [30, 33, 125, 460]. However, because scores can vary wildly (see e.g. [233]) mixture models (such as multivariate Gaussian) could be considered for inclusion in our principled approach to quantify such fluctuations. Aggregation or grouping of the scores may also be required to reduce the noise, but how to do this is an open question. Other areas include dynamic pruning of documents to avoid scoring documents which cannot make it into the top-$k$ [283] ranking, for collection fusion [57], for selecting the optimal number of relevant documents to use in pseudo-relevance feedback for each query [340] and combining the outputs of multiple search engines [287]. In the latter case, the results of evaluating a query by multiple different IR systems must be fused. As these IR systems may use different ranking models which assign scores in different ranges, exact modelling of the score distributions becomes important as scores are mapped to probabilities. As this restricts all scores of all IR systems to the interval $[0, 1]$, the results from different IR systems can be combined or fused. Our principled approach may be useful here, as even small performance improvements for large scale IR system may translate to a reduction in cost or time [225]. Correct modelling of term frequency distributions may also be used to e.g. assign probabilities to unseen terms in language modelling [318]. Here, using the "correct" statistical model rather than relying on assumptions regarding the distribution prevents terms from receiving inflated or deflated probabilities which may impact the final ranking. However, as we have shown (see Section 4.4.3), the difference between some models is small and the question of whether there is any *practical* difference between them for this task is not known to us.

Our results may also have applications when creating IR test collections. As this is both expensive and time-consuming, previous research has attempted to synthesise new collections by approximating one or more statistical properties of real life collections [189]. For example, to synthesize a representative Web collection with a link structure similar to a large Web collection, Gurrin and Smeaton [189] require that the distribution of incoming and outgoing links should be approximately power law-distributed. These requirements, however, could not be met when synthesizing new collections from the .GOV collection. One reason might be that the power law is *not* the correct distribution to use when modelling the distribution of incoming and outgoing links. Thus, in synthesizing the statistical properties of both documents and queries, using the best-fitting distribution may lead to more representative samples.

Overall, we see several potential uses of the approach proposed in this thesis. However, our approach and results have a number of limitations that we discuss next.

## 6.3 Limitations

Our results lend support to previous findings in IR e.g. [49, 50, 318, 422], but may also contradict many others e.g. [31, 118, 103, 104, 272, 358, 380, 409]. Here, contradiction means that we could not test the approximate power law-distributions reported in previous published research as the datasets used are either proprietary, not in the open domain (or otherwise not eligible for sharing), not owned by the author or require heavy pre-processing - the exact details of which we could not obtain. Thus, whether our tentative conclusions hold for these datasets as well remains unknown. Another limitation is that our results are predicated on the set of statistical models we consider which could be e.g. substituted for another set of statistical models or augmented/reduced. Adding more and/or specialised statistical models could change our results; for example, Arampatzis and Kamps [31] showed that the distribution of query lengths is best fitted by a Poisson-Power law mixture model, which may provide a better fit than either the inverse Gaussian or negative binomial (see Section 4.4.5). Hence, while we state in the conclusion of Chapter 4 that our approach is effective, this claim is predicated on the set of statistical models used. Furthermore, the set of statistical models used consists of both discrete *and* continuous models. While this facilitates testing a wider range of statistical models, it may come at the expense of statistical accuracy [106] as continuous models should be discretised. Discretisation techniques (such as equal-width or equal-frequency intervals – see e.g. [192, 321]) may generate inappropriate division of the data as variations within intervals are ignored, and care must be taken if statistical properties of the original fitted distribution, such as its moments, are to be preserved. Hence, those of our results suggesting that a continuous model provides the best fit should be treated more carefully than a discrete model, as we have not investigated the consequences of comparing discrete and continuous statistical models. This was also a reason we chose to use the Yule–Simon (YS) and power law (PL) model in Chapter 5 instead of the Generalized Extreme Value. Finally, it is worth noting that our results are obtained using large samples. However, our approach to model selection can be very sensitive in the presence of small samples where minute perturbations or alternative selection procedure can result in another model [426]. Consequently, *model averaging* which resolves uncertainty regarding small samples by averaging model outputs to determine the best "ensemble" of statistical models to use [111], can be considered.

Our results from determining the best-fitting statistical model to the collection term frequencies of non-informative terms, showed that the YS, PL (iSearch only) and GEV were the best-fitting discrete and continuous model. In our experiments, we only used the YS on the basis that the data used was discrete. However, an important caveat is that scoring of terms (and, by extension, documents) using our ADR models is done using *normalised* term frequencies which (i) are continuous values and (ii) may not be best-fitted by a YS or PL model. In contrast, the GEV is not designed for discrete data (i.e. the raw collection term frequencies), but for continuous values. Using the GEV, however, requires three parameters to be estimated and prior work with ranking models [120, 346] involving statistical models with $k \geq 2$ parameters have fixed $k-1$ parameters and concentrates on a single parameter, with no intuition behind why this parameter is chosen. As we are not aware of any work that states which parameter of the GEV is most important, we refrained from adopting this approach. Post-hoc experiments using MLE to estimate the parameters of the GEV for the normalised term collection frequencies show, however, that the GEV obtains consistently better results compared to using the YS and PL model (whose parameter - either $p$ or $\alpha$ - is also estimated using MLE) on all datasets and all performance measures. Furthermore, the MLE approach for both models consistently performed worse when using the parameter settings in Table 5.6, suggesting that the models are sensitive to the parameter values in the statistical model.

Overall, there are several aspects of our work that potentially limit our results and their impact. These limitations make up several avenues for future work which we discuss in Chapter 7.

# 7 Conclusions

Assumptions regarding the distribution of properties of datasets are frequently made in information retrieval (IR). These assumptions are often made on the basis of mathematical convenience or ad hoc methods which may not accurately quantify the distributions, thus leading to sub-optimal modelling/solutions. As a practical alternative to such methods, this thesis investigated the extent to which a principled approach to statistical model selection can be useful to IR. The principled approach combines a widely used approach to parameter estimation with statistically principled tests to determine which statistical model, from a set of candidate statistical models, best quantifies, or fits, some data sample.

We investigated the merit of our principled approach from a theoretical and practical perspective. From the theoretical perspective, we used our principled approach to guide the derivation of new ranking models. These ranking models integrate the best-fitting statistical model to the collection term frequency distribution of non-informative terms on a per-collection basis. Experimental evaluation showed that these ranking models offered retrieval performance gains comparable to strong baselines across several collections. From the practical perspective, we investigated whether our approach could lend support to the distributional assumptions regarding properties of IR datasets that are often made in contemporary IR research. Specifically, we limited the scope of our investigation to power laws which are frequently reported in contemporary IR research. We find that, for most properties of the IR datasets investigated, such distributional assumptions cannot be corroborated by our principled approach.

## 7.1 General Conclusions

This thesis contributes knowledge on the topic of statistical model selection for IR in the following two ways:

1. New insights into the validity and limitations of commonly held assumptions about how properties of IR datasets are distributed. Specifically, we contribute knowledge that many commonly held assumptions regarding the distribution of properties in IR datasets, cannot be supported when using our principal approach to statistical model selection.

2. New ranking models that adapt to the distribution of the data they process. Specifically, we contribute knowledge that because these ranking models "closely" quantify the distribution of the data, they (i) are optimised for individual collections and (ii) better conform to the theory on which they are based and (iii) can be used to improve ranking.

Taken together, our conclusions indicate that principled model selection in IR is an area worth of more research, as outlined below.

## 7.2 Research Directions for Future Work

Below are given possible research directions for future work. We include both short-termed and long-termed directions.

### 7.2.1 Discretising Continuous Models

The results presented in this thesis are directly linked to the statistical models used in our principled approach. However, because this set consists of both discrete and continuous statistical models and because data in IR are typically discrete, (i) fitting a continuous model to discrete data and (ii) comparing a continuous and discrete model using the statistical tests in our approach is theoretically flawed. Consequently, before applying our principled approach to other areas of IR, we should have discrete versions of all continuous models used. While such "versions" would inevitably be approximations, the benefit is that the same set of models could be used independently of whether the data was continuous or discrete. A recent paper by Chakraborty [101] offers discrete versions to several of the continuous statistical models we considered. Once these new sets of discretised models are created, the results in this thesis would need to be re-examined to validate our findings.

### 7.2.2 Automatic Selection of ADR Models

Our results in Chapter 5 show that despite adapting to each collection, our ADR models are not consistently the best performing. For example, on ClueWeb12 cat. B. our ADR models are almost consistently outperformed by the baselines (when stop words are not removed), whereas on ClueWeb09 cat. B, our ADR models are almost consistently the best. This is interesting as the ADR models and the baselines are derived from the same basic framework, with only relatively minor differences between them. For example, the only difference between the standard DFR ranking models used here and the ADR models is the choice of the statistical model used. This suggests that some term/document/collection statistics change substantially from e.g. ClueWeb09 cat. B. to ClueWeb12 cat. B (and/or iSearch), which our ADR models cannot accommodate despite adapting to processed data. Identifying the exact term/document/collection statistics that affect which ADR model is found to be the best for a given collection, would facilitate selecting the best possible ADR model *without* having to identify the non-informative terms and determine the best-fitting statistical model to the distribution of these.

### 7.2.3 Statistical Dispersion for Term Weighting

Scoring documents based on the best-fitting statistical model on a per-term basis, may be done using statistical dispersion which calculates the deviation of observations from an expected value (e.g. the mean, mode or median). That is, given a term $t$ and a list of within-document frequencies $X_t = \{f_{t,d_i}\} : i = 1,...,n$ of $t$ in some collection, the statistical dispersion may be defined as [305]:

$$SD_t = \mathbb{E}[\{x - \mathbb{E}(X_t) | x \in X_t\}] \qquad (7.1)$$

A term with high dispersion is one with high variability which, according to Shannon [388], is proportional to the uncertainty and hence the informativeness of the term. Assuming the expectation is e.g. the mean, knowing the "correct" statistical model is important to accurately model the variability of each term. Notice, that Eqn. 7.1 requires two applications of our principled approach to account for the two expectations.

### 7.2.4 Statistical Models of Non-Informative Terms

In deriving our ADR models, we found the Yule–Simon, power law and Generalized Extreme Value model to provide the best fits to the distribution of collection term frequencies of non-informative terms. These were the *same* models found by [344] when using all terms in the collection. There are at least two reasons for this. Firstly, our results are predicated on the set of models used (as discussed previously). Ignoring the difference between discrete and continuous models, it might be that the set if not sufficiently large. For example, only five of the models we consider are discrete and only two of those can be said to quantify "extreme" distributions. Thus, that the Yule–Simon, power law and Generalized Extreme Value model are found may be caused by lack of options, rather than by choice. Including additional models may resolve this to a degree (see also Section 7.2.1).

### 7.2.5 Distribution-specific Adjacency Labelling Schemes

The increasing need to process and store large-scale graphs in machine learning and data mining presents a critical challenge [182]. An adjacency labelling scheme assigns a bit string - a label - to each vertex so that a query between any two vertices can be deduced solely from their respective labels, and can be used (i) to store a graph implicitly in a distributed manner, (ii) in XML search engines [122], (iii) mapping services [3] and (iv) internet routing [250]. The objective is to minimise the size of the largest label. Power-law graphs, where vertices are assumed to be perfectly distributed by a power law distribution, are one particular family of graphs which have been extensively studied (see e.g. [117, 314] for an overview). Recent findings [343], however, show that for empirical graph-based datasets that are not power law distributed (but assumed to be so), the maximum label size is severely overestimated relative to the theoretical size. As a result, it would be interesting to investigate distribution-specific adjacency labelling schemes as they may result in superior performance for e.g. storing and processing large graph-based datasets.

# A Paper Abstracts

## A.1    Power Law Distributions in Information Retrieval

**Power Law Distributions in Information Retrieval**
Casper Petersen and Jakob Grue Simonsen and Christina Lioma
*ACM Transactions on Information Systems*,34(2):1–37,2016

Several properties of information retrieval (IR) data, such as query frequency or document length, are widely considered to be approximately distributed as a power law. This common assumption aims to focus on specific characteristics of the empirical probability distribution of such data, e.g. its "scale-free nature" or its long/fat tail. This assumption however may not be always true. Motivated by recent work in the statistical treatment of power law claims, we investigate two research questions: (1) To what extent do power law approximations hold for term frequency, document length, query frequency, query length, citation frequency and syntactic unigram frequency? (2) What is the computational cost of replacing ad hoc power law approximations with more accurate distribution fitting? We study 23 TREC and 5 non-TREC datasets and compare the fit of power laws to 15 other standard probability distributions. We find that query frequency and 5 out of 24 term frequency distributions are best approximated by a power law. All remaining properties are better approximated by the Inverse Gaussian, Generalized Extreme Value, Negative Binomial or Yule distribution. We also find the overhead of replacing power law approximations by more informed distribution fitting to be negligible, with potential gains to IR tasks like index compression or test collection generation for IR evaluation.

## A.2    Adaptive Distributional Extensions to DFR Ranking

**Adaptive Distributional Extensions to DFR Ranking**
Casper Petersen and Jakob Grue Simonsen and Kalervo Järvelin and Christina Lioma
*The 25th ACM International Conference on Information and Knowledge Management, 4 pages, (in press), 2016*

Divergence from Randomness models use probability distributions on non-speciality terms to produce rankings of search results. However, the choice of distribution and its parameters may significantly influence the effectiveness of the ranking model. We propose Adaptive

Distributional Ranking: a novel approach to select and exploit probability distributions to derive Divergence from Randomness models that are both theoretically principled and exhibit superior retrieval effectiveness. We evaluate the new DFR ranking models against the standard DFR ranking models, information ranking models, and a tuned language model baseline on three TREC collections. Our results show that our method (i) outperforms all DFR baselines, and (ii) produces results on par or better than the tuned language models baseline for all measures on several TREC and a ClueWeb collection.

## A.3   Brief Announcement: Labeling Schemes for Power-Law Graphs

**Brief Announcement: Labeling Schemes for Power-Law Graphs**
Casper Petersen and Noy Rotbart and Jakob Grue Simonsen and Christian Wulff-Nielsen
*Proceedings of the ACM Principles on Distributed Computing (PODC),(1):39-41, 2016*

A condensed version of the paper in Section A.4.

## A.4   Near Optimal Adjacency Labeling Schemes for Power-Law Graphs

**Near Optimal Adjacency Labeling Schemes for Power-Law Graphs**
Casper Petersen and Noy Rotbart and Jakob Grue Simonsen and Christian Wulff-Nielsen
*Proceedings of the 43rd International Colloquium on Automata, Languages, and Programming (ICALP),(2):564-574, 2016*

An adjacency labeling scheme labels the $n$ nodes of a graph with bit strings in a way that allows, given the labels of two nodes, to determine adjacency based only on those bit strings. Though many graph families have been meticulously studied for this problem, a non-trivial labeling scheme for the important family of power-law graphs has yet to be obtained. This family is particularly useful for social and web networks as their underlying graphs are typically modelled as power-law graphs. Using simple strategies and a careful selection of a parameter, we show upper bounds for such labeling schemes of $O\left(\sqrt[a]{n}\right)$ for power law graphs with coefficient $\alpha$, as well as nearly matching lower bounds. We also show two relaxations that allow for a label of logarithmic size, and extend the upper-bound technique to produce an improved distance labeling scheme for power-law graphs.

## A.5   Deep Learning Relevance: Creating Relevant Information (as Opposed to Retrieving it)

**Deep Learning Relevance: Creating Relevant Information (as Opposed to Retrieving it)**
Christina Lioma, Birger Larsen, Casper Petersen and Jakob Grue Simonsen
*Proceedings of the 1st international Workshop on Neural Information Retrieval (Neu-IR) hosted by the 39th ACM SIGIR conference on research and development in information retrieval. 6 pages, 2016*

What if Information Retrieval (IR) systems did not just retrieve relevant information that is stored in their indices, but could also "understand" it and synthesise it into a single document? We present a preliminary study that makes a first step towards answering this question. Given a query, we train a Recurrent Neural Network (RNN) on existing relevant information to that query. We then use the RNN to deep learn a single, synthetic, and we assume, relevant document for that query. We design a crowdsourcing experiment to assess how relevant the deep learned document is, compared to existing relevant documents. Users are shown a query and four wordclouds (of three existing relevant documents and our deep learned synthetic document). The synthetic document is ranked on average most relevant of all.

## A.6 Exploiting the Bipartite Structure of Entity Grids for Document Coherence and Retrieval

**Exploiting the Bipartite Structure of Entity Grids for Document Coherence and Retrieval**
Christina Lioma and Fabien Tarissan and Jakob Grue Simonsen and Casper Petersen and Birger Larsen
*Proceedings of the 2nd ACM SIGIR International Conference on the Theory of Information Retrieval (ICTIR),3(4):87-96,2016*

Document coherence describes how much sense text makes in terms of its logical organisation and discourse flow. Even though coherence is a relatively difficult notion to quantify precisely, it can be approximated automatically. This type of coherence modelling is not only interesting in itself, but also useful for a number of other text processing tasks, including Information Retrieval (IR), where adjusting the ranking of documents according to both their relevance and their coherence has been shown to increase retrieval effectiveness. The state of the art in unsupervised coherence modelling represents documents as bipartite graphs of sentences and discourse entities, and then projects these bipartite graphs into one–mode undirected graphs. However, one–mode projections may incur significant loss of the information present in the original bipartite structure. To address this we present three novel graph metrics that compute document coherence on the original bipartite graph of sentences and entities. Evaluation on standard settings shows that: (i) one of our coherence metrics beats the state of the art in terms of coherence accuracy; and (ii) all three of our coherence metrics improve retrieval effectiveness because, as closer analysis reveals, they capture aspects of document quality that go undetected by both keyword-based standard ranking and by spam filtering. This work contributes document coherence metrics that are theoretically principled, parameter-free, and useful to IR.

## A.7 Entropy and Graph Based Modelling of Document Coherence using Discourse Entities: An Application to IR

**Entropy and Graph Based Modelling of Document Coherence using Discourse Entities: An Application to IR**

Casper Petersen and Christina Lioma and Jakob Grue Simonsen and Birger Larsen

*Proceedings of the 1st ACM SIGIR International Conference on the Theory of Information Retrieval (ICTIR),*1(1):191-200,2015

We present two novel models of document coherence and their application to information retrieval (IR). Both models approximate document coherence using discourse entities, e.g. the subject or object of a sentence. Our first model views text as a Markov process generating sequences of discourse entities (entity n-grams); we use the entropy of these entity n-grams to approximate the rate at which new information appears in text, reasoning that as more new words appear, the topic increasingly drifts and text coherence decreases. Our second model extends the work of Guinaudeau & Strube [188] that represents text as a graph of discourse entities, linked by different relations, such as their distance or adjacency in text. We use several graph topology metrics to approximate different aspects of the discourse flow that can indicate coherence, such as the average clustering or betweenness of discourse entities in text. Experiments with several instantiations of these models show that: (i) our models perform on a par with two other well-known models of text coherence even without any parameter tuning, and (ii) reranking retrieval results according to their coherence scores gives notable performance gains, confirming a relation between document coherence and relevance. This work contributes two novel models of document coherence, the application of which to IR complements recent work in the integration of document cohesiveness or comprehensibility to ranking [60, 418]

## A.8 The Impact of using Combinatorial Optimisation for Static Caching of Posting Lists

**The Impact of using Combinatorial Optimisation for Static Caching of Posting Lists**

Casper Petersen and Jakob Grue Simonsen and Christina Lioma

*Proceedings of the Asian Information Retrieval Societies (AIRS),*9460(3):420-425, 2015

Caching posting lists can reduce the amount of disk I/O required to evaluate a query. Current methods use optimisation procedures for maximising the cache hit ratio. A recent method selects posting lists for *static* caching in a greedy manner and obtains higher hit rates than standard cache eviction policies such as LRU and LFU. However, a greedy method does not formally guarantee an optimal solution. We investigate whether the use of methods guaranteed, in theory, to find an approximately optimal solution would yield higher hit rates. Thus, we cast the selection of posting lists for caching as an integer linear programming problem and perform a series of experiments using heuristics from combinatorial optimisation (CCO) to find optimal solutions. Using simulated query logs we find that CCO yields comparable results to a greedy baseline using cache sizes between 200 and 1000 MB, with modest improvements for queries of length two to three.

## A.9    Comparative Study of Search Engine Result Visualisation: Ranked Lists Versus Graphs

Typically search engine results (SERs) are presented in a *ranked list* of decreasing estimated relevance to the user query. While familiar to users, ranked lists do not show any inherent connections between SERs, e.g. whether SERs are hyperlinked or authored by the same source. Such potentially useful connections between SERs can be displayed as *graphs*. We present a preliminary comparative study of ranked lists versus graph visualisations of SERs. Experiments with TREC data from the domain of web search and a small user study of 10 participants show that ranked lists result in more precise and also faster search sessions than graph visualisations.